# What can one learn about Fe-Cr alloys using Mössbauer spectroscopy?

Stanisław M. Dubiel[*]

Faculty of Physics and Applied Computer Science, AGH University of Science and Technology, Al. Adama Mickiewicza 30, PL-30-059 Kraków, Poland

## Abstract

Applications of the Mössbauer spectroscopy (MS) in the investigation of Fe-Cr alloys are reviewed. A high sensitivity of the hyperfine magnetic field to the presence of Cr atoms in the vicinity of the probe Fe atoms permits quantitative investigation of various aspects related both to the crystallographic as well as to the magnetic phase diagram of this alloy system. Concerning the former, presented is the relevance of MS for determining borders of the miscibility gap and kinetics of the phase decomposition, distinguishing between nucleation and growth and spinodal decomposition, identifying the sigma-phase and studying kinetics of its precipitation. Regarding the magnetic phase diagram, MS is useful for determining the Curie, the Néel and the spin-freezing temperature, hence studying paramagnetic→ferromagnetic, paramagnetic→antiferromagnetic and paramagnetic→spin-glass transitions. An effect of different heat treatments, strain and irradiation with various particles on a distribution of Cr atoms in the Fe matrix is demonstrated, too. For $\alpha$-FeCr alloys relevance of MS for determining changes in spin and charge densities at Fe-sites induced by neighboring Cr atoms is illustrated, as well as its usefulness in studying changes caused by a high-temperature sulphidation and oxidation. Concerning properties of $\sigma$-FeCr alloys the application of MS for determining the Curie and Debye temperature is reviewed. Application of MS to study an effect of magnetism on the lattice dynamics of Fe atoms in $\sigma$-FeCr is also exemplified. Last but not least, determining a magnetic texture and mechanical alloying is depicted.

*Stnislaw.Dubiel@fis.agh.edu.pl

# Table of Contests

## 1. Introduction

Fe-Cr alloys have been intensively investigated both experimentally and theoretically. The great interest in these alloys has been stimulated, on one hand, by their magnetic properties, and on the other hand, by their industrial importance. Concerning the former, magnetism of elements constituting this alloy i.e. ferromagnetism (FM) of Fe and ant ferromagnetism (AFM) of Cr are in opposition to each other. Substituting Cr into Fe-matrix gradually weakens the FM state what is reflected in a monotonic decrease of the Curie temperature, $T_C$. Adding Fe into the Cr-matrix causes a gradual drop of the Néel temperature, $T_N$. The question how does the magnetic phase diagram looks like in the range where the two form of magnetism disappear remained open for many years. Three scenarios were considered: (1) there was a critical concentration at which $T_C = T_N$, (2) there were two critical concentrations of Fe, $x_{FM}$, at which $T_C= 0K$, and $x_{AFM} < x_{FM}$ at which $T_N = 0$ K i.e. a paramagnetic (PM) state existed down to T=0 K, and (3) $T_C = T_N$ at T > 0 K i.e. there was a crossover range below the crossing temperature in a certain composition range. Its magnetic state was not clear. Consequently, in late 1970's and early 1980's the scientific interest in the magnetism of the $Fe_{100-x}Cr_x$ alloy system was focused on investigation of samples with compositions covering the puzzling range i.e. ~70 ≤ *x* ≤ ~90. Neutron diffractions and low temperature magnetization measurements carried out by Burke and co-workers constituted a milestone in solving the issue [1,2]. The authors revealed a spin-glass (SG) state for 81 ≤ *x* ≤ 84 i.e. in that composition range a PM→SG transition occurred with the spin-freezing temperature ranging between ~22 and ~28 K [2]. Furthermore, they found evidence on spin-clusters freezing in the AF phase for *~85 ≤ x ≤ ~90* and determined the freezing temperature [1]. This double-transition (PM→AFM→SG). phenomenon is termed as the re-entrant transition (PM→AFM→SG) was also observed by means of the Mössbauer spectroscopy in a Fe-Cr alloys containing 86.5 at.% Cr [3]. Concerning the low-temperature magnetism of the Fe-Cr alloys with the Cr concentration lower than the critical value of 81 at.%, it was regarded as "complex", because of such effects like pronounced curvature in Arrots plots [4], strongly field dependent magnetization [5], field cooling effects [6] or disappearance of well-defined spin-wave excitations al low temperatures [7,8]. Several propositions were put in order to explain the effects like: existence of "giant magnetic clusters" [9] or superparamagnetic clusters [6]. In turn, Nieuwenhyus et al. suggested that the Fe-Cr alloys with the concentration close to the critical one (~19 at.% Fe) were ferromagnetic at high temperature but underwent a transition to a spin-glass state as the temperature was decreased [10]. In other words, they postulated a double (re-entrant) transition PM→FM→SG. This idea has turned out to be correct as proves the presently published phase diagram [11]. Also the in-field Mössbauer spectroscopy contributed to this interpretation [12,13]. Should additionally be mentioned that the low temperature magnetic structure of the Fe-Cr alloys containing ~5 to 31.5 at.% Fe was studied by neutron diffraction and magnetic methods [14]. The magnetic phase diagram constructed by these authors resembled the presently accepted one as far as the two double transitions viz. from the ferromagnetic and from the antiferromagnetic phase to a ground magnetic state is concerned. However, the ground state on the FM side was called "asperomagnetic" and that on the AFM side was named "antiasperomagnetic". Furthermore, a spin-glass state was admitted to exist in a very narrow concentration range between the two ground states [14].

Technological interest in the Fe-Cr alloys follows from their importance in the steel making industry. Since these alloys containing ~10 at.% Cr or more do not corrode, the alloys constitute the major ingredient of various types of stainless steels (austenitic, ferritic and duplex). Enhanced resistance to even a high temperature corrosion, excellent mechanical properties like strength, good weldability and exceptional resistivity to a neutron swelling cause that Fe-Cr – based steels have been used as excellent construction materials for various devices (boilers, heat-exchanges, pipes, etc.) that work at service not only at elevated temperatures but also in aggressive environments. In such conditions their useful properties deteriorate as a result of aging and/or irradiation damage. In particular they become brittle and more prone to corrosion. The reasons for the embrittlement can be understood in terms of the crystallographic phase diagram of Fe-Cr viz. the existence of the so-called miscibility gap (MG) and that of the sigma ($\sigma$) phase field. The former is related with the lack of solubility between Fe and Cr in a wide concentration range (~10-90 at.% Cr) and at temperatures lower than ~800 K. This reason for the brittleness is known as the "475$^o$ C embrittlement". Annealing at temperatures between ~800 K and ~1100 K may result in precipitation of $\sigma$. Numerous experimental and theoretical studies have been devoted to the issue and, as a result, various versions of the crystallographic phase diagram were proposed [15 – 19]. The emphasis has been rather focused on the Fe-rich side as industrially more relevant. However, a wide dispersion of the experimental data does not allow to uniquely distinguishing between various versions of the predictions [20,21] what justifies and stimulates further experimental investigations. Noteworthy, the majority of the experimental data concerning the borders of MG were delivered by the Mössbauer spectroscopy and the atom probe tomography [20-22]. Precipitation of $\sigma$ has also detrimental effect on mechanical and corrosive properties of the Fe-Cr alloys and therefrom produced steels. Consequently, a great body of relevant papers can be found in the literature. Mössbauer spectroscopy was applied frequently as the relevant techniques, in particular to study the effect of temperature, composition and addition of a third element on the kinetics of the $\alpha \rightarrow \sigma$ phase transformation.

This paper reports the author's choice of the applications of MS in the study of Fe-Cr alloys. Presented and discussed are both issues relevant to the magnetic and crystallographic phase diagrams, as well as those depicting physical properties of these alloys e. g. the effect of Cr atoms on Fe-site spin- and charge-densities, Debye temperature and effect of magnetism on lattice dynamics. Also addressed is the effect of high-temperature corrosion on the Fe-Cr alloys.

## 2. Mössbauer spectroscopy

Mössbauer spectroscopy (MS) is technique based on the Mössbauer effect [23]. Its extremely high resolution permits studying hyperfine interactions i.e. interactions between nucleus and electrons. Changes in energy levels caused by these interactions are of the order of $10^{-7} - 10^{-9}$ eV. Since the Mössbauer's discovery in 1957, MS has proved to be a very useful experimental tool in various branches of science like solid state physics, metallurgy, mineralogy, biophysics to name just few. As a resonance method its applications are limited to isotopes at which the effect occurs e. g. $^{57}$Fe, $^{119}$Sn, $^{153}$Eu and investigated samples have to be a solid state. The Hamiltonian of the hyperfine interactions, $\mathcal{H}$, consists of three terms:

$$\mathcal{H} = \mathcal{E}_0 + \mathcal{E}_2 + \mathcal{M}_2 \qquad (1)$$

Where $\mathcal{E}_0$ represents the Coulomb interaction between the nuclear charge and the and that of electrons, $\mathcal{E}_2$ is the interaction between the electric quadrupole moment of the nucleus and an electric-field gradient at the nuclear site, and $\mathcal{M}_2$ describes the dipole magnetic interaction between the magnetic moment of the nucleus, **μ**, and the magnetic field at the nuclear site, ***B,*** called the hyperfine magnetic field.

In a Mössbauer spectrum each of the type of the hyperfine interactions is represented by a corresponding spectral parameter viz. isomer shift, *IS*, for $\mathcal{E}_0$, quadrupole splitting, *QS*, for $\mathcal{E}_2$ and hyperfine field, *B*, for $\mathcal{M}_2$.

The magnitude of *IS* is given by the following formula:

$$IS \propto [R_e^2 - R_g^2][\rho_a(0) - \rho_s(0)] \qquad (2)$$

Where $R_e$ and $R_g$ are the radii of the excited and the ground-state of the nucleus, respectively, and $\rho_a(0)$ and $\rho_s(0)$ are the densities of s-like electrons at nucleus in the absorber and in the source of the $\gamma$-rays, respectively.

In the case of the $^{57}$Fe isotope, $R_e < R_g$, and consequently the value of *IS* is negative if $\rho_a(0) > \rho_s(0)$. It is customary to refer to isomer shifts relative to *IS* of metallic iron whose isomer shift as an absorber is arbitrarily assigned the value zero.

The isomer shift is the spectral parameter characteristic of each Mössbauer spectrum, and its change measures an effective variation of the charge-density of s-like electrons at nucleus (the value of *IS* may change not only due to a real change of $\rho_a(0)$, but also due to a change of the density of *d*-like electrons – so-called screening effect). If the hyperfine monopole interaction is the only one in an absorber, the Mösbauer spectrum has a form of a single line called singlet – see Fig. 1a.

The quadrupole splitting, *QS*, is characteristic of samples whose crystallographic symmetry is lower than regular. This type of the hyperfine interaction gives an opportunity to detect changes in crystal structure, local atomic arrangement and lattice defects. As a consequence of the quadrupole interaction energy levels of the ground and the excited nuclear states can be split. In the case of the Mössbauer effect observed with the $^{57}$Fe isotope the simplest spectrum has a form of two lines (doublet) with equal intensities – see Fig. 1b. In absorbers with the regular crystal structure this kind of the hyperfine interactions is absent (no electric field gradient).

In magnetic samples or nonmagnetic ones placed in an external magnetic field, the ground and the excited nuclear energy levels are equally split (so-called Zeeman Effect). The 57Fe site Mössbauer spectrum in this case consists of six equally-spaced six lines (sextet) having the mirror symmetry. For isotropic absorbers the relative ratio between the lines 1, 2 and 3 is equal to 3:2:1 – see Fig. 1c.

This kind of the hyperfine interaction is useful for observation magnetic and crystallographic phase transitions, determining transitions temperatures, studying an effect of local atomic environment and magnetic texture.

The hyperfine field, **B**, has several contributions, but in the case of iron and its alloys, the dominant term is the so-called Fermi contact term, **B$_s$**, whose magnitude can be expressed by the following formula:

$$B_s \propto \sum_{k=1}^{4}[\rho(s_k^\uparrow) - \rho(s_k^\downarrow)] \qquad (3)$$

Where $\rho(s_k^\uparrow)$ and $\rho(s_k^\downarrow)$ are the spin-densities of s-like electrons at the nuclear site with the spin up (↑) and spin down (↓), respectively. The difference in these densities concerns both the core s-like electrons (1s, 2s, 3s) as well as the conduction electrons (4s). In other words, a change of the hyperfine field, *ΔB,* implies an effective change of the s-like electrons spin-density at the nuclear site.

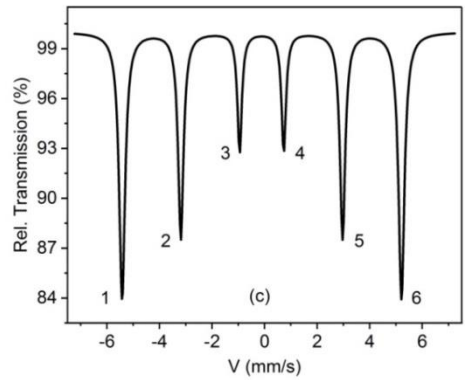

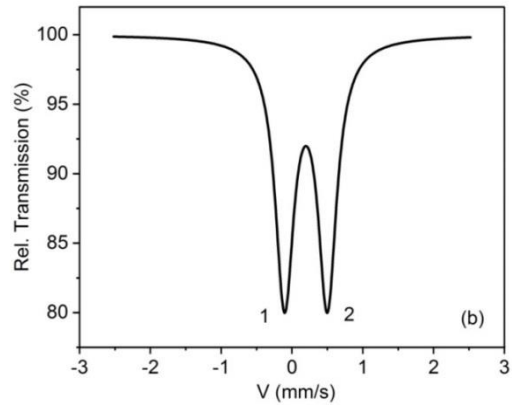

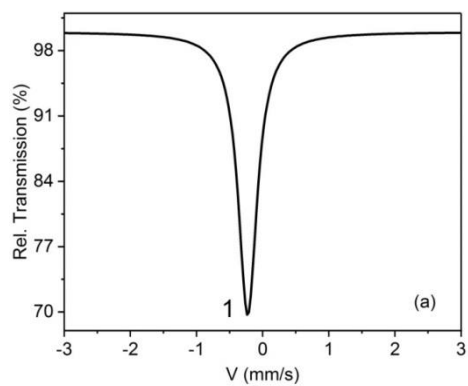

Fig. 1 Schematic picture of the $^{57}$Fe-site Mösbauer spectra representing the three types of the hyperfine interactions: (a) monopole electric, (b) quadrupole electric and (c) dipole magnetic.

Thus the whole Fermi contact term can be thought of as a sum of two contributions viz. one due to polarization of the of the core electrons spins, **$B_{CP}$**, and second due to polarization of the conduction electrons spins, **$B_{CEP}$**. Consequently, one can treat the hyperfine field as a sum of two terms:

$$\boldsymbol{B} = \boldsymbol{B}_{CP} + \boldsymbol{B}_{CEP} \tag{4}$$

According to various theoretical calculations $B_{CP} < 0$ while $B_{CEP} > 0$. As $B < 0$, the dominant contribution comes from the $B_{CP}$ term [24-27].

## 3. Effect of Cr atoms on the hyperfine field

The first topic studied in the Fe-Cr alloys with MS depicted the influence of Cr atoms present in the vicinity of Fe atoms on the Fe-site hyperfine field, *B*, and on the isomer shift, *IS* [28-32]. From these studies it was evident that the value of *B* was reduced by Cr atoms relative to its value in a pure Fe. The question to be solved was how many coordination shells had to be taken into account and what was the magnitude of the reduction of *B* by Cr atoms situated in different neighbor shells. According to Stearns [28] the effect of Cr atom in the third coordination shell (3NN) was significantly smaller than that in the first (1NN) and the second (2NN) shell. Consequently, a two-shell model (1NN+2NN) had been adopted in the analysis of the spectra. Another argument in favor of such model was the fact that the maximum number of possible atomic configurations is equal to 63, whereas in the case of the three-shell model (1NN+2NN+3NN) this number increases to 812! In practice, one takes into account only these configurations, (m,n), whose probability expected for the random distribution of Cr atoms, P(m,n) ≥ ~1% (m - number of Cr atoms in 1NN, n – number of Cr atoms in 2NN). This condition significantly reduces the number of the configurations to be considered, hence the number of free parameters. For example, for the Fe$_{100-x}$Cr$_x$ alloy with x=1.3, 3.3 and 4.8 it is enough to take into account, 3, 6, 7 configurations, respectively. The results reported in [28-32] were not in line with each other, as they could be classified into three following groups:

(a) $\Delta B_1 > \Delta B_2$, (b) $\Delta B_1 = \Delta B_2$, (c) $\Delta B_1 < \Delta B_2$ where $\Delta B_1 = B(1,0) - B(0,0)$ and $\Delta B_2 = B(0,1) - B(0,0)$.

For example according to Wertheim [30] $R = \Delta B_1 / \Delta B_2 = 1.19$, according to Cranshaw R=1.42 [31]. On the other hand, Sauer and Reynik [29] and also Vincze and Grüner [32] postulated that R=1. Finally, Stearns found R=0.77 – 0.90, depending on the Cr content [28].

The main reason for the lack of consistence in the R-value follows from the fact that all these authors investigated so-called diluted Fe-Cr alloys i.e. with a low content of Cr. In binary Fe-A alloys with a low concentration of A atoms, *x*, P(m,n;x) – values do not differ much for similar values of m and n. In addition, the actual distribution of Cr atoms might be different in Fe-Cr samples studied by different authors, so the assumption on randomness made in the fitting procedure might not be true. Consequently, good quality fits were achieved with different values of *B(m,n)*, what resulted in different R.

Dubiel and co-workers carried out MS studies on a series of $Fe_{100-x}Cr_x$ alloys with 1.3 ≤ x ≤ 14.9 [33,34]. The recorded spectra, some of them are displayed in Fig. 2, were analyzed with four models viz. (a) M12R, (b) M12. (c) M1=2 and (d) M123.

In models M12 and M12R it was assumed that only Cr atoms situated in the first two coordination shells had measurable effect on *B*. The only difference between them was that in M12R the random distribution of Cr atoms in the iron matrix was assumed i.e. the probabilities of different atomic configurations, P(m,n;x) were calculated from the binomial distribution and kept constant in the fitting procedure. In the M12 model the P(m,n;x) - values were treated as free parameters. The M1=2 was also the two-shell model but with an assumption that $\Delta B_1 = \Delta B_2$. Finally, in M123 the first three coordination shells were taken into account in order to find out the value of $\Delta B_3$. The quality of all fits was measured in terms of the $\chi^2$ – test. It had turned out that all spectra could have been well and consistently fitted with models M12 and M12R, with slightly lower values of $\chi^2$ for M12. This means that the distribution of Cr atoms in the studied alloys war pretty random. For all fits $\Delta B_1 > \Delta B_2$ was found and their ratio, R, is shown in Fig. 3. Noteworthy, the values of $\Delta B_1$ and $\Delta B_2$ turned out to hardly depend on the Cr concentration and their average values were: $\langle \Delta B_1 \rangle$ = - 31.7 kGs and $\langle \Delta B_2 \rangle$ = - 22.2 kGs for the analysis with M12 and $\langle \Delta B_1 \rangle$ = - 31.5 kGs and $\langle \Delta B_2 \rangle$ = - 22.1 kGs for M12R [33,34].

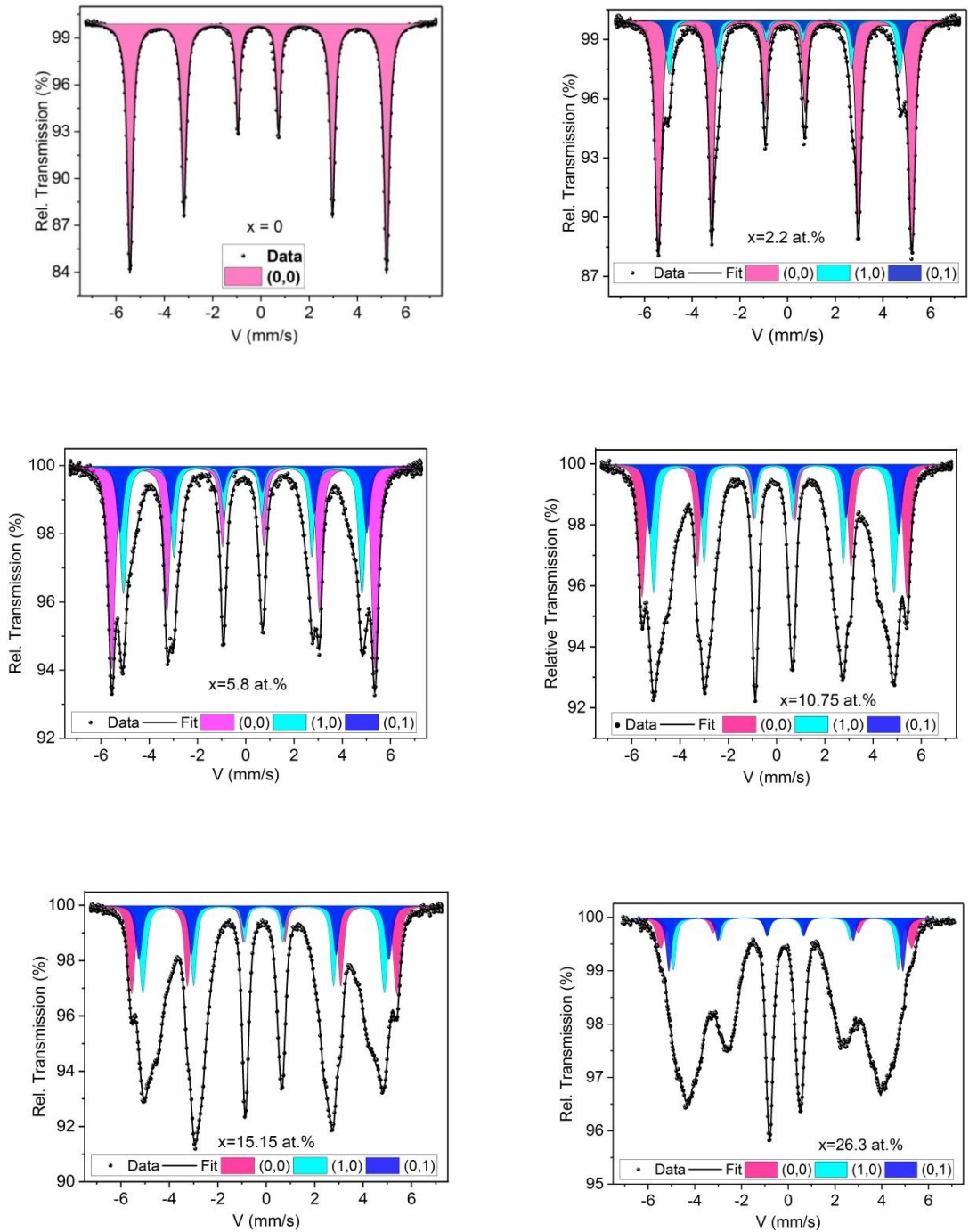

Fig. 2 $^{57}$Fe Mössbauer spectra recorded at 295 K in a transmission mode on a series of $Fe_{100-x}Cr_x$ alloys with different values of x. Sub-spectra associated with (0,0), (1,0) and (0,1) atomic configurations are highlighted.

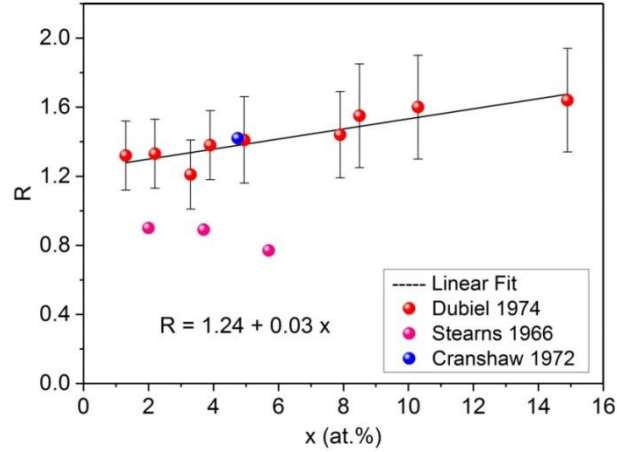

Fig. 3 Plot of R=$\Delta B_1/\Delta B_2$ versus Cr concentration in $Fe_{100-x}Cr_x$ alloys made by the author of this paper based on the results reported in papers listed in the legend.

Concerning the model M1=2, it was possible to obtain good fits up to x=3.9 at.%. For higher concentration the quality of fits was not acceptable (large values of $\chi^2$). With the model M123 only two spectra were analyzed viz. for 1.3 at.% Cr and 3.3 at.% Cr, yielding for $\Delta B_3$= - 9.9 kGs and - 8.2 kGs, respectively. These figures agree with the corresponding values reported by Stearns [28].

To simplify the analysis of the spectra i.e. to reduce the number of free parameters an additivity assumption is often made. Accordingly, if $\Delta B_k$ is a change of *B* by a single atom situated in the *k*-th coordination shell, than the total change of *B*, $\Delta B$, corresponding to a particular atomic configuration ($n_1, n_2, \ldots n_N$) is given by:

$$\Delta B(n_1, n_2, \ldots n_N) = \sum_{k=1}^{N} n_k \Delta B_k \qquad (5)$$

Applying the additivity formula to the M12 model, the hyperfine field at Fe nuclei having *m* Cr atoms in 1NN and *n* atoms in 2NN, *B(m,n;x)*, can be expressd as follows:

$$B(m,n;x) = B(0,0;x) + m \cdot \Delta B_1 + n \cdot \Delta B_1 \qquad (5a)$$

It should be mentioned that the analysis of Fe-X spectra based on the additivity formula was introduced by Stearns [28] and shortly after applied by Chandra and Schwartz [35].

Its validity was *explicite* verified by Dubiel and Krop [33] on a series of Fe-Cr alloys containing up to ~15 at.% Cr. The authors found that it was fulfilled to the accuracy ±10%. Noteworthy, this way of the spectra analysis measured on various binary alloys of Fe has become a standard. In the case of the Fe-Cr alloys it was successfully applied even for the alloy with 45.5 at.% Cr [36], yielding $\Delta B_1$ = -32.5 kGs and $\Delta B_2$ = -20.5 kGs in a very good agreement with the corresponding values found by the analysis of much less concentrated spectra [33,34,37- 40].

## 4. Distribution of Cr atoms

### 4.1. Probabilities of atomic configurations

As shown in Section 3, the hyperfine field is sensitive to the presence of Cr atoms in the vicinity of the probe nuclei. In practice, one uses the two-shell model, M12, in the analysis of the spectra which permits determining the probabilities of various atomic configurations, P(m,n;x). The obtained in this way P(m,n;x)-values can be next compared with the corresponding probabilities calculated for the random distribution, $P_r$(m,n;x), given by the binomial distribution:

$$P_r(m,n;x) = \binom{8}{m}\binom{6}{n} x^{m+n}(1-x)^{14-m-n} \qquad (6)$$

The comparison can give information whether or not the real distribution of atoms agrees with the distribution expected for the random case. It is well know that the real distribution depends on a thermal history of alloys, so MS can be successfully used to study an effect of a heat treatment on the probabilities of various atomic configurations. Examples of such applications of MS can be found e.g. in [33,39].

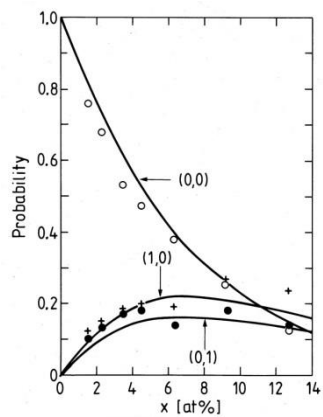

Fig. 4 Probabilities of (0,0), (1,0) and (0,0) atomic configurations in $Fe_{100-x}Cr_x$ alloys versus Cr content, x. Solid lines represent their values expected for the random distribution while symbols stand for the values determined from the analysis of the spectra in terms of the two-shell model. The plot made based on the data from [33].

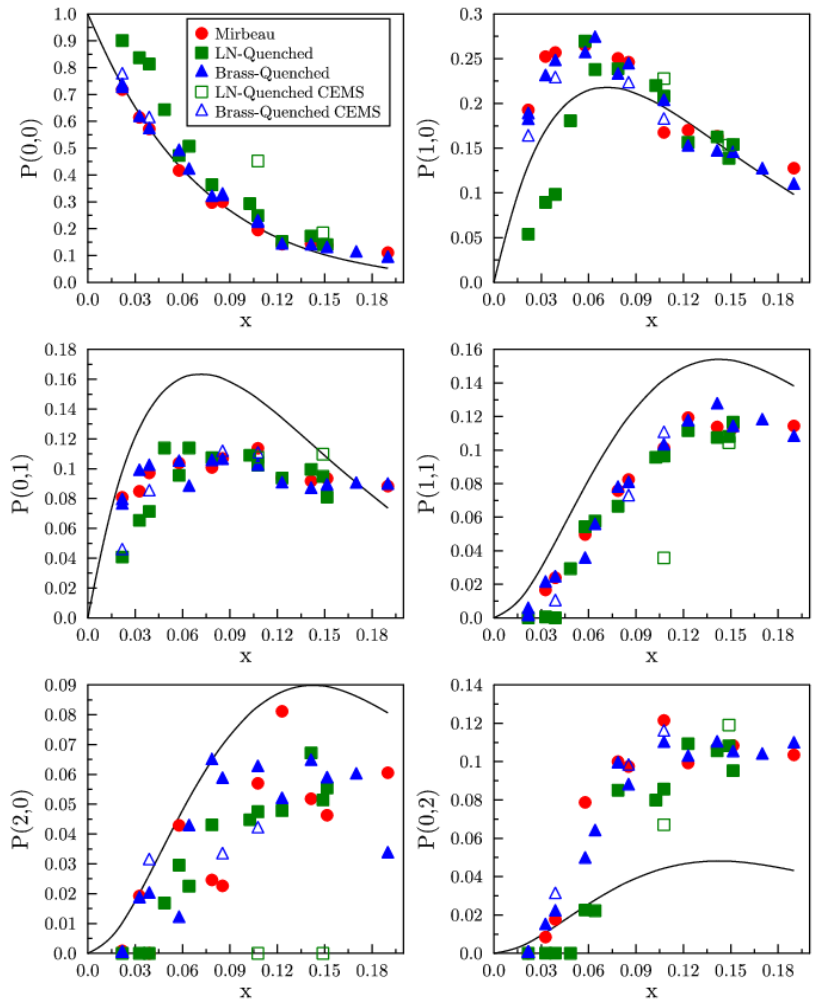

Fig. 5 Comparison between probabilities of finding Cr atoms in six various configurations, P(m,n), as obtained from the analysis of the Mössbauer spectra measured on the $Fe_{100-x}Cr_x$ alloys (symbols) and those expected from the binomial distribution (solid lines). Various symbols represent different heat treatments. The plot made based on the data reported in [40].

Figure 5 proves that in all cases the real distribution of Cr atoms is not random, despite all the investigated samples were quenched (into LN or onto block of brass). Full symbols represent the data found with the transmission-mode measurements (~30 μm thick sample) and open symbols stand for the data obtained with the conversion-electrons-mode measurements (~2 μm pre-surface zone). In particular, probabilities of (0,1), (2,0) and (1,1) configurations are significantly lower than expected for the random distribution. The opposite trend is seen for the (0,2) configuration.

Changes in the probabilities of finding Cr atoms around the Fe probe can also be brought about by strain or irradiation. For instance, Shabashov and co-authors using MS investigated Fe-Cr alloys containing 12.0 13.2 and 20.6 at.% Cr that underwent: (1) various plastic deformations and (2) 5 MeV electrons irradiation [41]. They calculated probabilities of finding 0, 1, 2, 3 or 4 Cr atoms in the 1NN+2NN neighborhood of the Fe probe atoms and found that the probabilities are characteristic of a given deformation process and, in general, deviate from the corresponding values expected from the binomial distribution. The effect of the electron irradiation was studied on the samples with the lowest and the highest content of Cr. In both cases meaningful effect was observed viz. the concentration of Cr atoms within the two-shell neighborhood around the Fe probe decreased by ~1 % in the 12.0 at.% sample and by ~2 % in the 20.6 at.% one [41].

**4.2. Short-range ordering**

The results presented in Fig. 5 clearly demonstrate that Cr atoms in Fe-Cr alloys are non-randomly distributed within the two-shell volume around Fe atoms. In other words they show a short-range ordering (SRO). This effect can be quantitatively described in terms of the SRO parameters, $\alpha_1$, $\alpha_2$ and $\alpha_{12}$ also known as the Warren-Cowley's parameters:

$$\alpha_k = 1 - \frac{P_A^k}{c_A} = 1 - \frac{P_B^k}{c_B} \qquad (7)$$

Where $c_A$ and $c_B$ are concentrations of atoms A(Fe) and B(Cr), respectively, and $P_A^k$, $P_B^k$ denote the conditional probabilities of having B(Cr) and A(Fe) atoms as neighbors in the *k*-the coordination sphere.

Analysis of the Mössbauer spectra in terms of the M12 model, as described above, gives a sound possibility for determining $\alpha_1$, $\alpha_2$ and $\alpha_{12}$ based on the following formulas:

$$\alpha_1 = 1 - \frac{\langle m \rangle}{\langle m_r \rangle} \qquad (8a)$$

$$\alpha_2 = 1 - \frac{\langle n \rangle}{\langle n_r \rangle} \qquad (8b)$$

$$\alpha_{12} = 1 - \frac{\langle m + n \rangle}{\langle m_r + n_r \rangle} \qquad (8c)$$

Where $<m_r>=8 \cdot x$, $<n_r> = 6 \cdot x$ and $<m_r+n_r>=14 \cdot x$ i.e. these are the average number of Cr atoms in 1NN, 2NN and 1NN+2NN, respectively, for random distribution.

$$<m> = \sum_{(m,n)} P(m,n) \cdot m \qquad (9a)$$

$$<n> = \sum_{(m,n)} P(m,n) \cdot n \qquad (9b)$$

$$<m+n> = \sum_{(m,n)} P(m,n) \cdot (m+n) \qquad (10c)$$

For the random distribution of atoms all three SRO-parameters are zero. Positive value of $\alpha_k$ indicates an ordering while negative value of $\alpha_k$ means clustering.

### 4.2.1. Effect of heat treatment

### 2.2.1.1. Bulk samples

Dubiel and Cieslak investigated by means of MS an effect of different heat treatments on SRO-parameters in a series of $Fe_{100-x}Cr_x$ alloys with $2 \leq x \leq 19$ [40]. Based on the analysis of the measured spectra, and formulas (8a) – (8c) they calculated the SRO parameters. The results are displayed in Fig. 7.

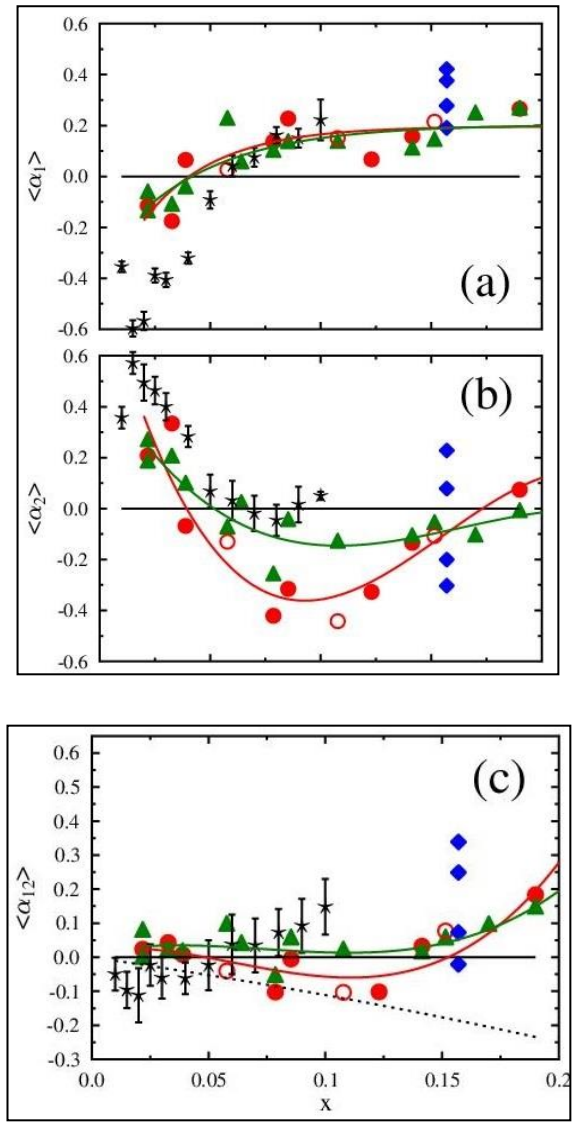

Fig. 7 SRO-parameters calculated from the Mössbauer spectra recorded on a series of $Fe_{1-x}Cr_x$ alloys that underwent three different heat treatments: T1 (triangles), T2 (circles) and T3 (diamonds) [40]. Asterisks are for the data from Ref. 42. Solid lines are guide to the eye, while the dotted line stands for the theoretical limit.

The data presented in Fig. 7 gives evidence that the SRO-parameter is characteristic of a given coordination shell. For a given shell it sensitively depends on the applied heat treatment, but neither for 1NN nor for 2NN the distribution is random. Interestingly, the behavior of $\alpha_1$ is quasi-opposite than that of $\alpha_2$ and at x≈0.05 there is a change of sign in both parameters. Worth noting, this concentration coincides pretty well with the one at which the enthalpy of the bcc phase formation in Fe-Cr changes its sign [20].

Concerning the SRO-parameter averaged over both shells, $\alpha_{12}$, its values are close to zero for the solution-treated samples (T1) with x ≤ ~0.13 and are positive for higher *x*-values.

**2.2.1.2. Quenching: Surface versus bulk**

It has been a general believe and practice that the so-called solution treatment i.e. an isothermal annealing at high temperature followed by a rapid cooling (quenching) results in a homogenous distribution of atoms. Theoretical calculations predict that magnetism has a crucial effect on the phase diagram of the Fe-Cr alloy system and related phenomena [43-50]. Short-range order magnetic correlations that exist above the Curie temperature are very important [48, 49], and they may be responsible for a non-random distribution of atoms at high temperatures at which the annealing process is performed. Consequently, the following quenching freezes this non-random distribution of atoms. Indeed, the results reported for a series of Fe-Cr alloys with Cr content up to 19 at.% gave evidence that the solution treatment did not resulted in a random distribution of Cr atoms within the bulk volume of these alloys [39]. The same series of alloys was used to study the effect of quenching into different media (water, liquid nitrogen, block of brass) on a distribution of Cr atoms in the pre-surface zone (up to ~0.3 μm thick) [51]. The spectra were recorded at room temperature both in the transmission (TRANS) as well as in the conversion electrons (CEMS) modes.

The CEMS spectra recorded on the samples quenched into water and LN were significantly different than the corresponding ones recorded in the TRANS mode.

It is evident from the CEMS spectra shown in Fig. 8 that the surface/ presurface zone is multiphase. Similar effect was revealed for the samples quenched into water. On the other hand, the TRANS spectra are characteristic of a single metallic phase viz. Fe-Cr. However, the concentration of Cr in the metallic phase of the samples quenched both into water and LN was smaller than that of the untreated samples. Obviously during quenching some amount of Cr atoms had diffused into surface/presurface zone and oxidized forming magnetite, hematite and wüstite. The relative amount of the oxides was found to depend on the quenching medium and for a given medium on the alloy composition.

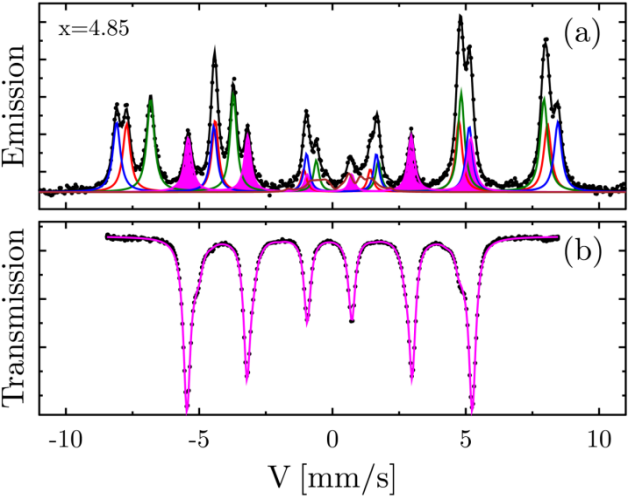

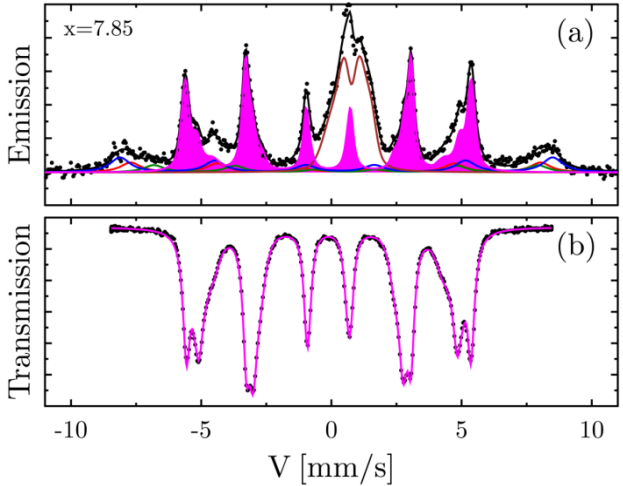

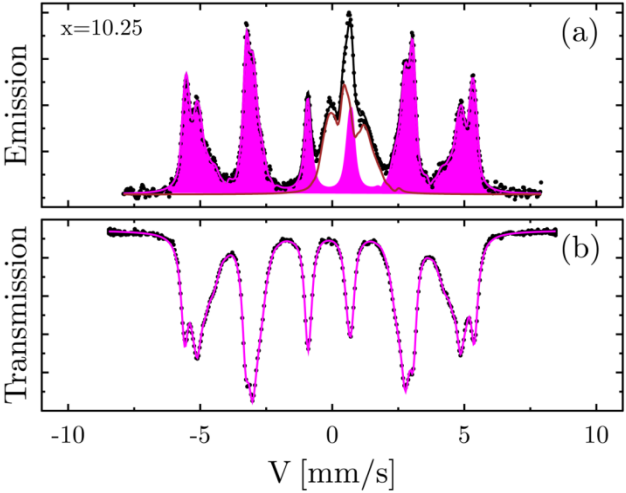

Fig. 8 Examples of the spectra recorded on the $Fe_{100-x}Cr_x$ samples with x=4.85, 7.85 and 10.25 quenched into LN. In each of the three panels the CEMS spectrum is labelled as (a) and the TRANS spectrum with (b). Purple subspectrum indicates the metallic phase [51].

The effect of the Cr depletion in the metallic phase can easily be deduced from the spectra presented in Fig. 9 where a third spectrum, designed as (c), was added viz. the one recorded on a given sample before it was quenched.

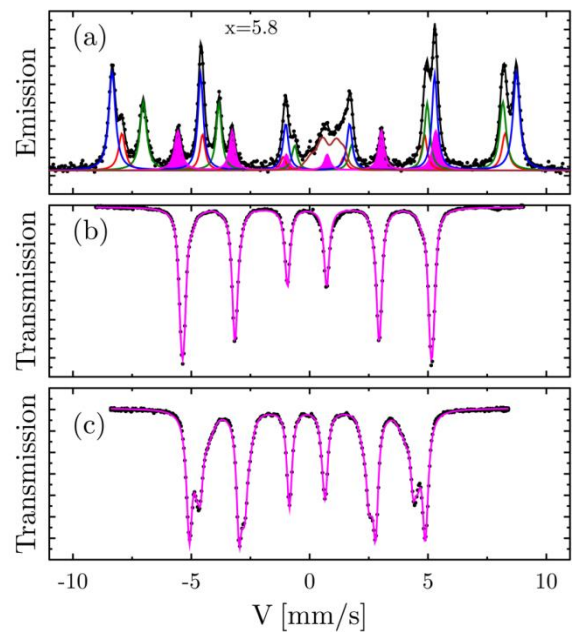

Fig. 9 $^{57}Fe$ spectra recorded at room temperature on the $Fe_{100-x}Cr_x$ sample with x=5.8. The spectra (a) and (b) were recorded on the sample quenched into water, while the spectrum (c) on the same sample but prior to the quenching. Note that the spectrum (b) looks like the one for a pure Fe [51].

In contrast, the surface/pre-surface zone of the alloys quenched onto the block of brass was single-phase for all samples. This can be clearly recognized in the spectra displayed in Fig. 10.

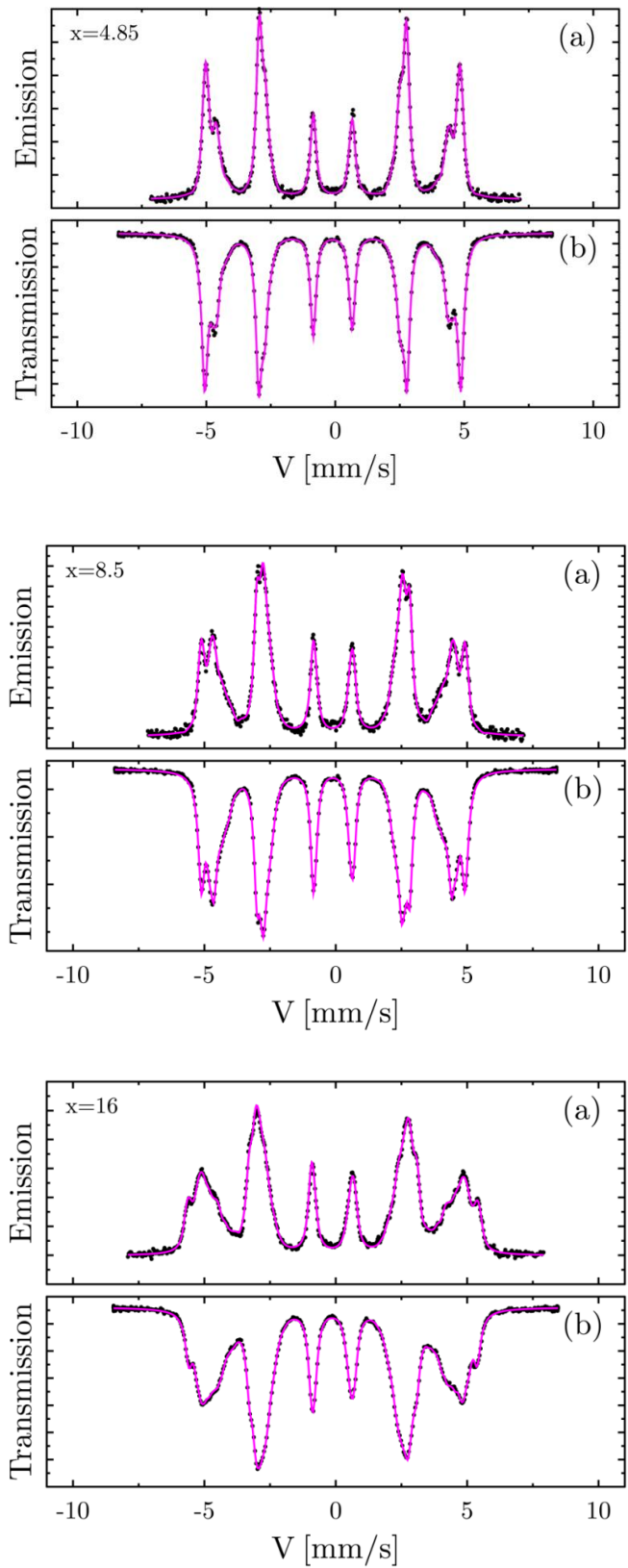

Fig. 10 Examples of the spectra recorded on samples with x=4.85, 8.5 and 16.0 quenched ono brass. In each of the three panels the CEMS spectrum is labelled as (a) and the TRANS spectrum as (b) [51].

In order to shed more light on the quenching-induced changes in the distribution of Cr atoms in the studied Fe-Cr alloys the authors calculated the SRO-parameters in a way outlined in section 4.2. Thus obtained results are displayed in Fig. 11.

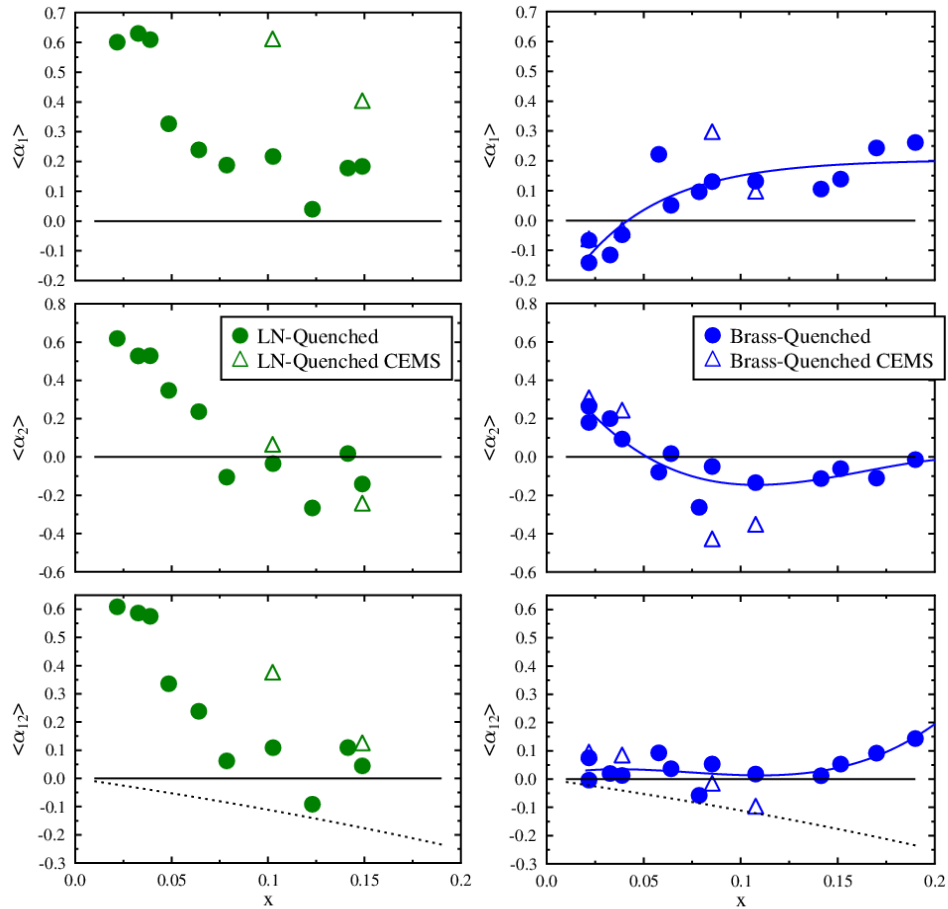

Fig. 11 The SRO-parameters $<\alpha_1>$, $<\alpha_2>$ and $<\alpha_{12}>$ versus Cr content, $x$, for the $Fe_{100-x}Cr_x$ samples quenched into LN (left panel) and onto a brass block (right panel). Circles represent the data obtained from the spectra measured in the TRANS mode, while triangles are for those derived from the CEMS spectra. The dotted line shows the thermodynamically allowed limit [51].

The results shown in Fig. 11 clearly demonstrate that the random distribution of Cr atoms was achieved neither with the quenching into LN nor onto the block of brass. The quenching into LN (and water) resulted not only in the oxidation of the surface/presurface zone of the samples with $x \leq \sim 11$, but also in a significant depletion in chromium in the bulk. Formally, the latter is reflected by high values of the SRO parameters, particularly in the low-concentrated samples. Positive values of the SRO parameters mean that the actual number of Cr atoms around the probe Fe ones is smaller than that expected for the random distribution

The samples that underwent the quenching onto the block of brass were not oxidized. However, also here the distribution of Cr atoms is not random neither in 1NN nor in 2NN. The values of $<\alpha_1>$ are for low concentration positive while those of $<\alpha_2>$ are negative. This means that in 1NN there are less Cr atoms than expected for the random case, whereas in 2NN there are more Cr atoms. However, the difference shrinks with $x$ for both shells reaching zero at $x \approx 5$. This means that at this concentration the distribution of Cr atoms is perfectly i.e. in both coordination shells random. For $x \geq \sim 5$ $<\alpha_1>$ is positive indicating deficit of Cr atoms relative to the value expected for the random distribution. On the contrary, $<\alpha_2>$ is negative meaning an excess of Cr atoms. The values of the SRO averaged over both shells, $<\alpha_{12}>$, is close to zero for $x \leq \sim 14$. Thus taking into account both shells one can say that the distribution of Cr atoms is random in the Fe-Cr alloys quenched onto the block of brass provided the content of Cr is not larger than ~14 at.%. For higher concentration the positive value of $<\alpha_{12}>$ is in line with a tendency of Cr atoms to form clusters.

The mentioned above reduction of the Cr content in the samples quenched into LN and water can be quantified by determining the content of Cr in these samples. For this purpose one can use the dependence of the average hyperfine field, $<B>$, on Cr content. The dependence turned out to be linear for $x$ up to ~50 [36,52], hence the knowledge of $<B>$ permits a unique determination of $x$.

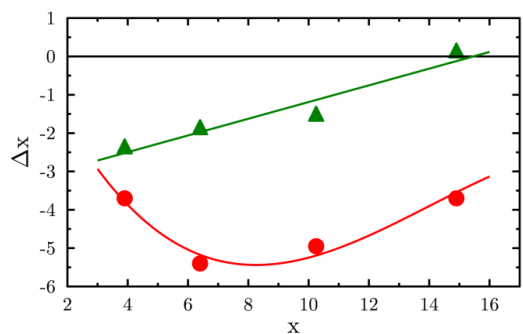

Fig. 12 Chromium depletion, $\Delta x$, in the metallic Fe-Cr phases of the pre-surface zone (circles) and the bulk (triangles) versus the initial Cr content, x, for the samples quenched into LN. Solid lines are guide to the eye [51].

As expected, the Cr depletion in the pre-surface zone is significantly larger than the one in the bulk.

**2.2.1.3. Isothermal annealing**

Investigation of the effect of isothermal annealing on the redistribution of Cr atoms in the Fe-Cr alloys is of a great interest as it is related with two phenomena responsible for hardening and brittleness not only in the Fe-Cr alloys themselves, but also in ferritic/martensitic steels. These types of steels play an important role as structural materials used in various branches of industry. This role follows from their excellent properties like good resistance to a high-temperature corrosion, low swelling and high toughness. Accordingly they have been used in various branches of industry e. g. power pants (including nuclear ones), chemical and petrochemical industries to produce devices that work at service at elevated temperatures and often in aggressive environment. In the case of the nuclear power plants their lifespan is limited by a degradation of structural devices such as vessel and primary circuit owing to exposure to high temperature and radiation. The former causes a thermal aging and the latter radiation damage causing degradation of mechanical properties and enhanced susceptibility to corrosion. Precipitation of the Cr-rich $\alpha'$ phase is responsible for the degradation for the annealing temperature lower than ∼ 800 K [53-55]. However the upper limit of this temperature is not precisely known. For example Williams located it at 830 K [56] whereas according to Dubiel and Inden this critical temperature lies between 773 and 805 K [38]. The underlying phenomenon responsible for the precipitation of $\alpha'$ is known as the phase decomposition or segregation and the phase field in which the phenomenon occurs is known as the miscibility gap (MG). Mössbauer spectroscopy has proved to be a very successful tool for the study of both borders of GP: one related to the solubility limit of Cr in iron and the other with solubility limit of Fe in chromium. More detailed information on this issue is given in Section 5.1.1.

Here in this section it will be shown how the SR-ordering changes with time of the isothermal annealing performed at different temperatures on the Fe-Cr alloys with different composition and metallurgical state [57-59].

Dependence of the SRO-parameters $\alpha_1$, $\alpha_2$ and $\alpha_{12}$ versus annealing time, $t$, as obtained from the spectra recorded on a $Fe_{84.5}Cr_{15.5}$ sample annealed at 688 K, are displayed in Fig. 13. Noteworthy, $\alpha_1 > 0$ in the whole *t*-range and it increases with *t* approaching saturation. Positive values of $\alpha_1$ mean that the number of Cr atoms in 1NN-shell is lower than expected for the random distribution. The *t*-dependence of $\alpha_2$ is similar but the initial value (*t*=0) is very negative, indicating an excess of Cr atoms in 2NN-shell. However in the course of annealing $\alpha_2$ increases reaching zero after ~2.5 days. For larger annealing times $\alpha_2$ becomes positive and eventually saturates. The SRO-parameter averaged over the two shells, $\alpha_{12}$=0 in the non-annealed sample, and with annealing time it increases to saturate after *t*≈15 days. Interestingly, the $\alpha_{12}$ parameter is linearly correlated with the average hyperfine field, <*B*>, as visualized in Fig. 13. The increase of <*B*> with $\alpha_{12}$ is caused by the decrease of the number of Cr atoms in the two-shell vicinity around the probe Fe atoms [36]. The decrease, in turn, is caused by the annealing-induced clustering of Cr atoms related to the phase decomposition (separation). More details are reported elsewhere [57].

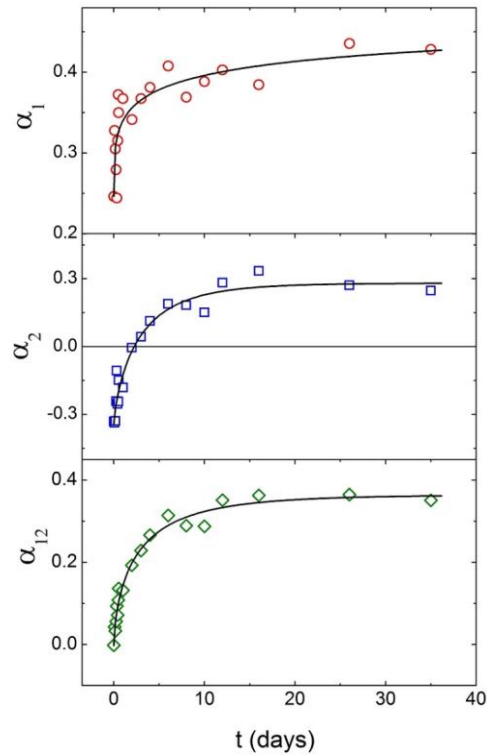

Fig. 13 Dependence of the SRO-parameters $\alpha_1$, $\alpha_2$ and $\alpha_{12}$ on the annealing time, *t*, as determined for the $Fe_{84.5}Cr_{15.5}$ sample isothermally annealed at 688 K [57]. The solid lines are the best-fits in terms of the Johnson-Mehl-Avrami-Kolmogorov equation.

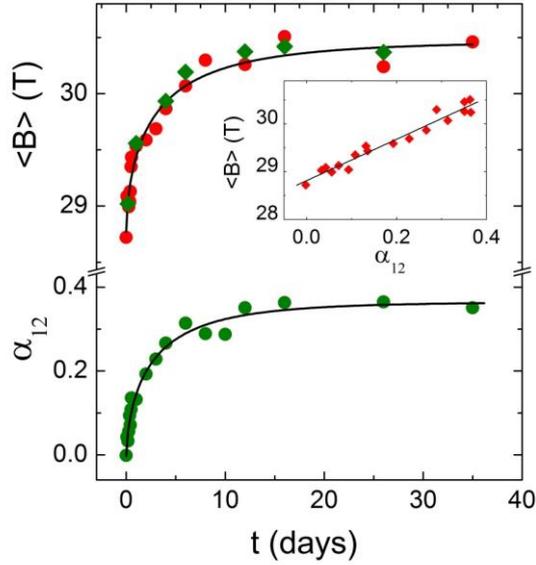

Fig. 14 Dependence of the average hyperfine field, <*B*>, and of the SRO-parameter, $\alpha_{12}$, on the annealing time, *t*, as found for the $Fe_{84.5}Cr_{15.5}$ sample isothermally annealed at 688 K [57]. The inset shows that the <*B*> - $\alpha_{12}$ relation is linear. The solid lines are the best-fits in terms of the Johnson-Mehl-Avrami-Kolmogorov equation.

The effect of the isothermal annealing on the SRO – parameters was studied in a strained and strain-relaxed sample of $Fe_{84.5}Cr_{15.5}$ [58]. Examples of the spectra recorded at RT in the TRANS mode alongside with the corresponding distributions of the hyperfine field are displayed in Fig. 15.

The p(B)-curves visualize the effect of the annealing better than the spectra themselves. Namely, the p(B)-curve for the annealed sample is shifted to higher fields. The shift corresponds to an increase of <*B*> by ~5 kGs [58], what implies a decrease of the number of Cr atoms in the vicinity of the Fe probe atoms or decrease of the local Cr content by ~2 at.%.

The process of the redistribution of Cr atoms that takes place during the annealing could be quantitatively followed by the annealing time dependence of $\alpha_1$, $\alpha_2$ and $\alpha_{12}$ as visualized in Fig. 16.

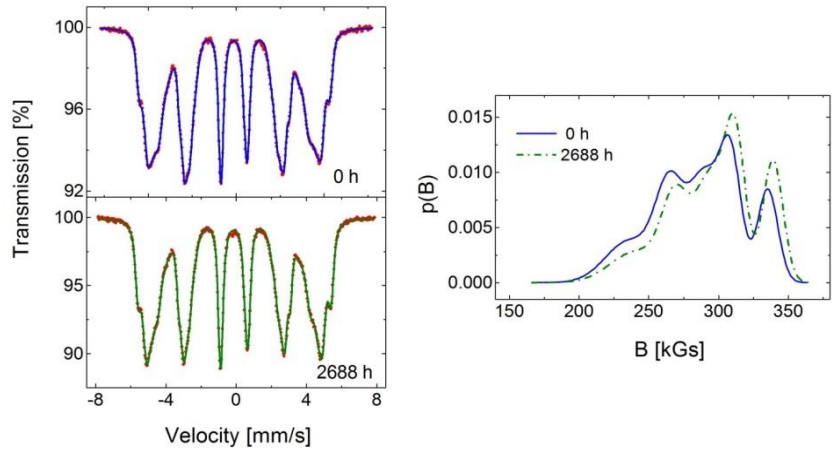

Fig. 15 $^{57}$Fe spectra recorded at RT on non-annealed (0 h) and annealed at 675 K (2688 h) strained sample, and corresponding hyperfine field distribution curves, p(B) [58].

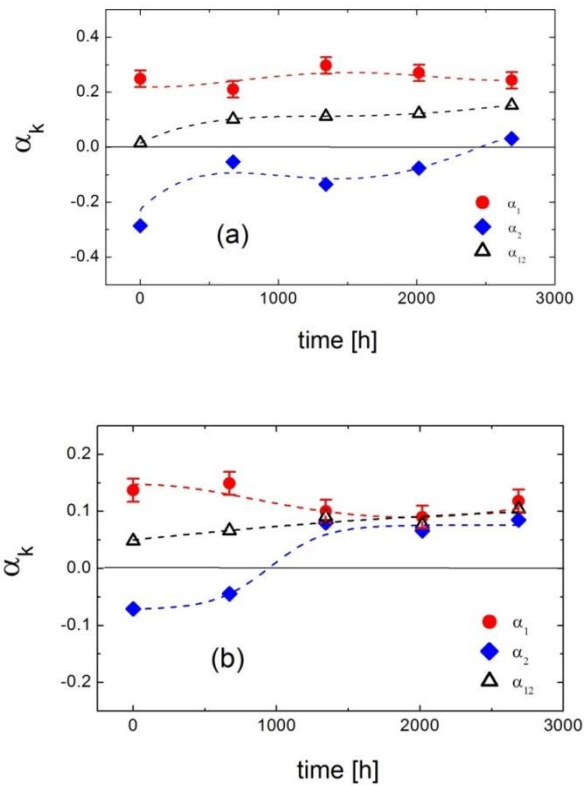

Fig. 16 SRO parameters versus the annealing time for (a) strained and (b) strain-free samples. The lines are the guide to the eyes [58]

Figure 16 gives evidence that the metallurgical state of the investigated alloy has been reflected in the behavior of the SRO-parameters. In particular, in the strain-free sample no difference exists between the parameters for the annealing time longer than ca. 1350 h. On the other hand, in the strained sample $\alpha_1$ hardly depends on the annealing time, whereas $\alpha_2$ slowly increases, yet for the longest annealing time it remains distinctly different than $\alpha_1$. Comparison of the $\alpha_{12}$ parameters for both samples displayed in Fig. 17 testifies to a clear difference in the distribution of Cr atoms within the first two neighbor shells in the strained and in the strain-free samples.

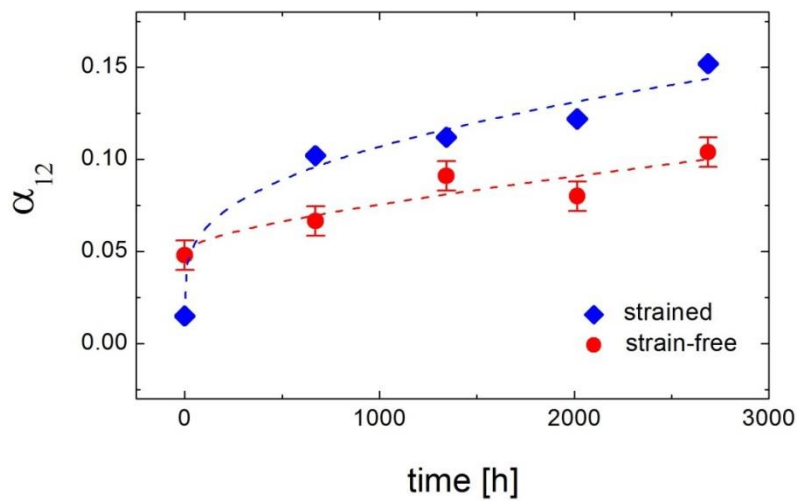

Fig. 17 Short-range parameter, $\alpha_{12}$, versus annealing time for the $Fe_{84.5}Cr_{15.5}$ sample. The lines represent the best fits to the data in terms of the JMAK-like equation [58].

The $\alpha_{12}$ - values for the strained sample are larger for all annealing times, except for the non-annealed state. This means that in the strained sample a degree of Cr atoms clustering is higher.

Dubiel and Żukrowski recently carried out Mössbauer-effect measurements on the Fe-Cr alloy containing 26.3 at.% Cr in search for the maximum temperature at which the Cr-rich $\alpha'$ phase could precipitate [59]. For this purpose the alloy was isothermally annealed at 832 K for a period of up to 1777 h. Examples of the spectra recorded at RT are displayed in Fig. 18.

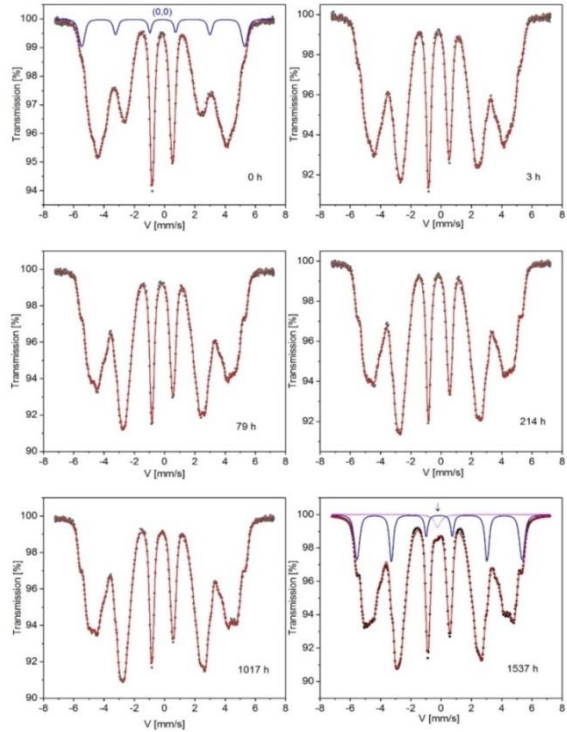

Fig. 18 Set of $^{57}$Fe Mössbauer spectra recorded at 295 K on Fe$_{73.7}$Cr$_{26.3}$. The spectra are labelled with the annealing time. The sub spectrum corresponding to the (0,0) atomic configuration is shown in the spectrum recorded on the untreated sample as well as in the one annealed for 1537 h. In the latter case the arrow indicates a single-line sub spectrum due to the σ-phase. The spectra adopted from Ref. 59.

The analysis of the spectra in terms of the two-shell model revealed that the annealing – induced process proceeded in two stages: the first stage for the annealing time up to ~350 h, followed by the second stage. The two-step nature of the process can be best seen in the behavior of the average hyperfine field – see Fig. 19.

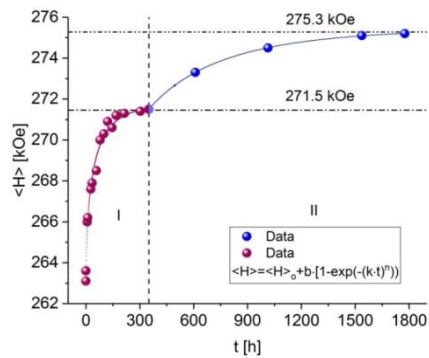

Fig. 19 The average hyperfine field, <H>, vs. annealing time, t. The solid lines represent the best-fits to the data in terms of the JMAK equation in range I and II, respectively [59].

This two-step process has been also reflected in the behavior of the SRO-parameter, $\alpha_1$, as shown in Fig. 20. The $\alpha_1$ - data in both ranges were successfully fitted to the JMAK-like equation yielding values of the kinetic parameters i.e. Avrami parameter, n, and the rate constant, k (shown in Fig. 21). The n-values are similar for both stages of transformation, whereas the k-value indicates that the process within the first step is faster than that in the second step.

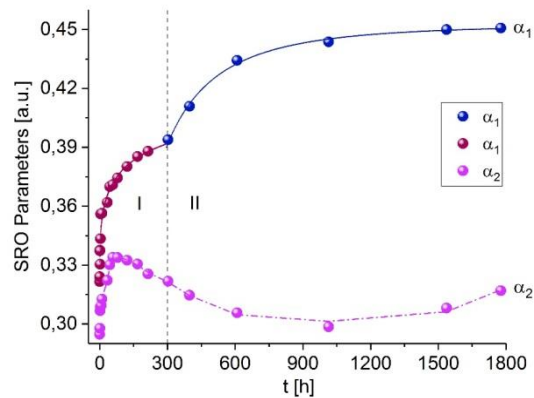

Fig. 20 SRO parameters $\alpha_1$ and $\alpha_2$ vs. annealing time, t. Solid lines stand for the best-fit of $\alpha_1$ data to the JMAK equation in the two ranges (I and II) separated by a vertical dash line. The broken line in $\alpha_2$ is the guide to the eye [59].

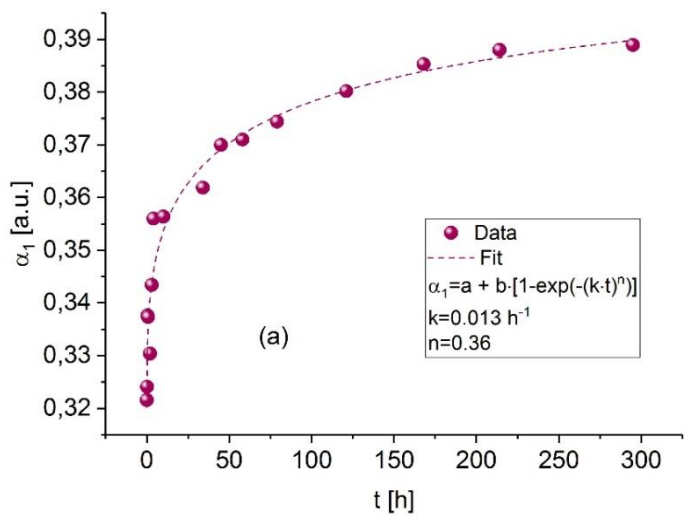

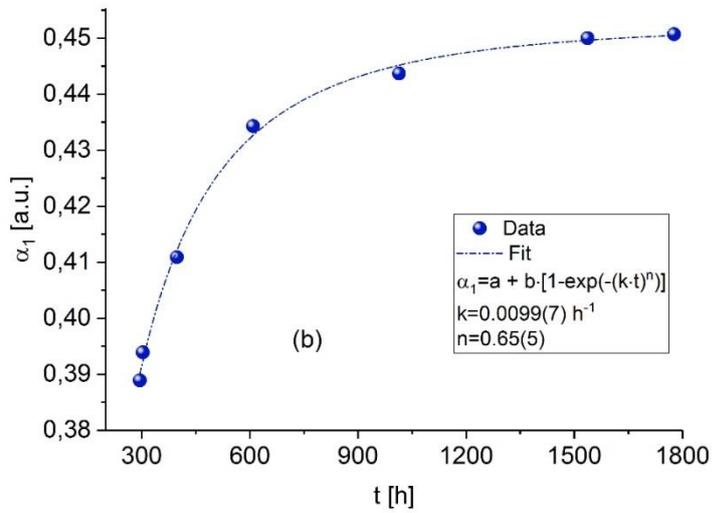

Fig. 21 SRO parameter $\alpha_1$ vs. annealing time, $t$, for the step: (a) first and (b) second. The lines are the best-fits of the data to the JMAK equation. Kinetic parameters $k$ and $n$ are displayed in the legends [59].

This two-step behavior means that the temperature of 832 K lies already above the upper limit of the miscibility gap. Noteworthy, theoretically predicted values for the upper-limit-temperature are ~785K [60], ~794 K [53] and ~796 K [55].

### 2.2.2. Effect of irradiation

Investigation of an effect of irradiation on the Fe-Cr alloys is of a great technological importance, as these alloys constitute the major ingredient of such structural materials like ferritic/martensitic and austenitic steels. Their stability and irradiation tolerance is a critical factor in determining the lifespan and safety of nuclear power plants. It is known that radiation accelerates precipitation of Cr-rich $\alpha'$ phase in the Fe-Cr alloys. Its presence causes hardening and brittleness. The deteriorating effect of the irradiation depends on many factors like kind of irradiating particles (neutrons, protons, electrons, heavy (Fe, Cr) or light (He, Kr) ions), their dose or the displacement per atom (dpa) rate, flux and temperature.

MS has proved to also be useful in this matter as the hyperfine field is sensitive to the local environment around the probe. Consequently, quite numerous MS studies relevant to the problem can be found across the literature.

Irradiation of the Fe-Cr alloys may also result in a redistribution of Cr atoms what can be quantitatively studied by means of MS, in particular, in terms of the SRO-parameters. The effect of He-ions was reported in [61-65], the effect of Fe-ions in [65, 66..], the effect of neutrons in [67] and effect of electrons in [68].

**2.2.2.1. He-ions**

Effect of irradiation on $Fe_{100-x}Cr_x$ ($x$=5.8, 10.75, 15.15) with 0.025 keV, 0.250 keV and 2.5 MeV $He^+$ to the dose of up to 30 dpa was studied by means of the CEMS technique [61-63].

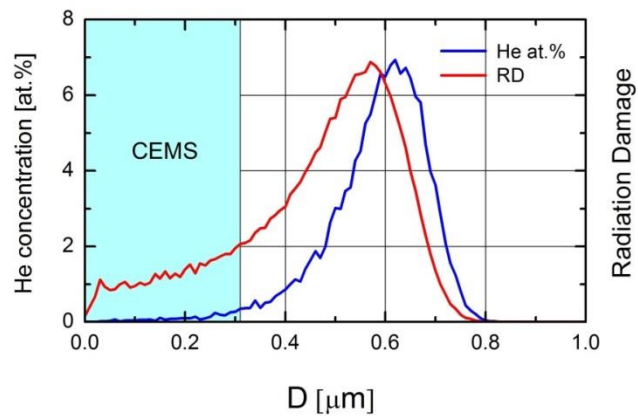

Fig. 22 Concentration and damage profiles versus depth, $D$, for the $Fe_{89.25}Cr_{10.75}$ sample irradiated with 0.25 MeV $He^+$ [64]. The pre-surface depth accessible to the CEMS measurements is indicated by a vertical band.

The effect of the energy of the He-ions as well as that of the implantation dose is of interest.

Examples of the CEMS spectra recorded on the non-irradiated and the irradiated side of the $Fe_{84.85}Cr_{15}$ sample can be seen in Figs. 23.

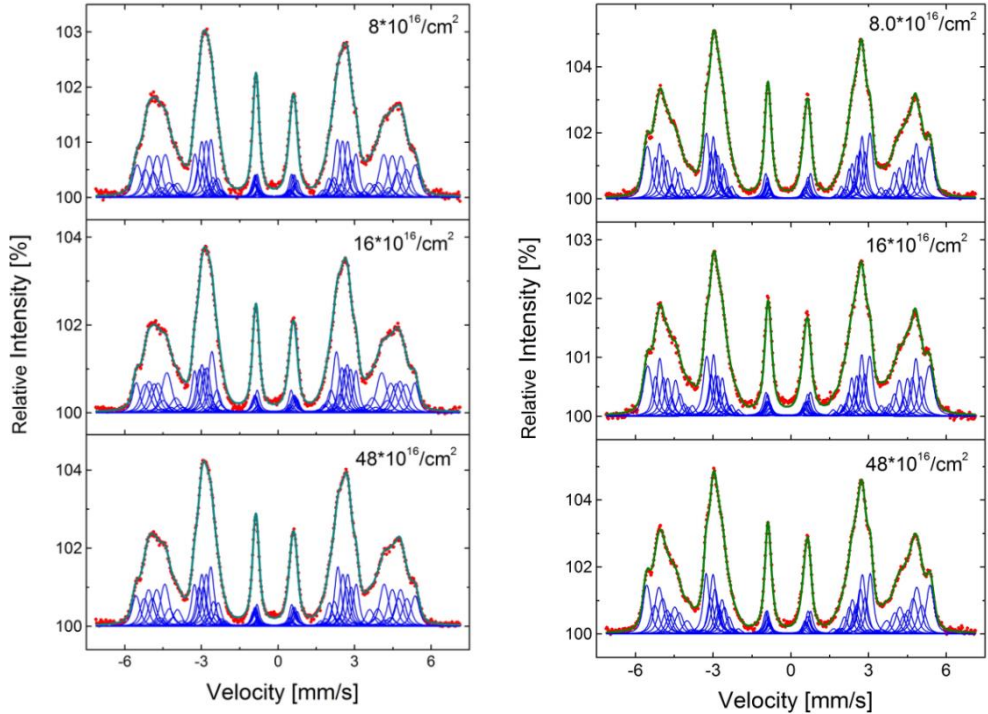

Fig. 23 CEMS spectra recorded at RT on $Fe_{84.85}Cr_{15}$: (left panel) non-irradiated side and (right panel) side irradiated with 0.25 MeV $He^+$ to different doses as labelled (or 5, 10 and 30 dpa). The effect of the irradiation is best seen in the outermost lines [63].

Figure 24 illustrates the dose-dependence of the SRO-parameters determined from the spectra recorded on both sides of the $Fe_{84.85}Cr_{15.15}$ [63].

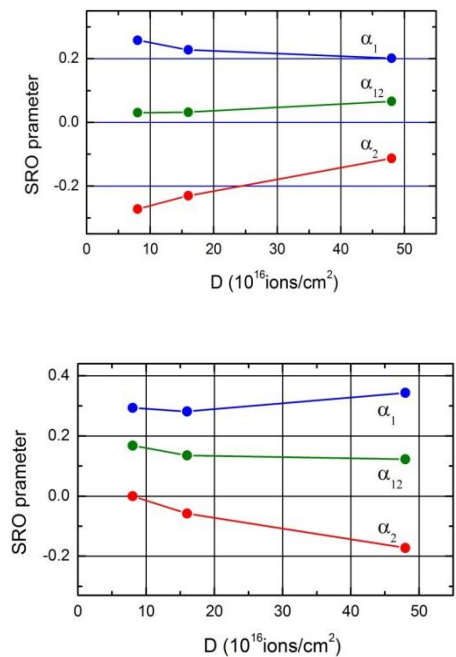

Fig. 24 SRO – parameters, $\alpha_1$, $\alpha_2$ and $\alpha_{12}$ determined from the CEMS spectra recorded at RT on the $Fe_{84.85}Cr_{15.15}$ sample irradiated with 0.25 MeV He-ions to doses 8, 18, 48·$10^{16}$ ions/cm$^2$ (or 5, 10 and 30 dpa, respectively). The lower plot illustrates the SRO-parameters determined for the irradiated side and the upper plot the ones determined for non-irradiated side of the sample [64].

The data shown in Fig. 24 illustrates the effect of the irradiation on the SRO-parameters. In order to get a quantitative insight into the effect, one has to consider a change in the values of these parameters caused by the irradiation. A pertinent plot is presented in Fig. 25.

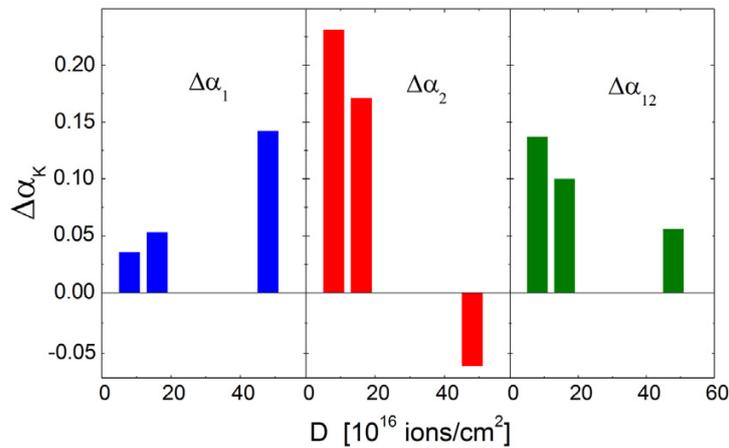

Fig. 25 Change of the SRO-parameters parameters, $\Delta\alpha_k=\alpha_k(IR)-\alpha_k(NR)$, versus the irradiation dose, D. The data depict the $Fe_{84.85}Cr_{15.15}$ sample irradiated with 0.25 MeV He$^+$ to the dose of 5, 10 and 30 dpa [63].

The data displayed in Fig. 25 evidently demonstrate that the behavior of $\Delta\alpha_1$ is different than that of $\Delta\alpha_2$. The former, being positive, increases with the dose. This means that upon the applied irradiation the number of Cr atoms in 1NN decreased with D. The trend in 2NN is opposite i.e. $\Delta\alpha_2$, being very positive for 5 and 10 dpa, decreased with D i.e. the number of Cr atoms in 2 NN increased with D upon bombardment. For D=30 dpa $\Delta\alpha_2$ is negative. The latter indicates that the number of Cr atoms in 2NN is greater than 0.91, the value expected in this shell for the case of the binomial distribution. In other words, Cr atoms were in this case replaced upon the irradiation from 1NN into 2NN. However, in total, the applied irradiation deceased the number of Cr atoms present in the 1NN+2NN vicinity around the Fe atoms ($\Delta\alpha_{12} > 0$). Interestingly, the strongest effect was found for the lowest dose and the smallest effect for the highest dose.

The influence of energy of He-ions on the distribution of atoms in $Fe_{100-x}Cr_x$ alloys with $x$=5.8, 10.75, 15.15 was investigated by means of CEMS by Dubiel et al. [64]. To this end the samples were irradiated to the dose of $12\cdot10^{16}$ ions/cm$^2$ (7.5 dpa) with 0.25 and 2.5 MeV He-ions. The revealed irradiation effect on the three alloys is shown in Fig. 26 where values of $\Delta\alpha_k=\alpha_k(IR)-\alpha_k(NIR)$, k=1,2,12 are displayed.

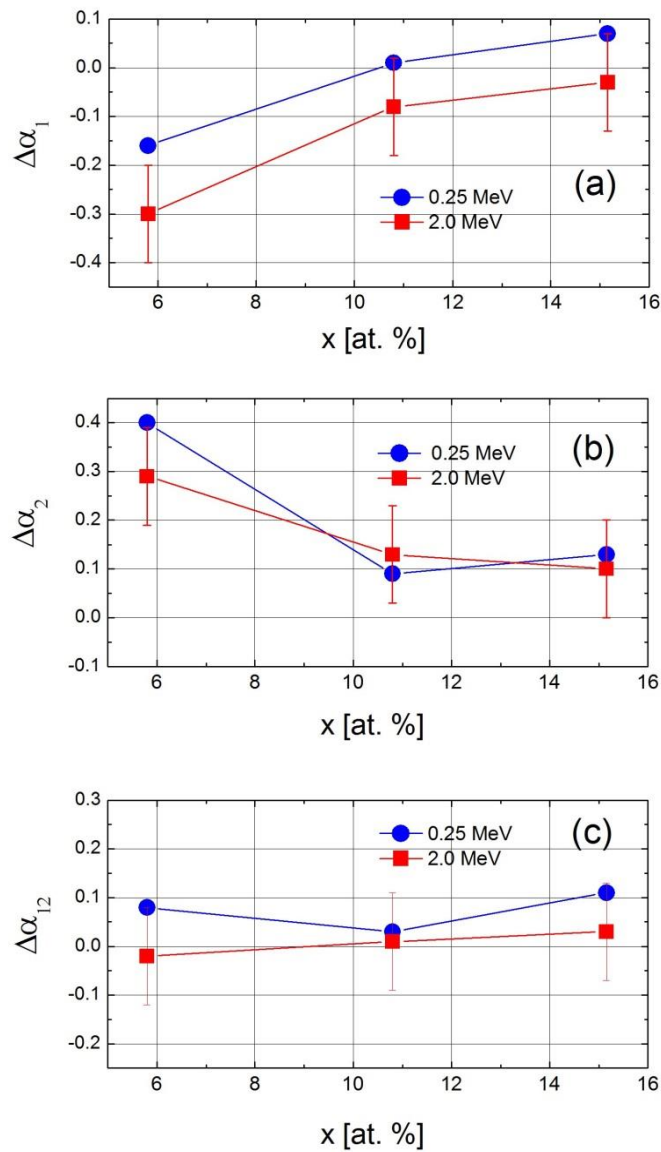

Fig. 26 $\Delta\alpha_k$-parameters for: (a) 1NN-shell, (b) 2NN-shell, and (c) 1NN+2NN-shells versus Cr content, $x$. The lines are to guide the eye [64].

It can clearly be seen in Fig. 26 that the change of the SRO-parameters is characteristic both of the neighbor shell as well as of the Cr concentration, yet it hardly depends on the energy of He-ions. The concentration dependence of $\Delta\alpha_1$ and that of $\Delta\alpha_2$ has a quasi-mirror symmetry i.e. for $x$=5.75 $\Delta\alpha_1 < 0$ and $\Delta\alpha_2 > 0$ with about the same absolute value. This means that the number of Cr atoms increased in 1NN and simultaneously the number of Cr atoms in 2NN decreased by about the same amount. On the other hand for $x$=10.75 and 15.15 practically no change either in $\Delta\alpha_1$ or in $\Delta\alpha_2$ occurred. Taking into account both shells, the distribution of atoms hardly changed under the bombardment of the three alloys with 0.25 and 2.5 MeV He-ions.

Results on the effect of He-ions on the SRO in the $Fe_{84.85}Cr_{15.15}$ sample (the same origin as the one described above) were also reported in [65]. The authors studied the sample irradiated with 45 keV He-ions without and with application of an external magnetic field of 0.4 T. They found that the irradiation affected mostly $\alpha_2$ which upon irradiation with no magnetic field increased from -0.32 to -0.08. This means that the applied irradiation decreased the number of Cr atoms in 1NN. Consequently, the average SRO-parameter, $\alpha_{12,}$ increased from 0, for non-irradiated sample, to 0.10 indicating thereby a decrease of Cr atoms in the 1NN+2NN vicinity of the probe Fe atoms. Clustering of Cr atoms was suggested as possible mechanism underlying the observed effect. However, if the irradiation was performed in the field, a very small reduction in $\alpha_1$ and $\alpha_2$ was revealed, but $\alpha_{12}$ remained unchanged i.e. equal to zero.

### 2.2.2.1. Fe-ions

Dubiel studied an effect of 2 MeV $Fe^{3+}$ irradiation on the distribution of atoms in the $Fe_{89.25}Cr_{10.75}$ alloy in two metallurgical states: (a) strained and (b) strain-free. The irradiation was carried out in vacuum at 573 K to the dose of ~7 dpa [Data not published yet]. CEMS spectra were recorded *ex situ* at room temperature both on the irradiated (IR) as well as on the non-irradiated (NIR) side of the sample. The irradiation profile is shown in Fig. 27.

Analysis of the spectra yielded values of the SRO-parameters. The difference $\Delta\alpha_k=\alpha_k(IR)-\alpha_k(NIR)$, k=1,2,12, is displayed in Fig. 28. The effect of the irradiation is very weak. Only changes in $\alpha_1$ are statistically meaningful (error ±0.02) testifying to a weak decrease of Cr atoms in the 1NN shell. Some clustering of Cr atoms for the strain-free sample is not excluded ($\alpha_{12} \geq 0.2$).

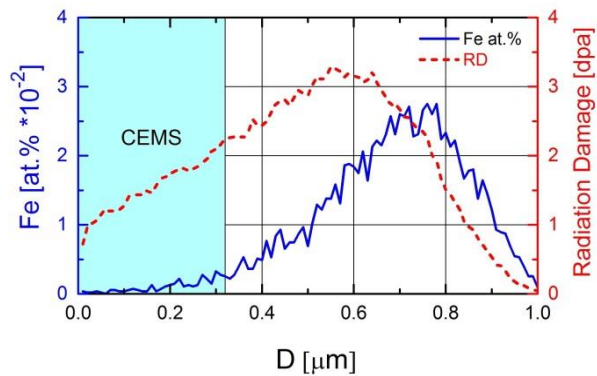

Fig. 27 Concentration and damage profiles versus depth, *D*, for the $Fe_{89.25}Cr_{10.75}$ sample irradiated with 2.0 MeV $Fe^{3+}$ [Data not published yet]. The pre-surface depth accessible to the CEMS measurements is indicated by a vertical band.

The analysis of the spectra yielded, among others, values of the SRO-parameters which are presented in Table 1.

Table 1.

SRO-parameters determined from the spectra recorded *ex situ* on both sides of strained and strain-free samples of $Fe_{89.25}Cr_{10.75}$ irradiated with 2 MeV $Fe^{3+}$. NIR denotes the non-irradiated and IR the irradiated side.

|  | Strained | | Strain-free | |
|---|---|---|---|---|
| SRO | NIR-side | IR-side | NIR-side | IR-side |
| $\alpha_1$ | -0.012 | 0.036 | 0.037 | 0.074 |
| $\alpha_2$ | -0.177 | -0.196 | -0.048 | -0.041 |
| $\alpha_{12}$ | -0.069 | -0.063 | 0.001 | 0.024 |

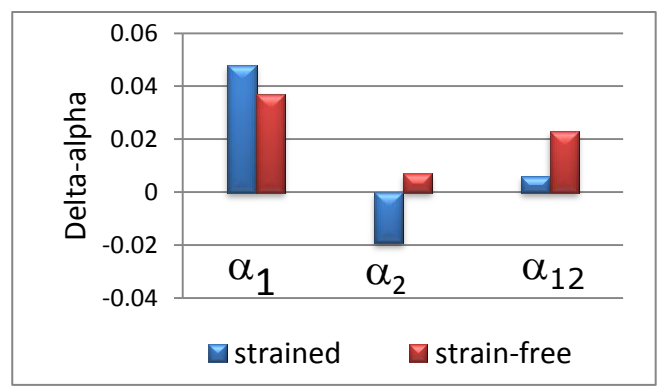

Fig. 28 The difference in the SRO parameters, Delta-alpha=$\alpha_k$(IR) - $\alpha_k$(NIR), k=1, 2, 12, caused by the irradiation with $Fe^{3+}$ ions. The plot made using the data from Table 4.

Small changes in the distribution of Cr atoms caused by the applied self-ion irradiation were revealed. The major change was found for $\alpha_1$ both in the strain and strain-free samples. Its increase under irradiation is equivalent to a decrease of Cr atoms in 1NN-shell. In 2NN-shell small increase of Cr atoms was revealed in the strained sample, whereas in the strain-free sample almost negligible decrease occurred. In total, the applied irradiation decreased the number of Cr atoms within the two-shell vicinity of Fe atoms. This effect, stronger for the strain-free sample, can be regarded as an indication of clustering of Cr atoms.

The effect of Fe-ions on the SRO-parameters in a $Fe_{85}Cr_{15}$ alloy was also studied by Garcia-Cortes et al. [65]. The investigated sample was irradiated with 1 MeV $Fe^+$ to two doses viz. 3 and 22 dpa. Noteworthy the irradiation was performed also in external magnetic field of 0.4 T. By analyzing the CEMS spectra in terms of the two-shell model they determined <$\alpha_1$> to be positive and <$\alpha_2$> to be negative for the irradiated sample. However, in the in-field irradiated sample <$\alpha_1$> was 0.1 smaller and <$\alpha_2$> by 0.1 larger. Consequently, <$\alpha_{12}$> remained unchanged and close to zero.

### 2.2.2.2. Neutrons

The effect of neutron irradiation on the SRO parameters in a $Fe_{92}Cr_8$ alloy was studied by Huang and co-workers [67]. Two irradiations were performed: at 363 K viz. to the dose of $5.3 \cdot 10^{-3}$ dpa and 0.2 dpa, and one at 573 K to the dose of 0.015 dpa. The spectra were measured in the TRANS mode before and after irradiations. Analysis of the spectra yielded values of <$\alpha_1$>, <$\alpha_2$> and <$\alpha_{12}$>. In particular, under the 0.2 dpa@363 K irradiation. <$\alpha_1$>, decreased from ~0.1 to ~0 and <$\alpha_2$> increased from ~-0.20 to ~-0.16. The average SRO-parameter, <$\alpha_{12}$> was found to be negative both in the non-irradiated sample (~0.03) as well as in all irradiated in which <$\alpha_{12}$> had more negative values. The strongest effect of the irradiation was found in the sample irradiated at 573 K, for which the value of <$\alpha_{12}$> ≈ -0.11. This means the number of Cr atoms in the two-shell neighborhood of the probe Fe was enhanced by the irradiation.

**2.2.3. Effect of strain**

Dubiel and Żukrowski investigated an effect of strain on the distribution of Cr atoms in a $Fe_{89.15}Cr_{10.75}$ alloy in two different metallurgical states viz. (1) a heavily strained by cold rolling and (2) a strain-relaxed. Three samples of each state were measured. Details on the samples can be found elsewhere [68].

The samples were measured at RT by means of MS. Recorded were conversion electrons (CEMS mode) on each side of the samples, labelled as A and B. The CEMS spectra contain information on up to ~0.3μm thick pre surface zone. An example of the spectra measured on the strained and the strain-free samples can be seen in Fig. 29. Optically, a difference between the two spectra can be hardly seen. However, hyperfine field distribution curves of the hyperfine field derived from the two spectra exhibit pronounced difference.

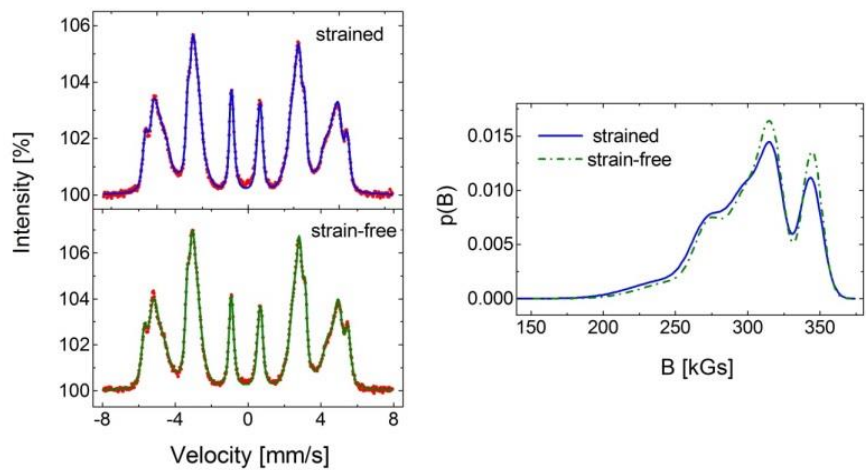

Fig. 29 $^{57}$Fe-site CEMS spectra recorded at RT on strained and strain-free samples of the $Fe_{89.15}Cr_{10.75}$ alloy. The right-hand panel shows the distributions of the hyperfine field derived from the spectra [69].

The spectra were analyzed by means of the M12 model, and using necessary parameters and formulas (8a) to (8c), the SRO parameters were calculated. They are presented in Fig. 30.

A clear difference between the strained and strain-free state is reflected in the values of the SRO-parameters. The most significant contrast takes place in $\alpha_1$ which is positive for the strain-free sample and negative for the strained one. The values of $\alpha_2$ are negative in both cases, but in the strained sample the degree of the short-range order is much higher On average i.e. within the volume of the first two coordination shells the distribution of Cr atoms is fairly random in the strain-free sample, whereas a high degree of the short-range order exists in the strained sample.

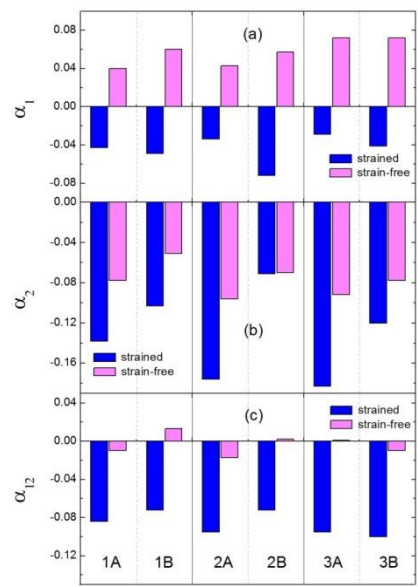

Fig. 30  SRO parameters for the three samples of the investigated alloy. The letters A and B label both surfaces of each sample [69].

Kozlov et al. investigated an effect of a warm plastic deformation induced by high pressure torsion [70]. Their measurements were carried out in TRANS mode on Fe-Cr alloys containing 6, 9.4 and 13.2 at.% Cr. They determined the SRO-parameter averaged over 1NN+2NN shells, $<\alpha_{12}>$, and found that (a) the applied deformation accelerated the SRO-ordering and (b)  $<\alpha_{12}>$ - parameter showed an inversion i.e. it was negative for $x$=6, weakly positive for 9.4 and positive for 13.2.

Shabashov et al. studied an influence of a severe deformation on SRO-ordering in Fe-Cr alloys containing 12, 13.2, 20.6 at. % [68]. The deformation was caused by a cold and warm HTP method, filing and milling. Despite the measured spectra were analyzed in terms of the two-shell model with the additivity condition, the authors did not distinguish between the particular shells in discussing the effect on a redistribution of Cr atoms. The observed changes were expressed in probabilities of atomic configurations with 0, 1, 2, 3 and 4 Cr atoms present within the first-two shells as well as the average local concentrations. In any case, the reported results gave evidence that the clustering of Cr atoms took place in the studied samples.

**5. Phase diagram**

Mössbauer spectroscopy has also been successfully applied in the study of issues pertinent to the crystallographic and magnetic phase diagrams what has been reflected in a great body of published papers relevant to the issue.

**5.1. Crystallographic**

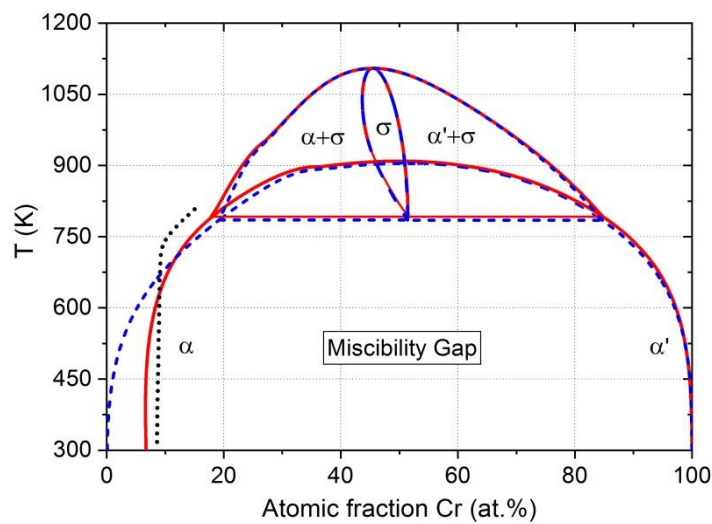

Fig. 31 Calculated crystallographic phase diagram of Fe-Cr adopted from Ref. [53]

Mössbauer spectroscopy has turned out to be suitable tool to study quantitatively the following issues related to this diagram: (a) borders of the miscibility gap (MG) related with the solubility limit of Cr in iron (Fe-rich border of MG) and the solubility limit of Fe in chromium (Cr-rich border of MG), (b) the maximum temperature of MG, (c) kinetics of the phase decomposition (separation) responsible for the occurrence of MG, (d) borders of the metastable part of MG. (e) kinetics of $\alpha \rightarrow \sigma$ phase transformation and (f) borders of the $\sigma$-phase field.

## 5.1.1. Borders of MG

The borders of MG or solubility of Cr in the Fe matrix and solubility of Fe in the Cr matrix can be determined by means of MS with the accuracy ≤ ± 1%. Applicability of MS to study this aspect of the phase diagram is based on the fact that at room temperature the Fe-rich (α) phase is magnetic, hence the corresponding spectrum has a form of sextet, while the Cr-rich (α') phase is at RT paramagnetic, thus its spectrum has a form of singlet. Normally, after the phase decomposition has been completed the two phases co-exist and the spectrum contains both magnetic and paramagnetic components as illustrated in Fig.32.

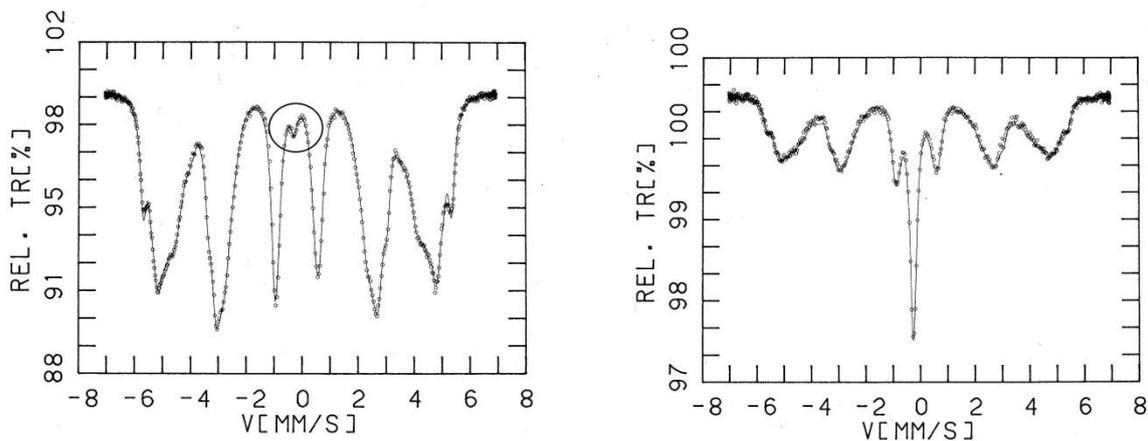

Fig. 32 $^{57}$Fe Mössbauer spectra recorded at RT on: (left-hand side) $Fe_{79.2}Cr_{20.8}$ alloy annealed at 773 K for 35500 h, and (right-hand side) $Fe_{29.4}Cr_{70.6}$ alloy annealed at 733 K for the same period [38]. The singlet is associated with α'.

Probably the first Mössbauer spectroscopic study related to the issue was that by Yamamoto who investigated four Fe-Cr alloys containing between 20.1 and 46.6 wt.% Cr and aged at 773 K for 150 h [71]. Based on the values of *<B>* reported by Yamamoto and the *<B(x)>* relationship given by Dubiel and Inden [38] one can determine the values of the Cr content in the Fe-rich phase as 7.3 ± 3.6 at.% for the least Cr-concentrated sample and 12.4 ± 3.6 at.% for the most concentrated one. Solomon et al. studied an effect of aging on several Fe-Cr alloys at 748 K in relation with the so-called "475°C embrittlement" and underlying mechanism(s) [72] . They did not determined *explicite* quantities directly related to the borders of MG, but determined values of the isomer shift of the paramagnetic line present in the spectra measured on the annealed samples. They ascribed this line to the Cr-rich α' phase. Based on the *IS(x)* relationship given in [38] one can estimate the concentration of Cr in that phase, hence the solubility limit of Fe in Cr as lying between 90.8 and 94.5 at.% at 748 K. Similarly, using the *<B(x)>* relationship given in the same paper, one finds that the concentration of Cr in the Fe-rich phase falls in the range 12.7 - 15.3 at.%. More systematic study relevant to the problem was carried out by Kuwano [73]. He investigated a series of Fe-Cr alloys containing between 10.2 and 71.1 at.% Cr. The alloys were annealed at temperatures between 748 and 808 K for up to 3000 h. From his measurements he found that: (1) the Fe content in the Cr-rich lied between 10 ± 1 at. % at 748 K and 13 ± 1 at. % at 808 K, and (2) the concentration of Cr in the Fe-rich phase was between 85±2 and 81±2 at.%. This trend in the narrowing of the width of MG with increasing temperature or increase of the solubility limits is, at least qualitatively, in line with theoretical predictions shown in Fig. 31. The process of the phase decomposition progresses with its characteristic kinetics and the rate of the evolution depends on the annealing temperature and the initial content of Cr in the sample. Consequently, long or very long annealing times are needed to achieve the final state of the decomposition and connected with this state correct value of the MG borders. Dubiel and Inden investigated Fe-Cr samples with the record-high annealing times viz. between $3.5 \cdot 10^4$ and $10^5$ h at 733 K and 773 K [38]. Consequently, one can assume that the decomposition process was terminated after performing such long annealing. Based on the analysis of the measured spectra the authors concluded that the concentration of Cr in the Fe-rich phase was between 12 and 12.9 at. % in the 15 wt. % Cr sample annealed at 733 K, between 14 and 14.3 at.% in the 20.8 wt.% Cr sample annealed at 773 K and between 15.7 and 16.5 at.% in the 70.6 wt.% sample annealed at 733 K. The content of Fe in the Cr-rich sample for the latter sample was determined as 88.0 at%. Costa et al. determined the values of the figures in merit by performing measurements on coarse grained $Fe_{55.5}Cr_{45.5}$ and nano-crystalline samples of $Fe_{55.5-x}Sn_xCr_{44.5}$ (x=3.5 and 6.2) annealed at 748 h for up to 1500 h [74]. Their values for the Cr content in the Fe-rich phase lie between 14.6(6) and 15.0(6) at.% for the nano-crystalline samples and 15.7(6) at.% for the coarse grained one. In turn those for the Fe content in the Cr-rich phase are 88 – 89 at.% for the nano-crystalline alloys and 93(2) at.% for the coarse grained one. It is not clear whether or not the pretty large difference in the latter case is due to the presence of Sn or to the difference in the grain size.

The values of the Cr concentration in the Fe-rich phase can be also regarded as the solubility limit of Cr in the iron matrix. According to theoretical calculations these values should be larger at higher temperatures. The above-given values show some dispersion that can have various reasons among which the most obvious could be not-long-enough annealing time i.e. lack of termination of the phase decomposition process. Less obvious reason might be the metallurgical state of the investigated samples. The nano-crystalline samples were obtained by a mechanical alloying (milling), a process that brings about strain. Concerning the latter, Dubiel and Żukrowski have recently shown that the concentration of Cr in the Fe-rich phase was 11.6(4) at.% in the strained sample and 13.2(4) in the strain-free one [58]. Also irradiation of Fe-Cr alloys may change the solubility limits. The same authors revealed for the $Fe_{84.85}Cr_{15.15}$ alloy isothermally annealed at 688 K that the Cr concentration in the Fe-rich phase in the non-irradiated sample was 10.5(5) at.% against 13.5(2) at.% in the $He^+$ irradiated one [75]. Thus the irradiation increased the solubility of Cr in iron by 3 at.%.

In the last decade Dubiel and Żukrowski carried out a series of measurements aimed not only at determining the solubility limit of Cr in iron and that of Fe in chromium, but also the kinetics of the phase decomposition at various temperatures falling in the range 675 – 832 K was of interest [57-59,75,76]. The former aim was motivated by differences in the calculated phase diagrams of the Fe-Cr system, especially its part pertinent to the Fe-rich miscibility border [53,55,77]. Relevant example is shown in Fig. 33.

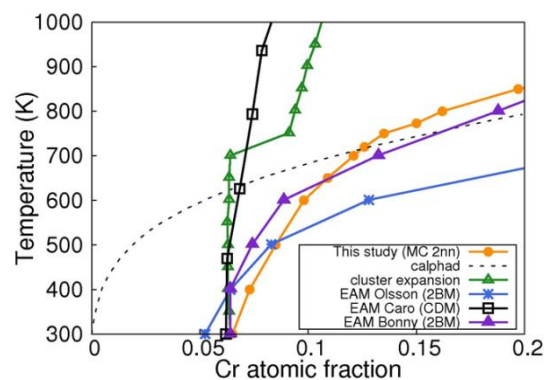

Fig. 33 Fragment of the crystallographic phased diagram of the Fe-Cr system presenting different calculated Fe-rich borders of MG [77].

A set of the spectra recorded on the $Fe_{84.85}Cr_{15.15}$ sample annealed at 688 K for different time is displayed in Fig. 34.

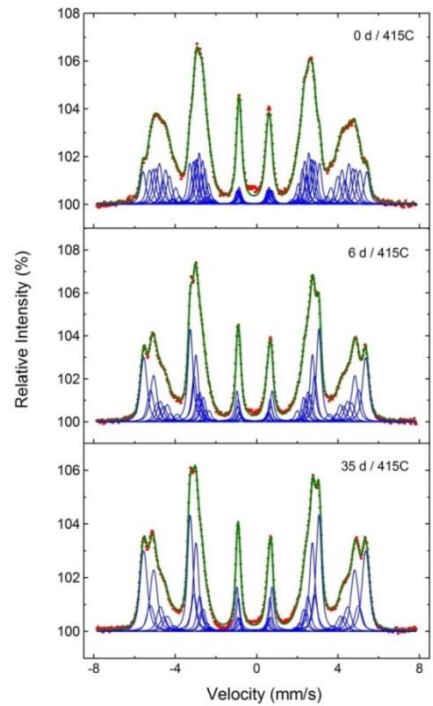

Fig. 34 CEMS spectra recorded at RT on the $Fe_{84.85}Cr_{15.15}$ sample annealed at 688 K for different time shown. Individual sub spectra corresponding to various atomic configurations taken into account are indicated [57].

The best-visible annealing-induced changes in the spectra can be seen in the outermost lines. The hyperfine field distribution curves, p(B), derived from the spectra visualize these changes even better – see Fig. 35.

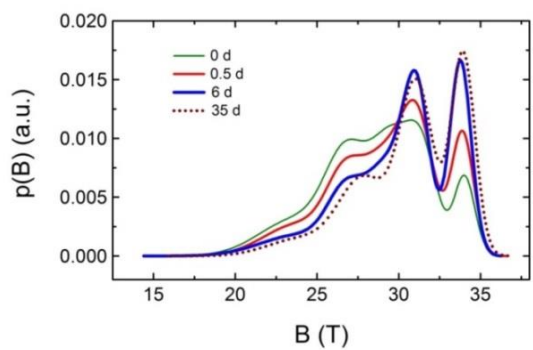

Fig. 35 *p(B)*-curves as obtained from the spectra analyzed with the HFD method for different annealing periods shown [57].

To notice is a shift of the distributions to the higher *B*-values and a growth of the peak centered at the maximum value of *B*≈33 T. This peak can be ascribed to those Fe atoms that in their neighborhood do not have Cr atoms. Thus its intensity can be compared with the probability of the (0,0) atomic configuration, P(0,0), obtained by fitting the spectra with the two-shell model described in section 3. Indeed, the relevant comparison presented in Fig. 36 justifies this interpretation.

As already mentioned values of the solubility limits can be regarded as correct provided the phase decomposition process has been terminated. In the case of studying the phase separation by means of the Mössbauer spectroscopy, the average hyperfine field, *<B>*, can be taken as the best indicator for that. Namely, *<B>* is linearly correlated with the number of Cr atoms in the vicinity of the probe Fe atoms, N viz. *<B>* decreases when N increases and *vice versa* [36]. As in the phase decomposition process Cr atoms tend to cluster to ultimately form the Cr-rich $\alpha'$ phase the number of Cr atoms neighboring the Fe probe decreases. The effect is reflected in the increase of *<B>* versus annealing time. A typical dependence of *<B>* on annealing time, *t*, is illustrated in Fig. 37.

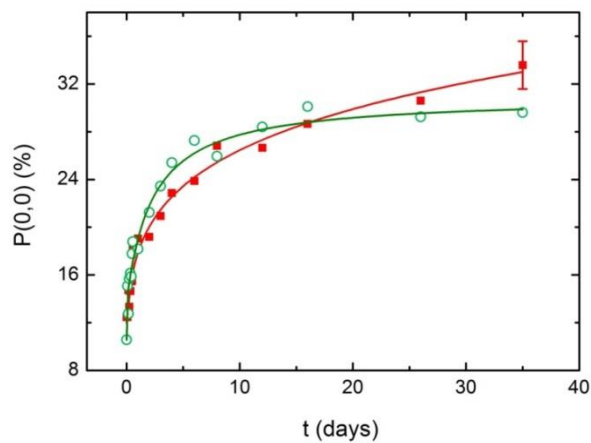

Fig. 36 Probability of the (0,0) atomic configuration (squares), P(0,0), as found by analyzing the spectra with the two-shell model, and the relative contribution of the ~33 T peak (circles) as determined using the hyperfine field distribution approach [57].

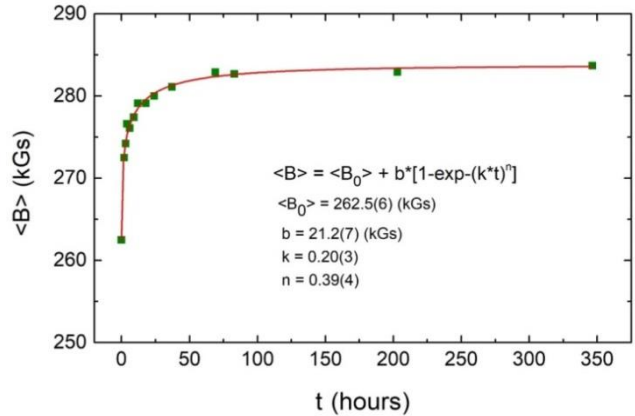

Fig. 37 The average hf. field, <B>, versus the annealing time, t, for $Fe_{73.7}Cr_{26.3}$. The best-fit of the JMAK equation to the data is shown as full line, and the best-fit parameters are shown, too. Avrami exponent is depicted by n and the rate constant by k [76].

Obviously, <B> saturates what can be regarded as the sign of the phase decomposition termination. The value of <B> in saturation is thus the average value of the hyperfine field in the Fe-rich phase. It can be next used to determine the concentration of Cr in this phase, hence the limit of Cr solubility in the Fe matrix. For that purpose one takes into account the linear correlation between <B> and Fe content, x, reported in several papers to hold [36, 38, 52, 78]. Figure 38 illustrates such correlation.

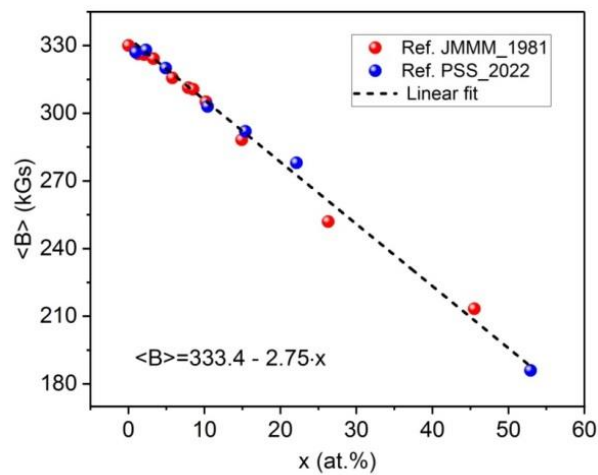

Fig. 38 <B> - x plot made by the author to illustrate the linear relationship for $Fe_{100-x}Cr_x$ alloys. The plot is based on the data reported in [38, 52].

On the other hand, the concentration of Fe in the Cr-rich α' phase, can be inferred based on the linear correlation between the isomer shift, *IS*, and Cr content in Cr-rich, hence paramagnetic Fe-Cr alloys. The correlation made using the data reported in [38] is displayed in Fig. 39.

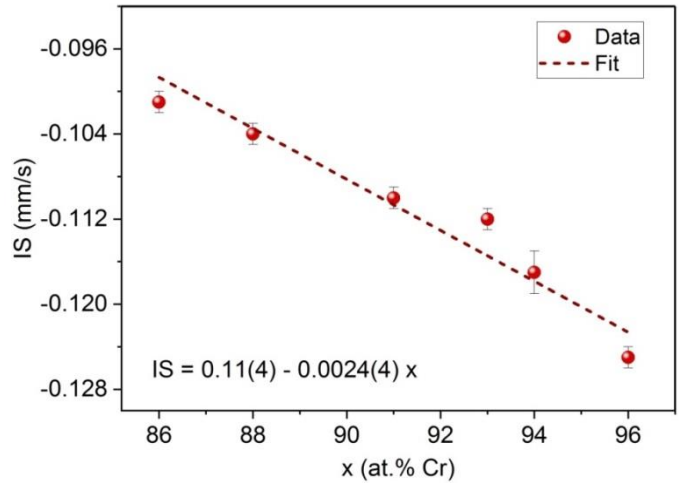

Fig. 39 Relationship between the isomer shift, *IS*, and the concentration of Cr in the Cr-rich (paramagnetic) Fe-Cr alloys. The plot made based on the data from Ref. 38.

Because the Cr content in the α' phase lies in the range ~80-90 at.%, the spectral component associated with α' is a singlet. It becomes visible in a spectrum provided its relative contribution is ≥ ~0.5 %. Consequently, in alloys with Cr concentration less than, say, 25 at.%, the statistical quality of the spectrum must be high in order to observe this singlet – see the left-hand spectrum in Fig. 32 or the bottom spectrum in Fig. 40. However, in alloys with a higher Cr content the intensity of the singlet cannot remain unnoticeable – see the right-hand spectrum in Fig.32 and/or spectra in [74] .

According to the measurements by Dubiel and Żukrowski the upper temperature limit for the occurrence of MG lies between 800 and 832 K. This temperature interval follows from that fact that the kinetics of the transformation due to the annealing at 800 K took place in one step i.e. it could have been correctly described in terms of only one JMAK equation [76]. This feature means that there is only one kind of the process, as expected to occur within the phase field of MG. Noteworthy, this type of the transformation process was also revealed for annealing at other temperatures lower than 800 K [57,58,75].

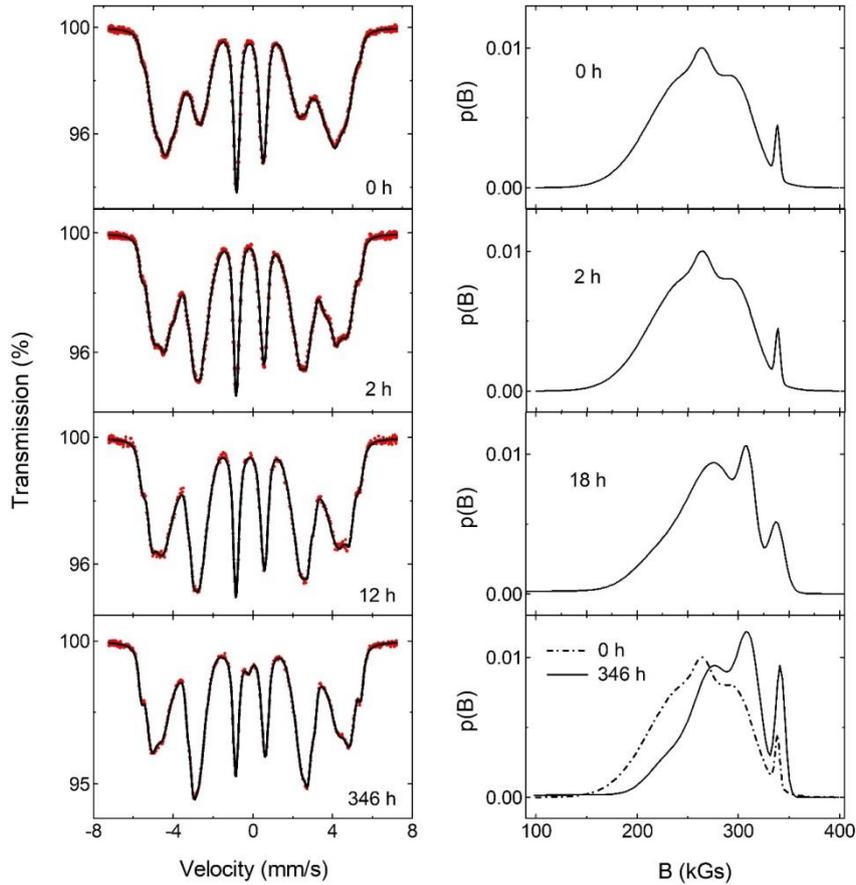

Fig. 40 (left panel) Set of $^{57}$Fe Mössbauer spectra recorded at RT on a Fe$_{73.7}$Cr$_{26.3}$ sample annealed at 800K for various periods shown. Note a single-line contribution in the central part of the bottom-most spectrum. It is associated with the Cr-rich α' phase. Distribution curves of the hyperfine field, p(B) derived from the spectra are shown alongside in the right panel. On the bottom plot a comparison is made between the extreme cases [76].

However, the kinetics of the transformation uncovered from the spectra measured on the Fe-Cr sample annealed at 832 K exhibited a two-step character [59]. The latter can be readily inferred from the dependence of <B> on the annealing time, as presented in Fig. 41. Consequently, the kinetics of the transformation had to be analyzed in terms of the JMAK equation with two sets of kinetics parameters.

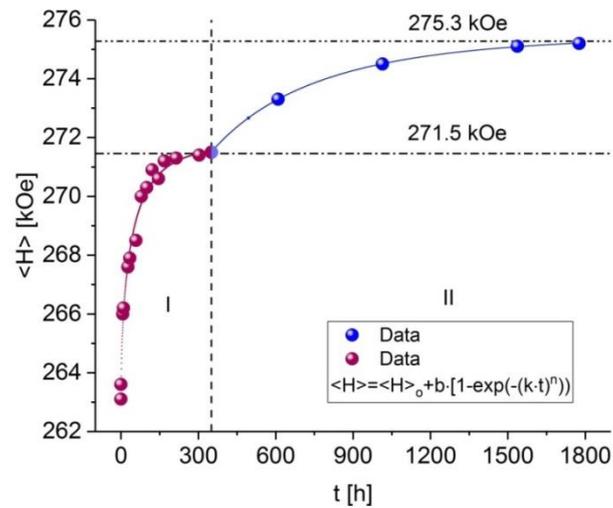

Fig. 41 The average hyperfine field, <H>, vs. annealing time, t. The solid lines are for the best-fits to the data in terms of the JMAK equation for range I and II, separately [59].

The single-like line present in two spectra recorded on the sample annealed for the longest periods (1577 and 1777h) was identified as due to the σ-phase [59]. This and the two-step character of the kinetics mean that the temperature of 832 K already falls above the MG field.

The value of <B> = 271.5 kGs (first step of the transformation) was used for determining the *locus* of the metastable MG at 832 K. Based on the value of 275.3 kGs (second step of transformation) and the plot shown in Fig. 38. The solubility limit of Cr in iron as 20.7 at.% at 832 K could have been determined.

### 5.1.2. Comparison with predictions

Mössbauer spectroscopic data depicting the borders of MG can be used for validation of relevant theoretical predictions. Such validation done by Dubiel et al. is presented in Figs.42-44.

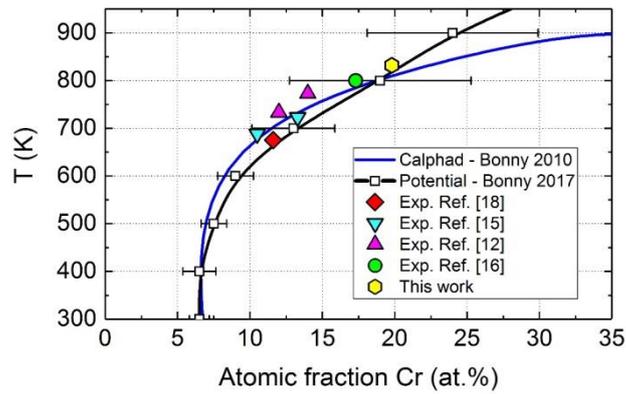

Fig.42 Comparison between two theoretical predictions concerning the Fe-rich branch of the MG [79] and experimental data obtained with the Mössbauer spectroscopy [76]. The size of the symbols is comparable with the error bar.

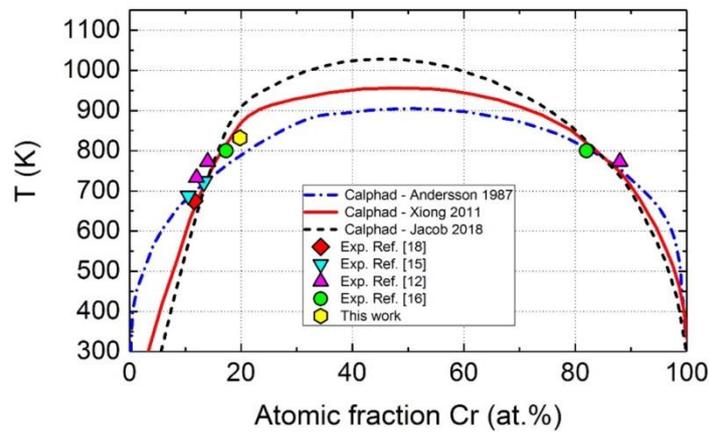

Fig.43 Comparison between three theoretical Calphad-based predictions concerning the full MG adopted from [55] and experimental data obtained with the Mössbauer spectroscopy [76]. The size of the symbols is comparable with the experimental error

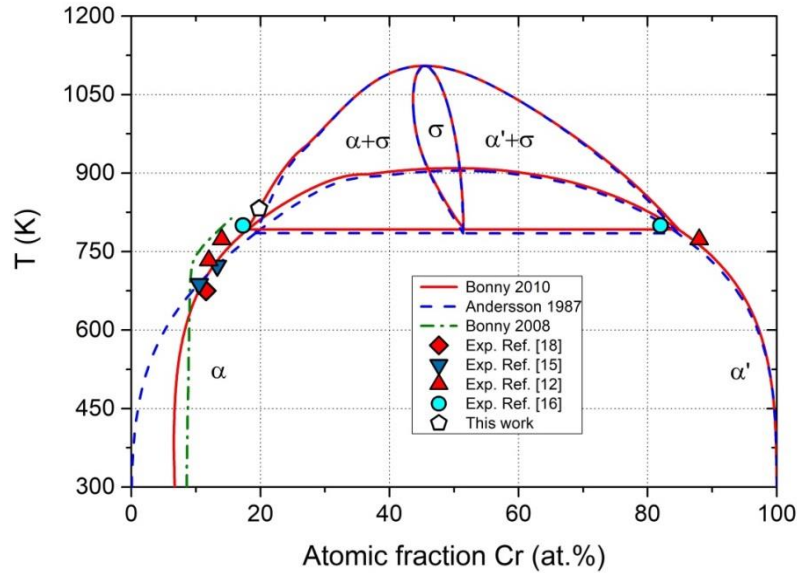

Fig. 44 Relation between experimental data obtained with the Mössbauer spectroscopy [76] and the calculated phase diagram of the Fe-Cr system adopted from [53]. Size of the symbols corresponds to experimental uncertainty.

The experimental data found by means of the Mössbauer spectroscopy are consistent with each other. Nevertheless the temperature interval over which they had been measured does not allow making the unique distinction between various predictions. As can be seen in Figs. 43 and 44 the most meaningful difference between the predictions exist for temperatures smaller than ~600 K, as well as higher than ~850 K. Measurements below the former temperature are not feasible because of an extremely low kinetics. However, those at $T \geq$ ~850 K are achievable and they should be carried out.

### 5.1.3. Kinetics of phase decomposition

There are two mechanisms responsible for the phase decomposition resulting in MG viz. (a) nucleation and growth (NG) and (b) spinodal (SP). The latter takes place in the central part of MG and the former outside SP regime on its both sides as schematically shown in Fig. 45. Positions of the border-lines for the SP are temperature dependent. For example, the experimentally determined spinodal regime lies within 36 and 70 wt.% Cr at 800 K [80]. However the exact borders between the NG and SP mechanism are not precisely known because the relevant experimental data are widely spread [81]. Some researchers claim that clear-cut lines between NG and SP modes of the phase decomposition do not exist [82-84]. This disagrees with the Mössbauer spectroscopic studies according to which the two modes of decomposition are distinguishable from each other [73,85,86]. Also model calculations of various quantities pertinent to the Mössbauer-effect performed for the two decomposition modes demonstrated that the distinction between NG and SP decomposition modes were feasible [87]. According to these calculations the most significant difference was predicted for the middle stages of the decomposition.

Analysis of the Mösbauer spectra recorded on isothermally annealed Fe-Cr samples at temperatures lower than the upper temperature limit for the existence of MG permits getting insight into both mechanisms. The decomposition process via SP was, in particular, studied by means of MS by de Nys and Gielen [88], Chandra and Schwartz [89] as well as by Kuwano [73]. De Nys and Gielen investigated a series of four $Fe_{100-x}Cr_x$ alloys with 20.1 ≤ x ≤ 49.6 annealed at 743 and 748 K. Based on the analysis of the Mössbauer spectra they concluded that the less Cr-concentrated alloy decomposed via NG and the most concentrated one by SP. Chandra and Schwartz studied the effect of aging at 748 K on Fe-Cr alloys with 24, 30, 37, 44 and 60 at.% Cr and concluded that the alloys with Cr content between 12 and 30 at.% decomposed by NG. In contrast, the SP mode of decomposition was pinpointed to the $Fe_{40}Cr_{60}$ alloy [89].

Kuwano systematically studied the Fe-Cr alloys containing between 21 and 63.9 at.% Cr in order to distinguish between the two mechanisms responsible for the phase decomposition. To this end the alloys were annealed at 748 K for different periods. According to him the distinction could be made by tracing the incubation time needed for a change of the Cr atoms distribution during the aging. The average hyperfine field, *<B>*, was used as an indicator of the incubation time. Based on the *<B>(t)* behavior he detected the incubation time for x=21, 36.3 and 63.9 and concluded that the NG mechanism was responsible for the decomposition in these alloys.

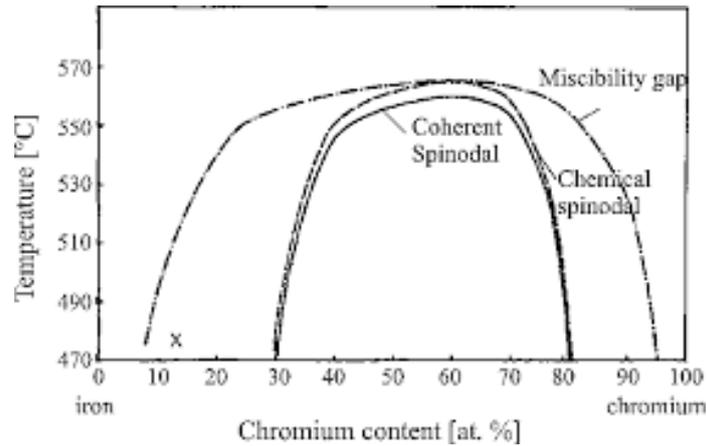

Fig. 45 Schematic presentation of the fragment of the Fe-Cr phase diagram showing the spinodal area within the miscibility gap [56].

On the other hand, no incubation time was detected in the alloys with the Cr content between 31.5 and 55.6 at.%, thus in these alloys the decomposition proceeded via the SP mechanism. Based on his study Kuwano concluded that the Fe-Cr alloys containing between 42 and 60 at.% Cr decomposed via the spinodal, while in the alloys with 21 to 36.3 at.% Cr the nucleation and growth mechanisms was active. He also suggested that there was no clear-cut border between the two mechanisms. Instead with increasing Cr content a gradual transition from NG to SP took place. Mössbauer spectroscopy was much more frequently applied to study the kinetics of the phase separation outside the spinodal field and, in particular, in the Fe-rich alloys, because such alloys are of a great industrial importance.

Kuwano also studied the kinetics of the phase separation in the $Fe_{63.7}Cr_{36.3}$ alloy that was subject of the isothermal annealing at various temperatures in the range 748 – 833 K [73]. He determined the activation energy, $E$, in two ways: (1) from the incubation time and (2) from the JMAK and Arrhenius equations. The former approach yielded $E$=204 kJ/mol, while the latter $E$=193(3) kJ/mol.

The best Mössbauer spectroscopic parameter for following the kinetics of the phase decomposition by NG turned out to be the average hyperfine field, $<B>$. Its typical dependence on the annealing time, $t$, is displayed in Fig. 37. The $<B>(t)$ data can be well fitted to the JMAK-like equation:

$$<B> = <B_o> + b[1 - \exp(-kt)^n] \qquad (10)$$

Where <$B_o$> is the value of the average hyperfine field for the non-annealed sample, $k$ is the rate constant, $n$ is the Avrami exponent, and $b$ is a free parameter.

The kinetic parameter $n$ gives information on the mechanism underlying the decomposition [90] and $k$, via the Arrhenius law, can be used for determining the activation energy, $E$, provided the kinetics was measured at two different temperatures, $T_1$ and $T_2$. For this purpose the following formula can be used:

$$E = \frac{T_1 T_2}{T_2 - T_1} k_B \ln(\frac{k_2}{k_1}) \qquad (11)$$

Where $k_B$ is the Boltzmann constant.

In particular, using the value of $k_2$ =0.3 and $T_2$ = 688 K reported in [75] and the taking into account the rate constants obtained for the strained and the strain-free samples annealed at $T_1$= 675 K [58] one could determine the values of $E$ for the sample in two different metallurgical conditions – see Table 2.

Table 2. The best-fit kinetic parameters $k$ and $n$ obtained by fitting the annealing time dependence of the average hyperfine field, <$B$>, to eq. (1), and the values of the activation energy, $E$, obtained with eq. (2) [58].

| Sample | $B_o$ [kOe] | b | k [h$^{-1}$] | n | E [kJ/mol] |
|---|---|---|---|---|---|
| Strained | 284.9(3) | 9.7(3) | 0.0011(1) | 2.49(7) | 1665.8 |
| Strain-free | 285.6(4) | 4.8(5) | 0.0028(1) | 2.49(3) | 1388.3 |

Evidently, the effect of strain has been reflected in the value of $k$ and, consequently, in that of $E$. The former indicates that the phase decomposition in the strain-free sample is almost 3-times faster. On the other hand, values of $n$ are the same and indicate that the phase decomposition in the studied samples has the same character viz. mixed nucleation and volume diffusion controlled growth [90]. It is worth mentioning that the $E$-values given in Table 1 are 6-7 times larger than the activation energy determined for $T_1$=668 K and $T_2$ = 723 K [75]. These figures clearly show how dramatically a decrease of $T$ slows-down the process of the phase decomposition caused by NG. This temperature effect also was reflected in the values of the rate constant reported in [75] viz. 0.3 s$^{-1}$ for $T$=688 K and 2.2 s$^{-1}$ for $T$=723 K.

The kinetics of the phase decomposition via the NG mechanism depends not only on temperature, metallurgical state (strain) but also on irradiation. For example the kinetics parameters found for the Fe$_{84.85}$Cr$_{15.15}$ sample annealed at 688 K were $k$=0.3 s$^{-1}$ and $n$=0.6(1) while those for the same sample but irradiated with 25 keV He-ions to the dose of 7.5 dpa were $k$=3.9 s$^{-1}$ and $n$=1.0(3) [75]. The values of the rate constant clearly demonstrate that the applied irradiation significantly accelerated the kinetics of the phase decomposition in the studied sample.

## 5.2. Sigma phase

It follows from the crystallographic phase diagram of the Fe-Cr systems that in its central part exists the sigma (σ) phase. The phase-field of σ lies between ~45-50 at.% Cr and ~800 and 1100 K. However, the low temperature limit for the occurrence of σ, being at the same time the upper temperature limit for MG, is not precisely known neither from experimental nor theoretical viewpoint. The σ-phase in the Fe-Cr system is regarded as an archetype because its crystallographic structure was for the first time identified in the Fe-Cr alloy [91]. In the tetragonal unit cell of σ there are 30 atoms distributed over five different lattice sites (A, B, C, D, E). All of them to different degree are occupied by Fe atoms. At room temperature σ-FeCr is non-magnetic, hence its Mössbauer spectrum consists of five doublets with various isomer shifts and relative contributions. The values of $QS$ of the doublets are small or moderate, so the RT spectrum is poorly resolved as shown in Fig. 46 making the analysis of the spectrum challenging and not unique. For example, Solly and Winquist analyzed the σ-phase spectrum in terms of three singlets [92], whereas Gupta et al. did so in terms of one singlet for the site E and two doublets; one for the sites B+C and another one for A +D [93].

The proper analysis of σ-phase spectra must be backed by measurements with neutrons (in order to determine the relative population of Fe atoms on particular lattice sites) and theoretical calculations giving information on charge-densities [94,95]. With this information correct analysis of the spectra can be successfully done [94].

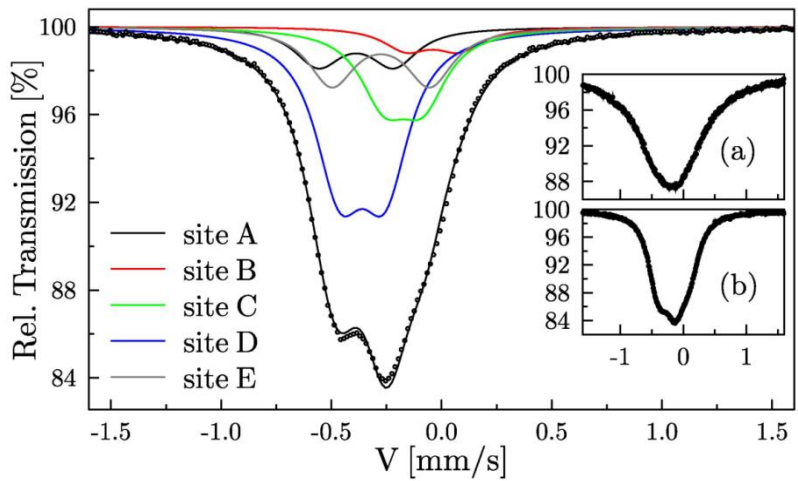

Fig. 46 Mössbauer spectrum recorded at RT on σ-FeCr alloy and fitted to five doublets corresponding to five lattice sites as denoted in the legend. The inset shows a comparison between the spectrum measured at 4 K (a) and at RT (b). The figure adopted from [94].

Interest in the σ-FeCr compounds has been two-fold: (a) scientific and (b) industrial. The former stems from its crystallographic complexity and interesting physical properties like low-temperature magnetism having a re-entrant character [96], and the latter due to possibility of σ precipitation in structural materials, like steels based on Fe-Cr alloys. The presence of even a small amount of σ drastically deteriorates their mechanical and corrosion properties. MS has proved to be suitable method to study various aspects related to σ, in particular, the kinetics of the α→σ phase transformation (σ precipitates by annealing α-FeCr alloys at ~800-1100 K).

### 3.2.1. Kinetics of α→σ

The kinetics of the σ-phase formation can be studied either *ex situ* or *in situ*. The former mode is based on the fact that (1) σ-phase is paramagnetic and α-phase is magnetic at RT – see Fig. 47, so the two phases can be easily distinguished from each other. In the *in situ* studies performed above ~600 K the quasi-equiatomic α-FeCr alloys are paramagnetic, so the Mössbauer spectrum has the form of a single line i.e. singlet. Yet, as illustrated in the right-hand panel of Fig. 47, it even optically differs from the spectrum of σ. The values of *IS* of both phases are also different. The singlet of α shown in Fig. 47 is centered at -0.53(1) mm/s, and the center of gravity (the average IS) of the σ spectrum at -0.63(3) mm/s. Both values are given relative to α-Fe at RT. By the way, the difference in the isomer shift values between the two phases is rather exceptionally large for metallic system. In total, the α→σ phase transformation can be quite precisely studied by means of MS. Noteworthy, the difference in the shape and in the isomer shift values between σ and Fe-rich α' can be also used for determining the lower temperature limit for the formation of σ.

The first known investigation of the σ-phase formation in a Fe-Cr alloy was carried out in 1955 by Baerlecken and Fabricius by means of a magnetization method [97]. The kinetics curves had S-shape characteristic of the NG process underlying the transformation mechanism. The description of the process was done with the Johnson-Mehl-Avrami-Kolmogarov (JMAK) equation:

$$A_\sigma(t) = 1 - \exp[-(k \cdot t)^n] \qquad (12)$$

Where $A_\sigma(t)$ is the amount of σ precipitated after time *t* of annealing, *n* is the Avrami exponent and *k* the rate constant. The rate constant is via the Arrhenius law related to the activation energy, *E*, by the following formula:

$$k = k_o \cdot \exp(-\frac{E}{R \cdot T}) \qquad (13)$$

Here *R* is the gas constant, *T* is temperature, and $k_o$ a prefactor.

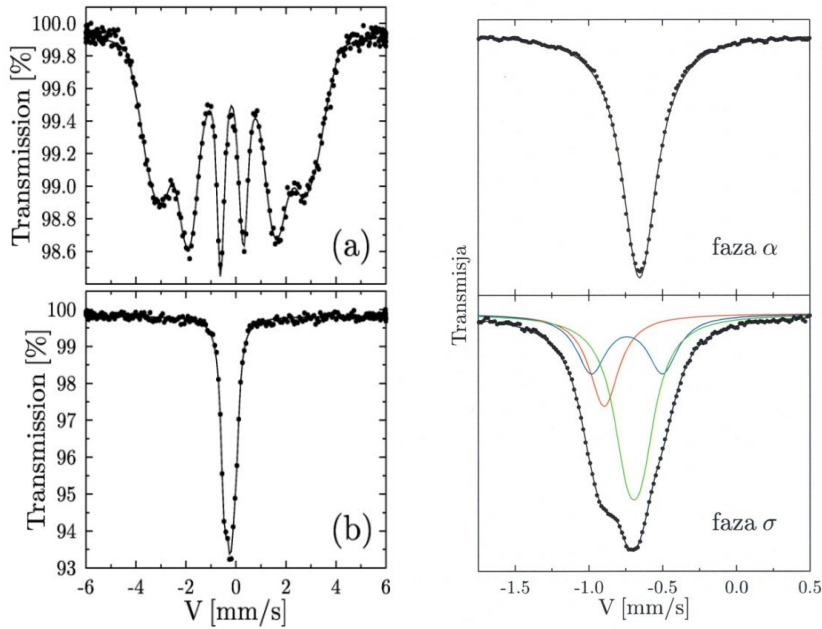

Fig. 47 (left-hand panel) Mössbauer spectra recorded *ex situ* at RT on (a) α-Fe$_{53.8}$Cr$_{46.2}$ and (b) σ-Fe$_{54.5}$Cr$_{45.5}$, and (right-hand panel) Mössbauer spectra recorded *in situ* at 973 K on the same samples [136].

Examples of the spectra measured with the *ex situ* method illustrating various stages of the α→σ transformation are presented in Fig. 48.

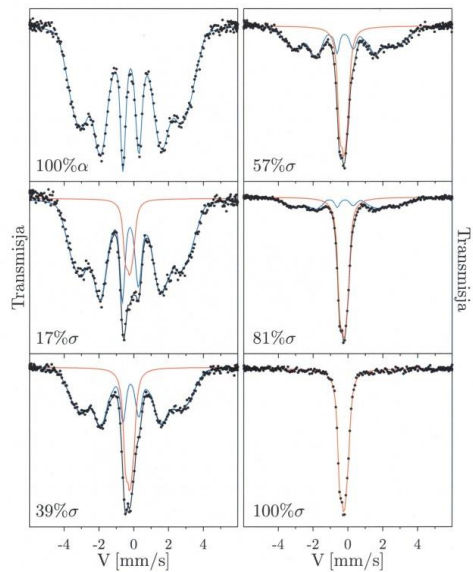

Fig. 48 Mössbauer spectra measured at RT on a sample of Fe$_{53.8}$Cr$_{46}$.0.3at.%Ti annealed at 973 K. Subspectrum due to σ is marked in red. The figure adopted from [98].

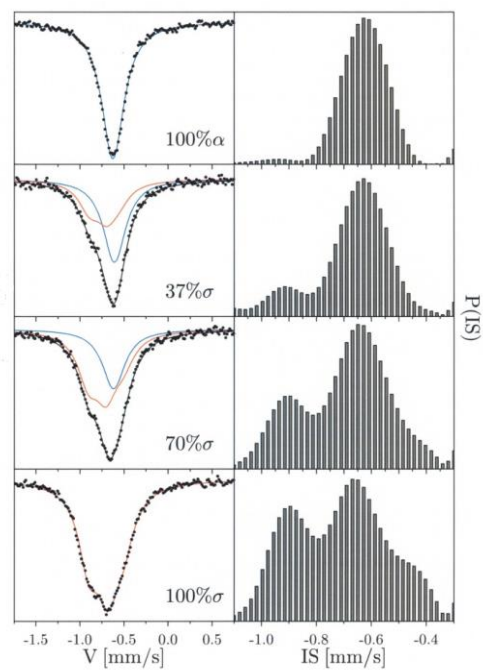

Fig. 49 (left-hand panel) Mössbauer spectra measured *in situ* on a sample of $Fe_{53.8}Cr_{46}$.0.3at.%Ti annealed at 973 K. (right-hand panel) Histograms of the distribution of the isomer shift derived from the spectra. The figure adopted from [99].

There are also two ways of following the kinetics of the precipitation of σ. First, the standard way i.e. from the relative spectral area and second, from the average isomer shift, <*IS*>, of the spectrum or its center of gravity (COG) [100,101]

The fraction of the precipitated σ-phase can be expressed by the following formula:

$$A_\sigma = \frac{f_\alpha \cdot S_\sigma + f_\sigma \cdot S_\alpha}{f_\alpha \cdot S_\sigma} \qquad (14a)$$

Where $f_k$ (k=α, σ) is the *f*-factor for α or σ, and $S_k$ is the spectral area.

Assuming equal detectability of the two phases i.e. $f_\sigma = f_\alpha$ the formula (14a) simplifies into the following one:

$$A_\sigma = \frac{S_\sigma + S_\alpha}{S_\sigma} \qquad (14b)$$

The validity of such assumption was verified and the authors concluded that values of the kinetics parameters obtained using formula (14a) were within error limit the same the ones determined based on formula (14b) [100].

The kinetics of the σ-phase precipitation can be studied with respect to the effect of temperature, metallurgical state of alloy, additive elements, etc.

The first known application of MS in the standard mode to study the α→σ kinetics was the one by Japa et al. [102]. The authors investigated qualitatively the transformation in a pure Fe-Cr alloy annealed at 923 K, 973 K, 1023 K and 1073 K. They found that the quickest transformation process took place at 973 K. More systematic and quantitative study of the effect of temperature on the α→σ phase transformation was carried out by Błachowski et al. [98]. Performing the *in situ* measurements and analyzing the spectra with two methods they determined the kinetics parameters in a pure and in a Ti-doped (0.3 at.%) $Fe_{53.8}Cr_{46.2}$ alloy. For the former the temperature range was 943 – 978 K, and for the latter between 923 and 1003 K. In the pure Fe-Cr alloy the fastest transformation took place at 978 K and in the Ti-doped one at 1003 K.

Kuwano et al. using MS and the classical approach determined for the first time the value of the activation energy, $E$ = 193 kJ/mol, for the formation of σ in the Fe-Cr system [73]. Costa and Dubiel investigated an effect of Sn on the kinetics in quasi-equiatomic Fe-Cr alloys [103]. They found that addition of Sn retards the formation process of σ, partly due to precipitation of Sn on grain boundaries known as the cradle for the σ precipitation. Waanders et al. using the classical method studied an effect of Mo, and they revealed that addition of 5.5 at.% Mo increased the activation energy from 200 kJ/mole for $Fe_{50.7}Cr_{49.3}$ to 283 kJ/mole for the Mo-doped sample [104]. The effect of Ti on the formation of σ at temperatures between 753 and 773 K was studied in the $Fe_{53.8}Cr_{46.2}$ alloy with various sizes of grains [99,100]. Addition of Ti up to ~1 at.% decreased the value of $E$ in the coarse-grained sample. Similar effect was found for the fine-grained sample for the addition of Ti up to ~1.2 at.%. Higher amounts of Ti had an opposite effect i.e. they increased the activation energy, hence retarded the process of the σ-phase formation.

Examples of the $A_\sigma(t)$ curves illustrating the effect of Ti on the α→σ phase transformation are displayed in Fig. 50.

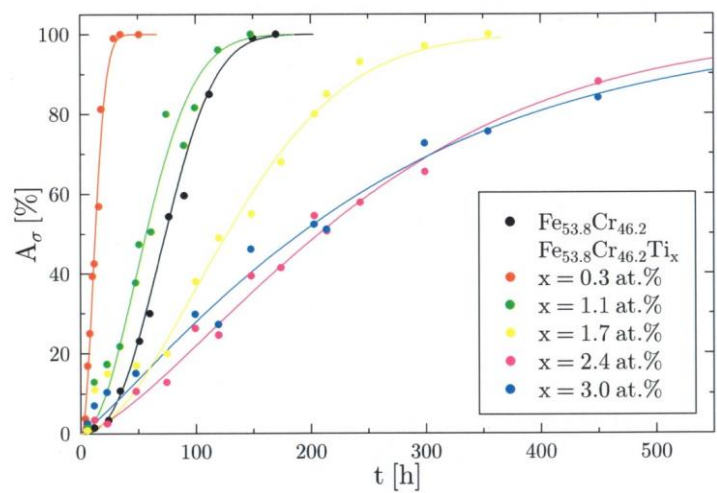

Fig. 50 S-shaped curves characteristic of the NG mechanism underlying the α→σ phase transformation. The plot illustrates a growth of the σ-phase fraction, $A_\sigma$, precipitated in pure and Ti-doped $Fe_{53.8}Cr_{46.3}$ alloys annealed at 973 K. The data

were derived from the spectra measured *ex situ* [98]. The solid lines are the best fits of the data to the JMAK equation.

The effect of Ti content on the activation energy in fine grained and coarse grained samples is illustrated in Fig. 51 where the difference, $\Delta E=E(\text{Ti-doped})-E(\text{undoped})$, is plotted versus Ti concentration.

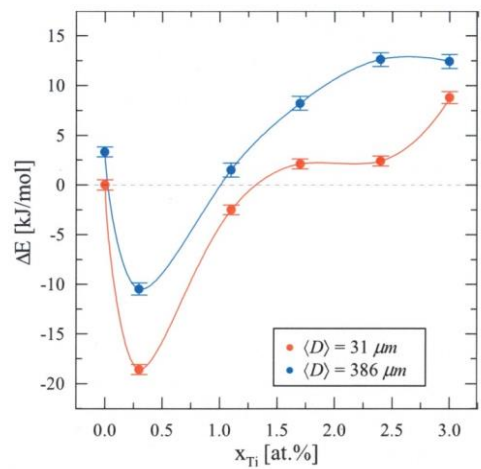

Fig. 51 Ti concentration dependence of the difference in the activation energy, $\Delta E$, for the formation of σ as determined in [99] for the crystalline Ti-doped $Fe_{53.8}Cr_{26.2}$ alloys with different size of grains, <D>.

Clearly the effect of Ti depends on its content in both samples with an inversion at ~1 at.% Ti in the coarse grained sample and at ~1.2 at.% Cr in the fine grained sample. In both samples containing less Ti than these critical values, the presence of Ti accelerates the formation of σ with the maximum rate at ~0.3 at.%. On the other hand, a retardation of the process was revealed for the samples containing more Ti than the critical values. The effect of addition of 0.2 and 1.0 at.% Al was also investigated both in fine grained, <D>=31 μm, and coarse grained, <D>=386 μm, Fe-Cr alloys. [105]. The authors of this study revealed that addition of 0.2 at.% Al hardly affected the transformation in both type of samples, whereas addition of 1 at.% Al significantly retarded the precipitation of σ. The increase of the activation energy for the coarse grained sample was 20.8(5) kJ/mole and 16.5(5) kJ/mole for the fine grained one.

The activation energy can be determined by different methods. Błachowski et al. based on the classical method of the spectra analysis determined values of $E$ using the following methods: (a) rate constant, (b) length of time between two different stages, (c) time to a given fraction, (d) maximum transformation rate and (e) two-temperature kinetics. In fact, the authors revealed that the actual value of $E$ depended on the method used for its determination. For instance, $E$ ranged between 197 and 260 kJ/mol in case of the $Fe_{53.8}Cr_{46.2}$ alloy and between 119 and 183 kJ/mol for this sample doped with 0.3 at.% Ti [98].

Cieslak et al. had shown that the kinetics of the $\alpha \rightarrow \sigma$ phase transformation could be quantitatively followed based on the temperature dependence of the average isomer shift (center of gravity) [100,101]. This method is advantageous especially in the tracing the kinetics from the spectra recorded *in* situ (because determining the center of gravity of a spectrum is easier and more precise than determining the spectral area of poorly resolved spectra). A comparison between the results obtained with the classical and the COG methods is displayed in Fig. 52.

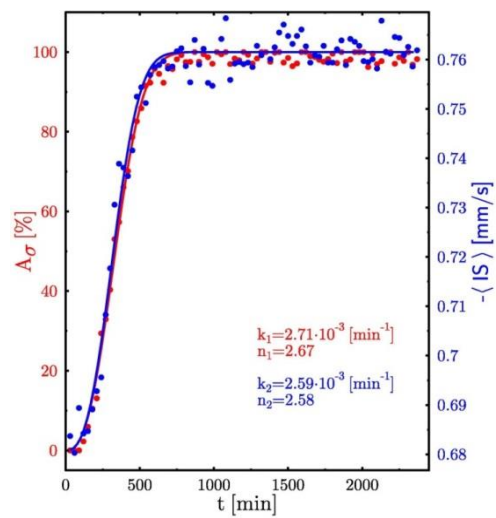

Fig. 52 Dependence of the σ-phase fraction, $A_\sigma$, (left-hand scale) and of the average isomer shift, <*IS*>, (right-hand scale) on the annealing time, *t*, found from the analysis of the spectra recorded on the $Fe_{54}Cr_{46}$0.1%Ti annealed at 973 K [101]. Solid lines represent the best fit of the data to the JMAK equation. The kinetics parameters obtained from the fits are also displayed and they agree well with each other.

Figure 53 illustrates several kinetics curves obtained using both methods in the analysis of the *in situ* measured spectra..

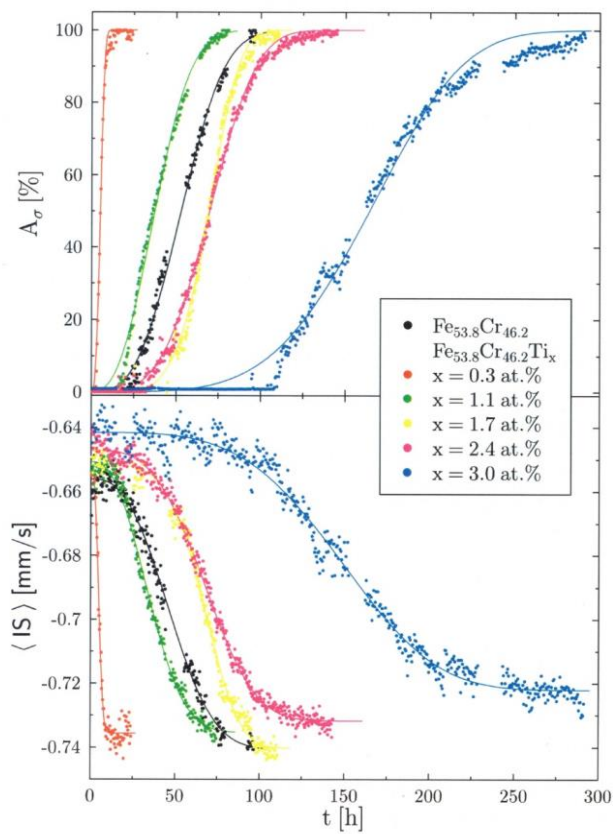

Fig. 53 Kinetics curves versus annealing time, *t*, obtained from the spectra recorded *in situ* on the $Fe_{53.8}Cr_{46.2}$xTi sample with various Ti content, *x*, annealed at 973 K. Two sets of the data are shown viz. obtained with the spectral area method (upper part of the plot) and with the average isomer shift (or COD) method (lower part of the plot). The solid lines are for the best fit of the data to the JMAK-equation [99].

The kinetics of the σ formation in bulk Fe-Cr alloys can be studied based either on the *ex situ* measured spectra or on the *in situ* recorded ones. The latter way of the following the α→σ transformation process was applied by Cieslak et al. [100]. The authors studied the $Fe_{54}Cr_{46}$0.1%Ti in form of a ~30 µm thick foil placed in a furnace. The spectra were recorded every 15 min during the vacuum annealing at *T*=873 and 973 K. They were evaluated using both methods mentioned above in order to compare them. Both $A_\sigma(t)$ and $<IS>(t)$ data derived from the spectra analysis were fitted to the JMAK-like equation yielding the kinetics parameters that are displayed in Table 3.

Table 3 Kinetics parameters *k* and *n* taken from Ref.[100]

| Method | T (K) | k·10$^3$ (min$^{-1}$) | n |
|---|---|---|---|
| Isomer shift | 873 | 2.71 | 2.7 |
| Spectral area | 873 | 2.59 | 2.6 |
| Isomer shift | 973 | 0.57 | 2.5 |
| Spectral area | 973 | 0.55 | 2.8 |

Agreement between the values of the kinetics parameters can be regarded as evidence in favor of equivalence of the two methods, but the $<IS>$-method is more accurate and does not suffer from a difference in the *f*-factor between $\alpha$ and $\sigma$. It can be also much quicker when using for determining $<IS>$ the center-of-gravity (COG) method recently outlined by Dubiel and Żukrowski [106].

It is also worth mentioning, regarding the kinetics of the $\sigma$-phase formation, that the ratio between the *f*-factor for $\sigma$ and $\alpha$ was determined as $f_\sigma/f_\alpha = 1.15$ [101]. However, the relative difference of 15% did not result in different values of the kinetics parameters [100].

**5.2.2. Stability**

There are few papers in which the $\sigma$-phase stability was investigated with MS. In particular, Dubiel and Inden reported that the $\sigma$-Fe$_{53}$Cr$_{47}$ alloy remained structurally unchanged after isothermal annealing for 18100 h at 783 K [38]. On the other hand annealing at *T* in the range 1093-1128 K resulted in the transformation of $\sigma$ into $\alpha$ [107]. The reversed transformation kinetics i.e. $\sigma \rightarrow \alpha$ was successfully described in terms of the JMAK equation. Based on the obtained values of the Avrami exponent, the authors concluded that there was a crossover in the mechanism of the transformation from heterogeneous nucleation of $\alpha$ at near-critical temperatures to homogenous nucleation at higher temperatures.

The effect of ball milling on two coarse grained and one nanometer grained σ-FeCr was studied by means of MS [108]. The grinding process performed in a protective Ar atmosphere up to 100 h resulted in a decomposition of α and an amorphous phase. The σ→α transformation process could have been successfully described in terms of the JMAK equations. The obtained kinetics parameters were for the coarse grained samples : $k$=0.054 and 0.142 h$^{-1}$ and $n$=0.95 and 0.76, and for the coarse grained samples and $k$=0.184 and $n$=0.83 for the nanometer grained sample. Thus the fastest transformation took place for the sample with the smallest grains. The difference in the kinetics of the coarse grained sample was caused by a different amplitude of milling was different.

Effect of irradiation with 2 MeV $Fe^{3+}$ ions performed at temperatures between 583 K and 973 K to the maximum dose of 12.5 dpa was investigated by means of the CEMS techniques on a σ-$Fe_{54.5}Cr_{45.5}$ sample [109].

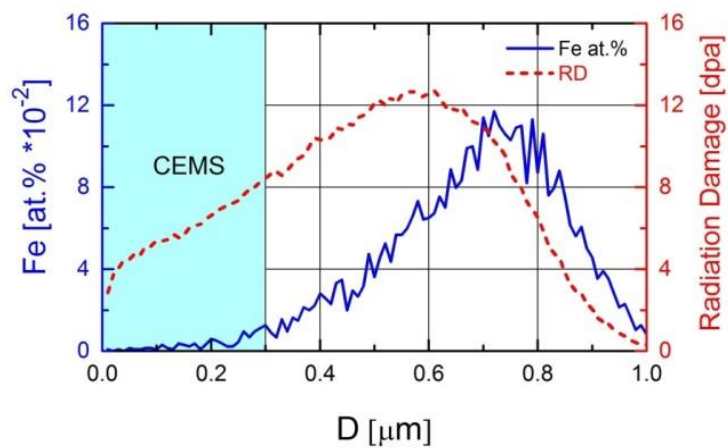

Fig. 54 Concentration of $Fe^{3+}$ and radiation damage profiles vs. depth, D, for the irradiated σ-$Fe_{54.5}Cr_{45.5}$ sample [109].

The authors showed that under the applied irradiation the tetragonal structure of σ remained unchanged. This can be qualitatively seen in Fig. 55 showing CEMS spectra before and after the irradiation. The shape of the spectrum recorded on the irradiated sample is optically the same as the one recorded on the non-irradiated sample.

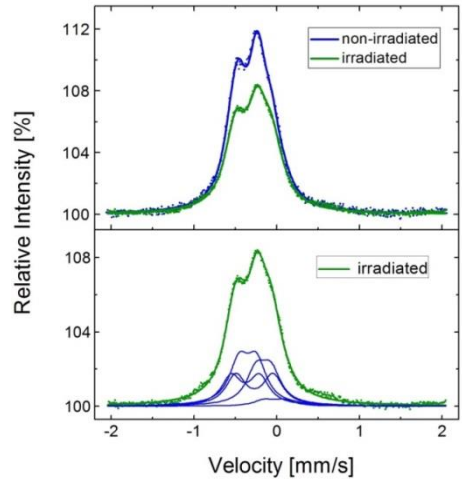

Fig. 55 CEMS spectra recorded at 295 K on non-irradiated and irradiated (573 K) sides of the σ-$Fe_{54.5}Cr_{45.5}$ sample. In the spectrum of the irradiated sample are indicated five doublets into which the measured spectrum was fitted [109].

However, the analysis of the spectra revealed differences in relative contributions of the five doublets. This was interpreted as evidence in favor of an internal redistribution of atoms on particular sites. The degree of the redistribution was found to be proportional to the initial number of Fe atoms on particular lattice sites. It was also negatively correlated with the temperature of the irradiation. Relevant plots can be seen in Figs. 56 and 57. A comparison of the upper and lower parts of the histograms displayed in Fig. 56 evidences the effect of the $Fe^{3+}$ ions irradiation.

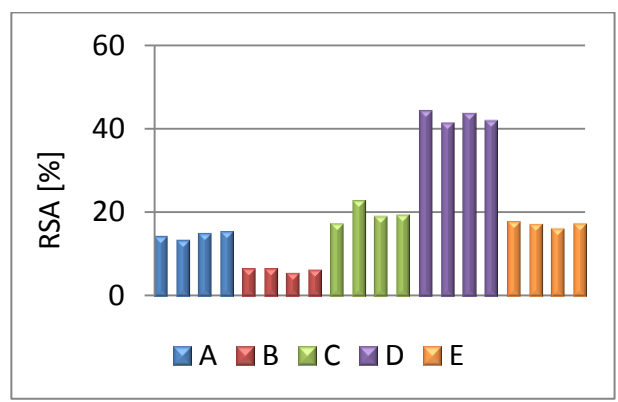

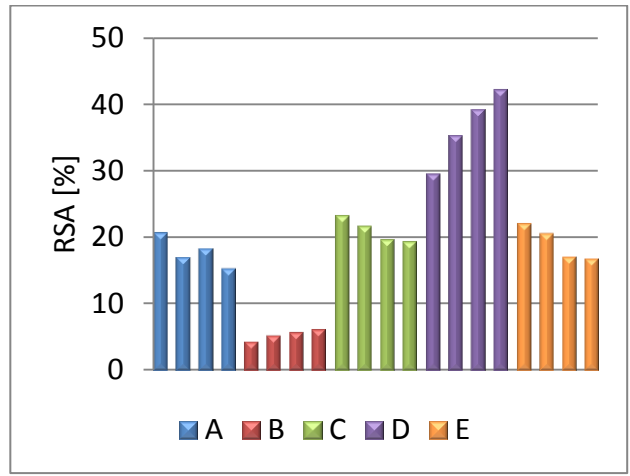

Fig. 56 Relative spectral area, RSA, as determined from the spectra recorded on the non-irradiated (upper panel) and irradiated (lower panel) sides of the σ-$Fe_{54.5}Cr_{45.5}$ sample irradiated at 573, 673, 746 and 973 K (from left to right for each sub lattice) [109].

Whereas in the non-irradiated part of the studied sample the distribution of Fe atoms on a given site hardly depends on temperature, the distributions derived from the spectra recorded on the irradiated side exhibit rather significant dependence on the irradiation temperature. The dependence is characteristic of the site viz. for A, C and E a decrease with temperature, while for B and D an increase took place. This different behavior is even better seen in Fig. 57 that illustrates the difference between the results shown in Fig. 56.

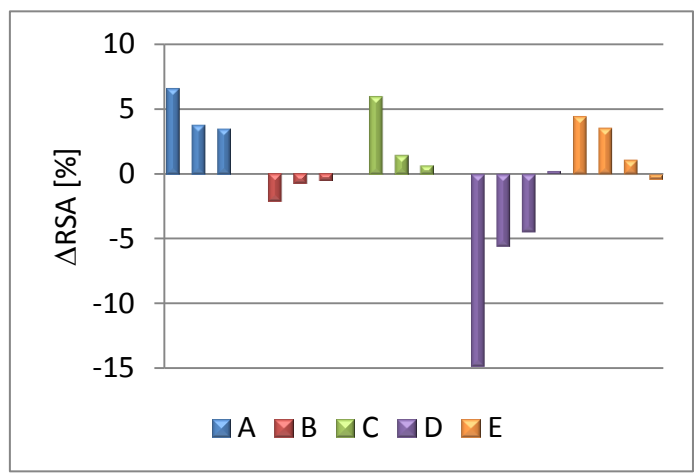

Fig.57 Difference in the relative spectral area, ΔRSA = RSA (IR)-RSA(NIR), for the irradiated (IR) and non-irradiated (NIR) sides of the sample [109].

Figure 57 gives clear evidence that the applied $Fe^{3+}$ ions irradiation caused a redistribution of Fe atoms in all five sub lattices. An increase of the number of Fe atoms occurred on A, B and C sites, whereas a decrease took place on the sites B and D. The strongest effect was determined for the site D. Interestingly, the effect gets smaller with the irradiation temperature for all sites. Noteworthy is also the fact that no redistribution was found in the sample irradiated at 973 K. This temperature coincides with the one at which the α→σ phase transformation proceeds at the fastest rate.

**5.2. Magnetic**

Magnetism of the Fe-Cr alloys was subject of numerous investigations including the Mössbauer spectroscopic ones. Despite the major contribution to the understanding the magnetism of these alloys and to the construction of the magnetic phased diagram came from magnetization and neutron diffraction experiments. Yet, the input of MS cannot be neglected.

The MS techniques is suitable for this purpose as its spectral parameters, in particular, the hyperfine field, *B*, are sensitive to the content of Cr – see section 3. Furthermore, in metallic iron the temperature dependence of *B* is very similar to the temperature dependence of magnetization [110]. Consequently, *B(T)* dependence can be used for determining the Curie, $T_C$, and, the Néel, $T_N$, temperatures. For example, Nemanich et al. using this approach found $T_C$ = 60(5), 160(5) and 260(1) K for the Fe-Cr alloys containing 20, 25 and 30 at.% Cr, respectively [111]. The authors emphasized that their $T_C$ – values agreed with the corresponding ones determined with the dc magnetization. Loegel and co-workers performed similar study on Fe-Cr alloys with 24 and 26 at.% Cr and arrived at $T_C$=110(5) K and 130(5) K, respectively [112]. The authors pointed out that these values were significantly smaller than the ones found with the magnetic permeability measurements. They suggested that a possible fluctuation in the local Cr concentration leading to a distribution of the local magnetic ordering temperatures could be the reason of the disagreement they revealed. Indeed, this idea was experimentally confirmed by Mössbauer spectroscopic measurements performed on $Cr_{75}Fe_{25}$ [113] and $Cr_{86.5}Fe_{13.5}$ alloys [114]. The real distribution of Fe atoms in the Cr matrix sensitively depends on the heat treatment applied for the preparation of samples. Indeed, such effect was observed by other researchers in the $Cr_{75}Fe_{25}$ alloy [115]. The dramatic effect of the heat treatment on $T_C$ of $Cr_{100-x}Fe_x$ alloys for x ≤ ~35 can be readily seen in Fig. 58. It seems that the value of $T_C$ can be used as a figure of merit with respect to the distribution of Fe atoms in the Cr-matrix viz. the lower the value of $T_C$ the higher the homogeneity degree in an alloy of a given composition.

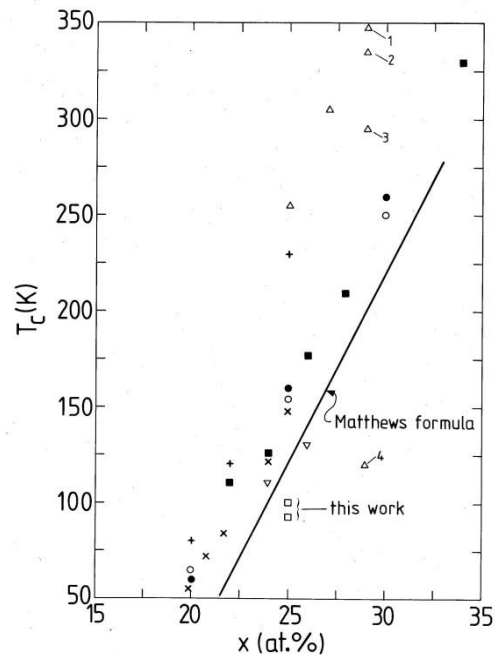

Fig. 58 Fragment of the magnetic phase diagram of the $Cr_{100-x}Fe_x$ alloy system showing the values of the Curie temperature, $T_C$, versus Fe content, x, for different heat treatments. The solid line was obtained based on the Matthews formula [116]. The meaning of various symbols is explained in [113].

The largest dispersion of the values of $T_C$ can be seen for the $Cr_{75}Fe_{25}$ and $Cr_{71}Fe_{29}$ alloys, for which $T_C$ lies between ~100 and ~255 K for the former and between ~120 and ~350 K for the latter.

The most freshly presented magnetic phase diagram of the Fe-Cr system valid for the Cr-rich range is shown in Fig. 59. Four phase fields can be distinguished in it viz. ferromagnetic (FM), antiferromagnetic (AFM), paramagnetic (PM) and spin-glass (SG) [117]. The SG is the magnetic ground state for the Fe content between ~10 and

~26 at.%. Its phase field extends both into FM and AFM phases. Consequently, there are three types of transitions to occur viz. PM→FM→SG, PM→SG and PM→FM→SG. The first and the third one are known as re-entrant transitions.

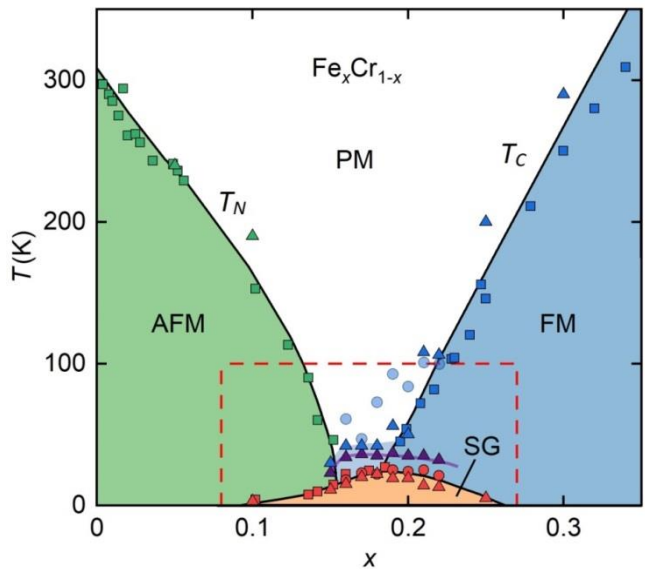

Fig. 59 Magnetic phase diagram of the Fe-Cr system adopted from [117]. Here *x* is the fraction of Fe.

Probably the most important contribution of MS to the magnetic phases diagram was related with the study of the SG state. The possibility of studying the latter with MS is based on the fact that the relative intensity of the 2-nd and the 5-th lines in the sextet, $I_{2,5}$, depends on an angle, θ, between the direction of the gamma-rays (usually perpendicular to the sample's surface) and that of the local magnetic moment, **μ**. Namely, $0 \leq I_{2,5} \leq 4$, depending on θ. In particular, for θ=0° $I_{2,5} = 0$, for θ=90° $I_{2,5} = 4$ and for randomly distributed orientations of **μ** $I_{2,5}=3$. Based on this feature to discover a FM → SG transition one has to carry out measurements in external magnetic field that should be strong enough to order all **μ**-vectors in the FM state parallel to the direction of the gamma-rays ($I_{2,5}=0$). One records then the spectra as a function of decreasing temperature. On entering the SG state, the directions of **μ** will no longer be aligned parallel to the gamma-rays what will be reflected in $I_{2,5}\neq0$. In this way the transition temperature into the SG state could be determined [118]. The pertinent illustration to this matter is presented in Fig. 59.

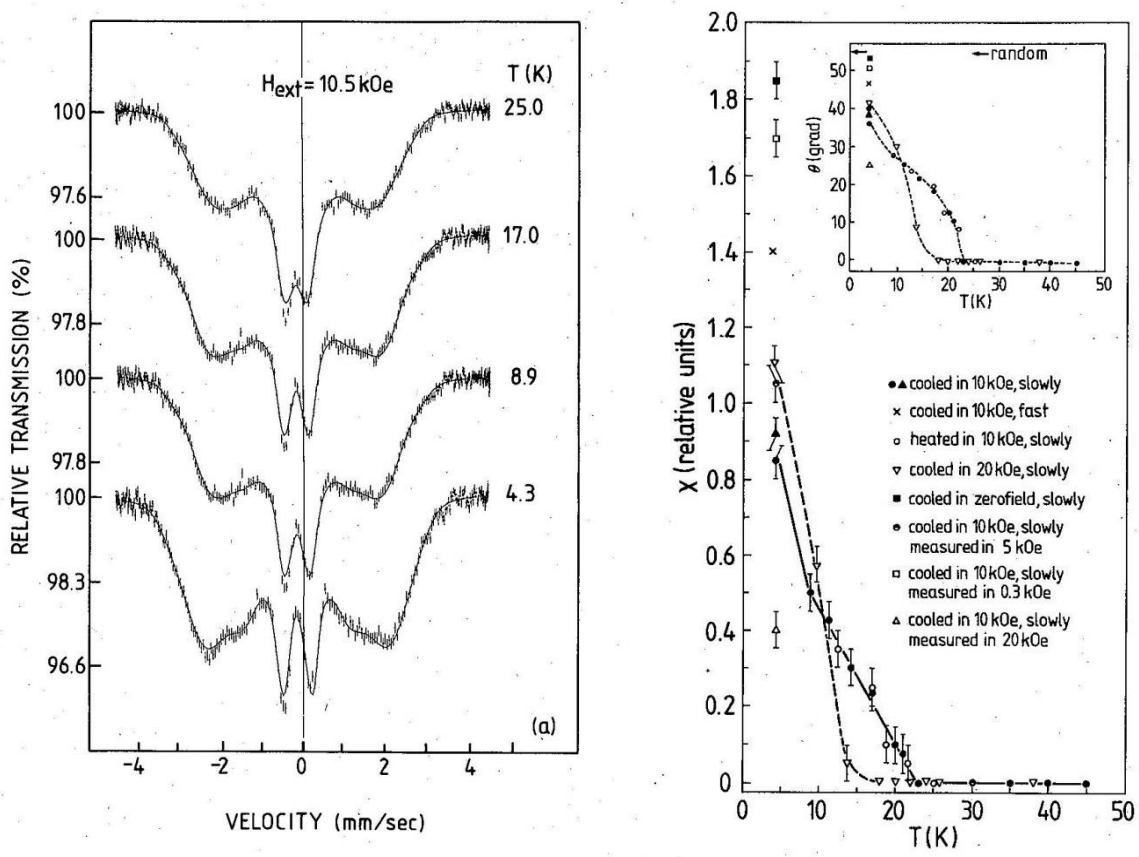

Fig. 60 (left-hand panel) $^{57}$Fe Mössbauer spectra recorded at different temperatures shown on a sample of $Cr_{75}Fe_{25}$ alloy placed in external magnetic field, $B_o$= 10.5 kGs (right-hand panel). The relative intensity of the second and the fifth line (here denoted per X) versus temperature for $T \leq 40$ K. The inset shows the corresponding behavior of θ. Two curves illustrate the behavior found for $B_o$ = 10 and 20 kGs. The figure reproduced from [118].

The transition from FM into SG state can also be detected by plotting a temperature dependence of the average hyperfine field, <B>(T). For a pure FM state, <B>(T) follows more or less the Brillouin function i.e. it saturates at low enough T. However in a system showing a re-entrant FM→SG transition, like $Fe_{25}Cr_{75}$, a deviation from the Brillouin-like behavior viz. an increase of <B> occurs at certain low T, $T_f$, regarded as a spin-freezing temperature. This increase can be understood in terms of freezing of the perpendicular component of spin in the SG state. In this way, the re-entrant transition PM→FM→SG was detected in $Fe_{25}Cr_{75}$ [113,117] and PM→AFM→SG in the Fe-Cr alloy with 13.5 at.% Fe [114]. Noteworthy, the authors of this reference used the Mössbauer-effect on both $^{57}$Fe and $^{119}$Sn nuclei. $^{57}$Fe-site MS was also used to study the Fe-Cr alloy with 17.5 at.% Cr [119], hence in the composition range where the PM→SG transition can be observed – see Fig. 59. Mössbauer spectra were measured at various temperatures between 4.5 and 296 K and also in external magnetic field of 7.5 T, 10.5 T and 13.5 T [119]. The authors determined the spin freezing temperature (marking the PM→SG transition) $T_{SG}$ = 32 K (from the linewidth) and 39 K (from the average hyperfine field). The in-field measurements gave evidence that correlations between magnetic moments existed at least up to $10 \cdot T_{SG}$.

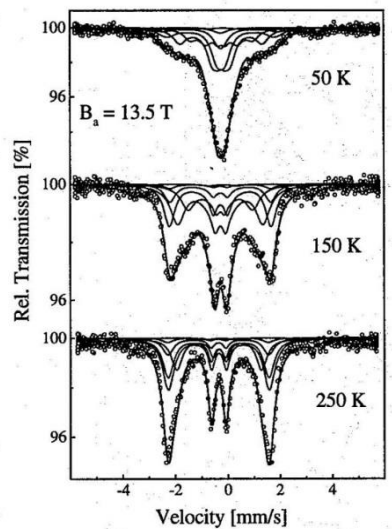

Fig. 60 $^{57}$Fe spectra recorded in the magnetic field of 13.5 T on the $Fe_{82.5}Cr_{17.5}$ at various temperatures shown [119].

Another important contribution of MS with respect to the magnetic phase diagram of the Fe-Cr system was the research in which an effect of an external magnetic field on the reentrant SG temperature was studied [120]. The obtained linear dependence was used to validate various mean-field theory models.

Worth mentioning, with respect to the magnetic phase diagram of Fe-Cr, is the investigation carried out at $^{119}$Sn atoms as probe [121]. The spectra were recorded on the samples with the composition covering the whole compositional range of the Fe-Cr alloys. The average hyperfine field derived from these measurements turned out to replicate, at least qualitatively, the magnetic phase diagram of the system. The position of the minimum in *<B>* coincided with the composition where the SG phase occurs.

## 6. Physical properties

### 6.1. Alpha phase

#### 6.1.1. Spin- and charge-density

Change in the hyperfine field reflects underlying change in the Fe-site spin-density. Similarly, change in the isomer shift can be associated with the change of the Fe-site charge-density. Due to the screening effect of the 3d electrons the changes are meant as effective changes. As discussed in paragraph 3, analysis of the Mössbauer spectra permits the precise determination of *ΔB₁* and *ΔB₂* and a fairly precise of *ΔIS₁* and *ΔIS₂*. These changes, due to the presence of one Cr atoms in 1NN (subscript 1) or in 2NN (subscript 2) coordination shell can be regarded as localized changes in the spin- or charge-density, respectively. The Cr concentration dependence of *B(0,0)*, *B(1,0)* and *B(0,1)* determined for the $Fe_{100-x}Cr_x$ alloys for $x \leq \sim 15$ is shown in Fig. 61.

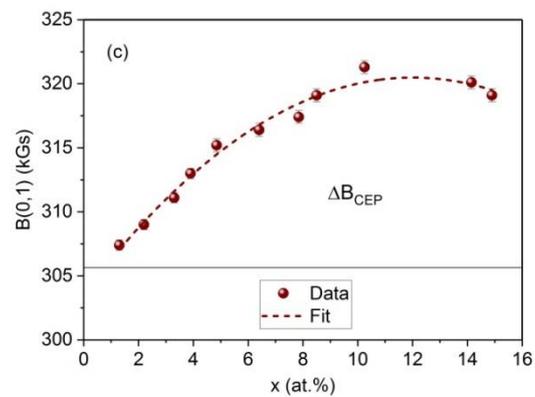

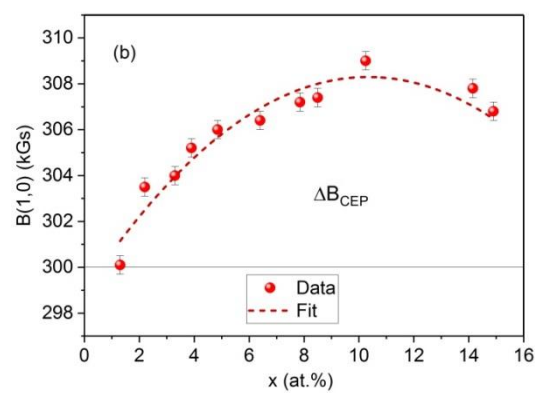

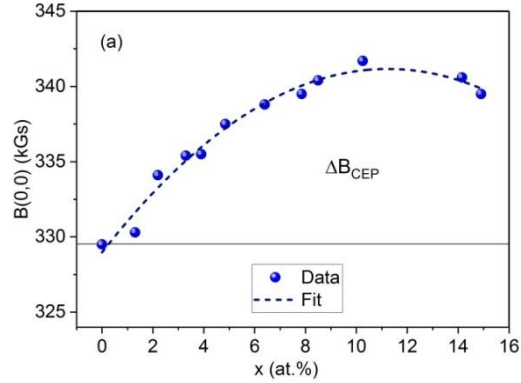

Fig. 61 Hyperfine field versus Cr concentration, *x*, for (a) *B(0,0)*, (b) *B(1,0)* and (c) *B(0,1)*. The dashed lines stand for the best-fit of data to a quadratic equation. The plots were made based on the data reported in [37].

Two characteristic features of the data shown in Fig. 61 should be perceived: (1) all three fields are concentration dependent, and (2) the dependence is configuration independent with a maximum at ~11 at.% Cr. The latter means that all *B*-values for configurations (1,0) and (1,0) are shifted relative to *B(0,0)* by the same value typical of the configuration viz. $\Delta B_1$ = -31.5 kGs and $\Delta B_2$ = -22.5 kGs, respectively. According to the author's interpretation, these values reflect localized changes in the Fe-site spin-density whereas the Cr concentration change in *B* similar for the three configurations was interpreted as due to the change in the polarization of the conduction electrons, hence in the $B_{CEP}$ term of the hyperfine field [37]. Noteworthy, the interpretation of the latter was confirmed by performing measurements at $^{119}$Sn diffused into the matrices of the Fe-Cr alloys. For nonmagnetic atoms like Sn, the whole hyperfine field measured at them can be due to the polarization of the conduction electrons, hence the $B_{CEP}$ term. If so, the hyperfine field recorded using the Mössbauer effect at $^{119}$Sn nuclei should have the same Cr concentration dependence as the $^{57}$Fe-site field shown in Fig. 61. The appropriate comparison of the data obtained from measurements performed on the two isotopes is shown in Fig. 62.

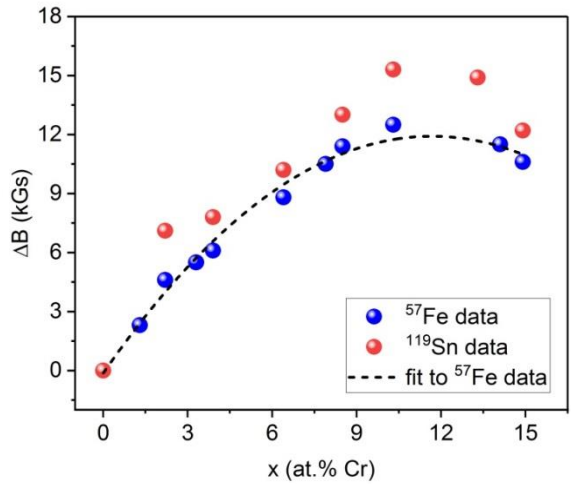

Fig. 62 Dependence of $\Delta B = B(0,0;x) - B(x=0)$ on the content of Cr, $x$, in $Fe_{100-x}Cr_x$ alloys as found from the spectra measured at $^{57}$Fe and at $^{119}$Sn. The dashed line is the best-fit of the $^{57}$Fe data to a quadratic equation. The plot was made using the data from [37,122].

Significant progress in the understanding Cr-changes in the electronic structure of iron was made by Dubiel and Żukrowski who investigated a series of Fe-Cr alloys with up to 45.5 at.% Cr [36]. They found liner correlations between: (1) *B(0,0)* and *IS(0,0)*, (b) *<B>* and *<IS>*, (c) *<B>* and *<m+n>*, (d) $\Delta B_1$ and $\Delta IS_1$ and (e) $\Delta B_2$ and $\Delta IS_2$. Based on these correlations they had managed to express the values of the hyperfine field and those of the isomer shift in terms of the underlying number of effective s-like electrons. In particular, from the *B(0,0)-IS(0,0)* correlation they concluded that a change of the conduction electrons polarization by 1 s-like electron was equivalent to a change of the hyperfine field by 1603 kGs. Noteworthy, this value of the hyperfine coupling constant agrees quite well with its theoretical estimations given in [25,26], but it is two times smaller than the theoretical values reported in [27,123]. From the *<B>-<IS>* correlation Dubiel and Żukrowski deduced that, on average, a unit change of the polarization of the s-like electrons caused change of *B* by 3277 kGs. Based on this value and the *<B>-<m+n>* correlation the authors calculated that one Cr atom per unit cell changes the effective spin polarization at Fe-site by 0.026 s-like electrons. Finally, the correlations between the local changes in *B* and *IS* permitted determination of the underlying changes in spin- and charge-densities in number of the s-like electrons. Namely, the change per one Cr atom situated in 1NN was estimated as 0.012 and as 0.007 per one Cr atom present in 2NN. Using the value of 1603 kGs/spin and taking into account the change in *B* found for any of the three atomic configurations shown in Fig. 61 one can calculate that the maximum increase i.e. the one at $x \approx$ 11 at.% corresponds to an effective increase of the spin-density in the conduction electrons by 0.007 s-like electron.

### 6.1.2. Dynamics of Fe atoms in Cr matrix

Dynamics of $^{57}$Fe atoms diluted in a metallic Cr was studied by Dubiel and coworkers [124]. It is known that even a single Fe atom embedded into the Cr matrix possesses quite large magnetic moment viz. $\mu_{Fe}$=1.4$\mu_B$ [125]. According to theoretical calculations, magnetic atoms have different effect on the spin-density waves (SDWs) than non-magnetic ones [126-129]. Namely, they pin the SDWs i.e. the amplitude of the SDWs at the Fe-site is significantly reduced. This pinning effect can be readily seen in the $^{57}$Fe Mössbauer spectra that have a shape of a singlet with a temperature dependent width [124]. Consequently, the $^{57}$Fe Mössbauer spectroscopy, contrary to the $^{119}$Sn one, is not suitable for studying the SDWs.

The authors of Ref. 124 figured out the dynamics of Fe atoms in chromium by determining the relative recoil-free fraction, $f/f_0 = \exp[-k^2 <\Delta x^2>]$, a quantity related to the relative average square amplitude of vibrations, $<\Delta x^2>$ (*k* is here the wave vector of the 14.4 keV gamma rays). Temperature dependence of ln(f/f$_o$) is shown in Fig. 63. Three ranges A, B and C were distinguished in the plot. In the range A the plot is linear, hence vibrations of Fe atoms are harmonic, while in contrast they are inharmonic in the ranges B and C.

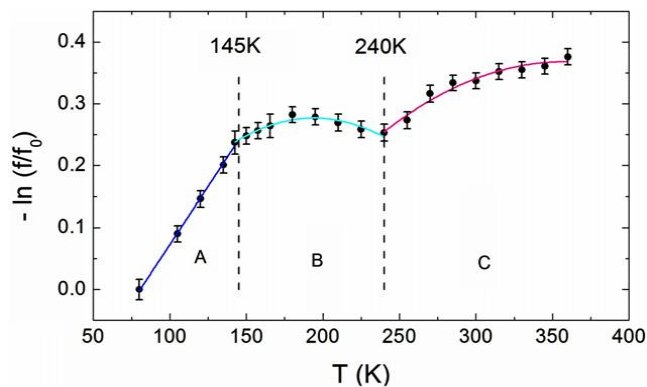

Fig. 63 Dependence of –ln(f/f₀) on temperature, T. The solid lines represent the best-fits of the data to eq. (7) [124].

The T-dependence of ln(f/f₀) was analyzed by fitting the data to the following equation:

$$ln\frac{f}{f_o} = -ln f_o + \frac{6E_R T}{T_D^2}(1 + \varepsilon T) \quad (15)$$

Where $E_R$ is the recoil energy, $T_D$ is the Debye temperature and $\varepsilon$ denotes the inharmonic coefficient.

The analysis of the data in terms of eq. (15) yielded $T_D$=190(2) K and ε=0 for A, $T_D$=155(2) K and ε=-25·$10^{-4}$ $K^{-1}$ for B and $T_D$=152(3) K and ε=-14·$10^{-4}$ $K^{-1}$ for C [124]. A relative binding force constant for Fe atoms of 0.0997 was determined for the harmonic regime of vibrations. This value means that the coupling between Fe atoms embedded into the Cr matrix is very weak at low temperature. Noteworthy, it is ~10-times weaker than the coupling between the host atoms themselves. The obtained values of $T_D$ are remarkably low in comparison with the $T_D$-value for pure Cr viz. 606 K at 0 K [130] and 424 K at 295 K [131]. They are also significantly smaller than the value of $T_D$=454(50) K determined based on the temperature dependence of <CS> measured in the range of 80-300 K using $^{119}$Sn as probe atoms in Cr [132]. The difference in the values of the Debye temperature of Cr found using $^{119}$Sn and $^{57}$Fe atoms as probe gives evidence that the dynamics of the two kinds of atoms is distinctly different. This likely has its source in a different coupling of the two types of probe atoms to the matrix.

### 6.1.3. Effect of sulphidation

Effect of a high-temperature sulphidation on Fe-Cr alloys was studied by means of TRANS and CEMS modes of MS [133-136]. The authors of [133-135] measured a series of $Fe_{100-x}Cr_x$ alloys with x ≤ 45.5 sulphidized in the atmosphere of $H_2$-$H_2S$ mixture at 1073 K. They found a critical Cr concentration of $x_{Cr}$ = 5(1) at.% below which the metallic phase became enriched in Cr and above it was enriched in Fe. In other words at $x_{Cr}$ there is a change of the preferential sulphidation process.

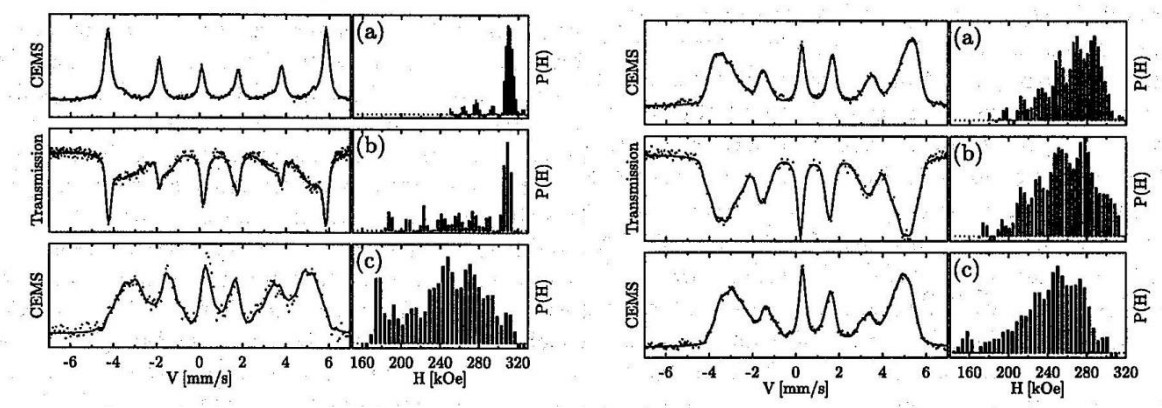

Fig. 64 (left panel) $^{57}$Fe Mössbauer spectra recorded at 295 K on the scale grown on the sulphadized Fe$_{97}$Cr$_3$ alloy: on the outer surface (a) CEMS mode, (b) TRANS mode and (c) on the inner scale (CEMS). Histograms of the hyperfine field distribution derived from the spectra are displayed on the right-hand side of each spectrum. The right panel illustrates the same but for the scale grown on the Fe$_{84.85}$Cr$_{15.15}$ alloy. Plots from Ref. 135.

Furthermore, the TRANS and CEMS modes of measurements were used to study sulphidation products i.e. scales and metallic cores in pure and Sn-doped quasi-equiatomic Fe-Cr alloys having $\alpha$ and $\sigma$ structure [136]. Several forms of iron present in the scales were identified like magnetic and non-magnetic (Fe,Cr)$_{1-x}$S, FeS$_{1+x}$ and $\alpha$-Fe (likely in form of small particles). Metallic cores were found to consist of an outer part (mantel) ~45 µm thick and of an inner part (inner core). Furthermore, the mantel was found to be enriched in Fe and Sn, and its ~0.2 µm thick presurface layer was composed of $\alpha$-Fe. The inner part of the core was entirely composed of $\alpha$-FeCr (in case of the $\alpha$-FeCr alloy) and of a mixture of $\alpha$-FeCr and $\sigma$-FeCr (in case of the $\sigma$-FeCr alloy).

### 6.1.4. Effect of oxidation

Oxidation induced changes in Fe$_{100-x}$Cr$_x$ alloys were investigated by means of MS in [137-139]. Results of measurements in the TRANS and CEMS modes carried out on a series of the alloys ($2 \leq x \leq 15$) exposed to air at RT were reported in [137]. The analysis of the spectra was performed only with the hyperfine distribution method. Based on the average values of *B*, *<B>*, obtained from the spectra measured in the two modes and on the *x*-dependence of *<B>*, no meaningful difference was detected between the composition in the bulk and in the surface-presurfce layer. The effect of the high-temperature (570, 870 and 1070 K) oxidation on two Fe-Cr alloys containing 3 and 6 at.% Cr was studied using the TRANS mode of MS [138]. The spectra were analyzed in terms of the M1=2 model i.e. the effect of Cr atoms, both on *B* as well as on *IS*, present in the first-two neighbor shells was assumed to be the same yet the additivity rule was imposed i.e. *X(n)=X(0) + n·ΔX, where X=B* or *IS, n* is the number of Cr atoms in the two-shell volume. The effect of temperature on the shape of the spectra was significant, as can be seen in Fig. 65.

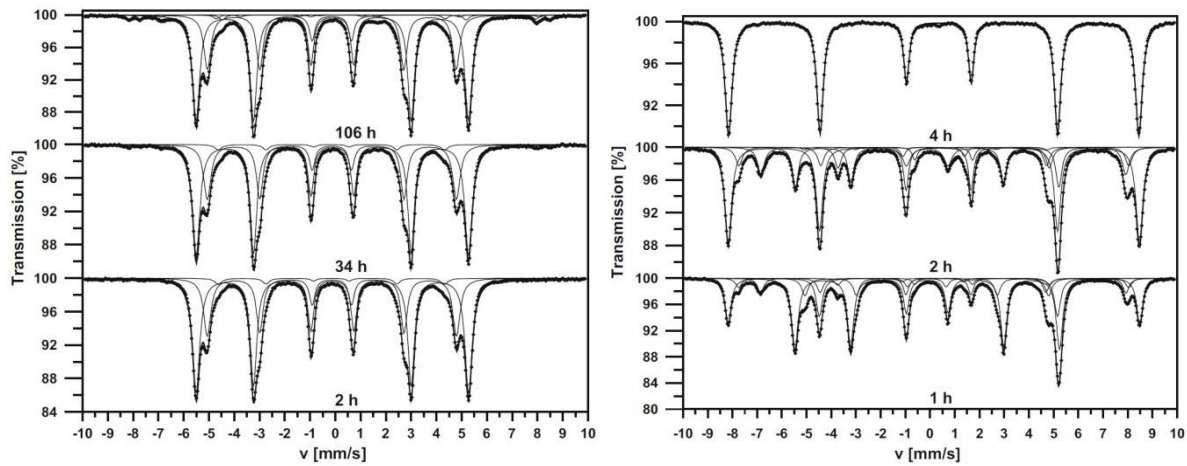

Fig.65 $^{57}$Fe Mössbauer spectra recorded on the $Fe_{97}Cr_3$ sample oxidized in air at 570 K (left panel) and at 1070 K (right panel) for different hours [138].

Based on the obtained results the author identified three phases containing Fe viz. metallic (Fe-Cr), α-$Fe_2O_3$ and $Fe_3O_4$ and determined their relative amounts in dependence of the oxidation temperature and duration. Furthermore, the oxidation kinetics at 870 and 1070 K was described in terms of the logarithmic law, and the activation energy of Fe atoms oxidation was determined as 44(2) kJ/mole for *x*=3 and as 0.60(3) kJ/mole for *x*=6.

The effect of oxidation in air at 870 K on the Fe-Cr alloys containing 12 and 15 at.% Cr was also a subject of study by means of the TRANS and CEMS Mössbauer spectroscopy [139]. Analysis of the spectra measured *ex situ* were performed in terms of the M12 model. The evaluation of the CEMS spectra gave evidence that besides the metallic phase (Fe-Cr) only one oxide viz. hematite was found. However, in the TRANS spectra recorded on the two alloys oxidized for longer time both hematite and magnetite was revealed.

## 6.2. Sigma phase

### 6.2.1. Curie temperature

For weakly magnetic systems like σ-FeCr, the magnetic ordering temperature (Curie temperature), $T_C$, can be determined by plotting the width of the spectrum at half maximum, G, as a function of temperature, $T$. The method is illustrated in Fig. 66.

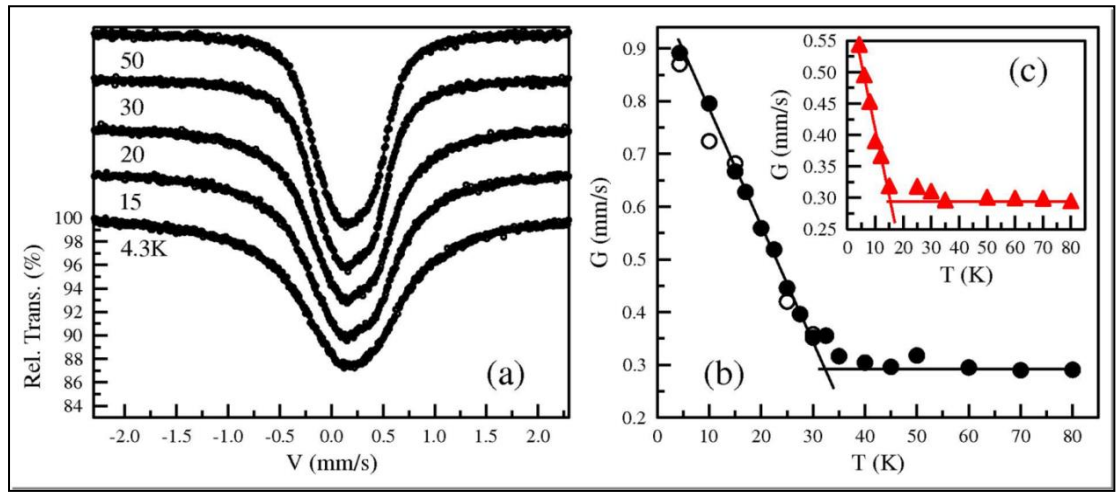

Fig. 66 (a) Mössbauer spectra recorded on σ-$Fe_{53.8}Cr_{46.2}$, at various temperatures in the vicinity of the magnetic ordering temperature, $T_C$, and the full line width of spectrum, G, versus temperature for (b) σ-$Fe_{53.8}Cr_{46.2}$ and (c) σ-$Fe_{52}Cr_{48}$. The solid lines are for the best linear fit to the data in the magnetic and in the paramagnetic phase. The interception of the lines marks $T_C$, equal to ~33 K and ~15 K for the two cases, respectively [140].

Values of $T_C$ for the σ-phase were determined in this way in the whole concentration range of its existence in the Fe-Cr system – see Fig. 66/67. The $T_C$ – values decline with $x$ at the rate of 7(2) K/at.%.

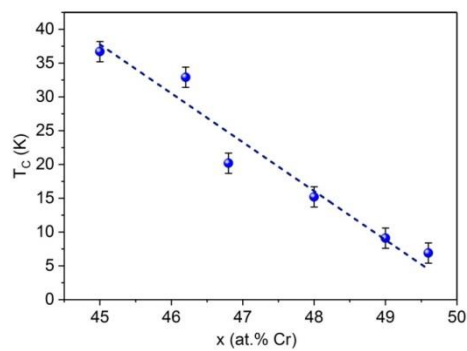

Fig. 67 Curie temperature, $T_C$, versus Cr content, $x$, in the σ-phase Fe-Cr alloys. The plot has been made using the data from Ref. [141]. The straight line is the best linear-fit to the data.

**6.2.2. Debye temperature**

Mössbauer spectroscopy offers a unique opportunity for determining the Debye temperature, $T_D$, in two ways. Firstly, from a temperature dependence of the center shift, $CS(T)$, and secondly, from the recoil-free fraction, $f$. The first method is more frequently used. The $T$-dependence of $CS$ is given by the following formula:

$$CS(T) = IS + SOD(T) \qquad (16)$$

Where $IS$ depicts the isomer shift (it hardly depends on $T$) and SOD(T) represents the so-called second-order Doppler shift (also known as temperature shift). Its temperature dependence in the Debye model can be expressed as follows:

$$SOD(T) = \frac{3k_B T}{2mc}\left(\frac{3T_D}{8T} - 3\left(\frac{T}{T_D}\right)^3 \int_0^{T_D/T} \frac{x^3}{e^x - 1}dx\right) \qquad (17)$$

Where $m$ is the mass of Fe atom, $k_B$ is the Boltzmann constant, $c$ is the speed of light, and $x=h\nu/k_B T$ ($\nu$ being the frequency of vibrations). The SOD-term is proportional to the mean-squared velocity of vibrations, $<v^2>$, hence to the kinetic energy of vibrating atoms. By fitting a set of $CS(T)$-data to eq. (17) one gains the value of $T_D$. In this way were determined values of $T_D$ for nano- and microcrystalline σ-Fe$_{100-x}$Cr$_x$ alloys in the whole range of the σ-phase existence [142]. The obtained results are displayed in Fig. 68.

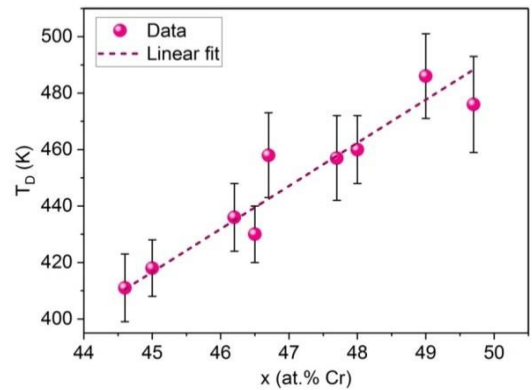

Fig. 68 Debye temperature, $T_D$, versus Cr content, $x$, in σ-phase Fe$_{100-x}$Cr$_x$ alloys. The plot has been made based on the data from Ref. [142].

It is evident from Fig. 68 that the Debye temperature increases linearly at the rate of 15(4) K/at.% with the concentration of Cr, *x*. Noteworthy, there is no difference in the values of $T_D$ between nanocrystalline and microcrystalline alloys [142].

### 6.2.3. Effect of magnetism on lattice vibrations

The influence of magnetism on lattice vibrations is routinely regarded as negligible. Such attitude follows from calculations of the standard theory of the electron-phonon interaction (EPI). This theory predicts that the figure of merit, so-called small adiabatic parameter $E_D/E_F \approx 10^{-2}$, hence it is very small [143 and references therein] where $E_D$ is the Debye energy, and $E_F$ is the Fermi energy. On the other hand, Kim demonstrated that the impact of the EPI on the spin susceptibility of metals could be increased by two orders of magnitude if exchange interactions between electrons had been included [143]. Consequently, the effect of the EPI on magnetic properties of metallic systems, and *vice versa*, should be much more important than the one predicted by the standard theory.

Good candidates for verification of the Kim's prediction are systems with the itinerant or delocalized magnetism. The magnetism of the σ-phase in the Fe-Cr alloy system is highly itinerant as follows from the Rhodes-Wohlfarth criterion (plot) [144-145]. The effect of magnetism on the lattice dynamics was studied with MS in σ-$Fe_{100-x}Cr_x$ alloys containing *x*=46 and 48 [140]. The authors investigated the effect of both the internal magnetism as well as that of an externally applied magnetic field. They revealed anomalies in the temperature behaviour both in the center shift, *<CS>*, as well as in the *f*-factor in both samples. The temperature dependence of the former is presented in Fig. 69.

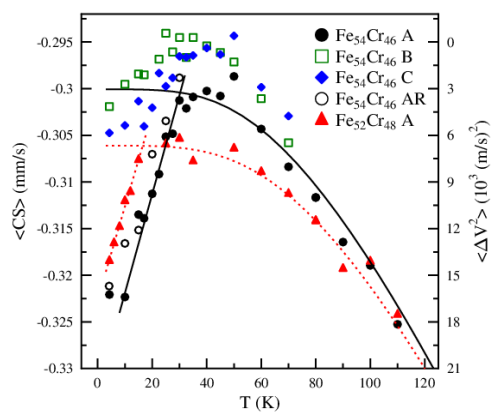

Fig. 69 Temperature dependence of the average center shift, <CS>, as determined for two σ-Fe$_{100-x}$Cr$_x$ alloys with *x*=46 and 48 using different fitting procedures (A, B, C). AR stands for the data obtained by fitting the spectra obtained with repeated measurements. Solid and dotted curves are for the best-fit of the data obtained with the procedure A to the Debye model [140].

A dramatic deviation from the behaviour expected from the Debye model can readily be seen for both alloys. The decrease of the data within the deviation is linear. The temperatures at which the straight lines cross the Debye-like curves agree well with the Curie temperature determined for the two samples [141]. The right-hand axis of Fig. 69 is scaled in the change of the average square velocity, <Δ$v^2$>, within the magnetic phase. Noteworthy, <Δ$v^2$> increases with decreasing temperature i.e. the kinetic energy increases. This behaviour must be of a magnetic origin as it contradicts the laws of thermodynamics. The maximum gain of the kinetic energy on lowering *T* from 35 K to 4 K was estimated to be ~4 meV [140]. The *f*-factor, in turn, is related to the average square amplitude of vibrations, <$x^2$>, hence to the potential energy in the case of harmonic oscillations. Its behavior was also anomalous on crossing the Curie temperatures in the investigated samples, and it evidenced a decrease in the potential energy of vibrations. The maximum estimated reduction of this energy was ~24 meV [140]. Thus the total energy of vibrations decreased by ~20 meV on cooling the samples from the Curie temperature down to 4 K. The effect can be termed as magnetism-induced softening of the lattice.

### 6.2.4. Charge-density

The average isomer shift in quasi-equiatomic α-FeCr alloys is negative (relative to α-Fe) what indicates an increase of the Fe-site density due to substitutional Cr atoms and can be understood in terms of a charge transfer from Cr onto Fe [36]. For a Fe-Cr alloy of the same composition but in the σ-phase <*IS*>$_\sigma$ is more negative by ~0.11 mm/s as shown in Fig. 70. This means that the Fe-site charge-density in σ is much greater than that in α and the increase is solely due to a change of the crystal unit cell from bcc to tetragonal. Making use of the relation between the isomer shift and number of s-like electrons in metallic iron systems [146], the value of -0.11 mm/s is equivalent to 0.06 s-like electrons.

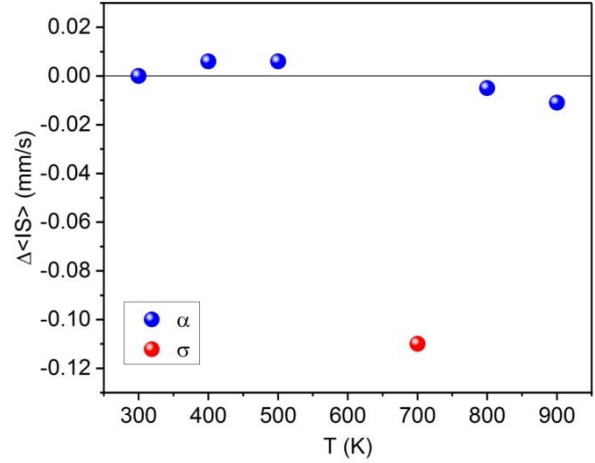

Fig. 70 Difference in the average isomer shift, $\Delta<IS>=<IS>_T -<IS>_Q$ for $Fe_{53.8}Cr_{46.2}$ isothermally annealed at vacuum for 5h at temperature $T$ in the range 300-900 K. The index $Q$ denotes the quenched sample. The sample annealed at 700 K was to 100% transformed into $\sigma$. The data reported in [36] were used to make this plot.

## 7. Miscellaneous

### 7.1. Magnetic texture

As already mentioned, the shape of the Mössbauer spectrum is sensitive to an angle, $\Theta$, between the direction of the gamma rays and that of the magnetization vector in a sample. Consequently, analyzing the spectra one can determine the value of $\Theta$ using the following the formula:

$$\Theta = arccos[(4-I_{2,5}/(4-I_{2,5}]^{1/2} \qquad (18)$$

Here $I_{2,5}$ denotes the relative intensity of the 2-nd and 5-th line in the sextet.

For polycrystalline samples $\Theta$ should be regarded as an average angle.

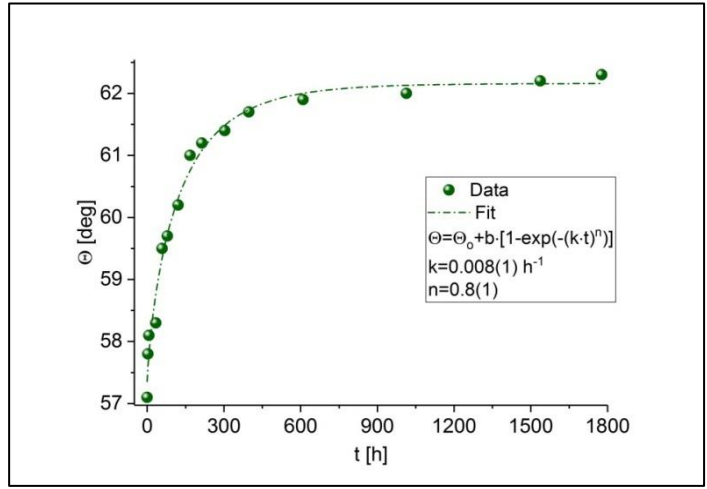

Fig. 71 Annealing time dependence of the average angle between the sample's magnetization vector and the normal to the sample's surface, $\Theta$. The best-fit to the data to the JMAK equation is shown as a dash-dot line. The best-fit kinetics parameters are displayed in the legend [59].

Figure 71 illustrates the $\Theta(t)$ dependence obtained from the analysis of spectra recorded on $Fe_{73.8}Cr_{26.2}$ isothermally annealed at 832 K for up to 1777 h [59]. It shows a saturation-like behavior, so the data could have been well fitted to the JMAK equation yielding the kinetics parameters $k$ and $n$ shown in the legend. This behavior reflects one process of the phase decomposition viz. nucleation and growth. The increase of $\Theta$ in the course of the annealing signifies a rotation of the magnetization vector towards the sample's surface. The maximum change of $\Theta \approx 5^o$. The effect can be associated with the preferential growth of magnetic domains during annealing.

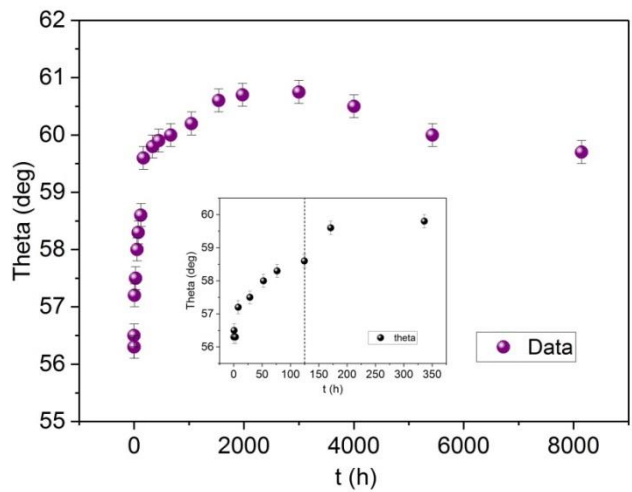

Fig. 72 Dependence of theta on the annealing time, *t*, derived from the spectra recorded on $Fe_{73.7}Cr_{26.3}$ isothermally annealed at 858 K for up 8144 h. The inset shows the behavior of theta within the first ~350 hours of the annealing [147].

Another example of the effect of annealing on $\Theta$ is displayed in Fig. 72. The plot again concerns the Fe-Cr alloy containing 26.2 at.% Cr but annealed at 858 K [147]. It can be easily noticed that in this case the $\Theta(t)$ behavior is more complex than the one shown in Fig. 71. Namely, $\Theta$ increases up to ~3000 h of annealing reaching the maximum of ~61° followed by a gradual decrease approaching ~60° for the maximum annealing time. This behavior reflects two different processes that occurred in the sample subjected to the annealing at 858 K viz. phase decomposition followed by a formation of σ [147]. The decrease of $\Theta$ was associated with the precipitation of σ because this phase starts to precipitate on grain boundaries, so the orientation of σ grains is random. Furthermore σ growth at the expense of the magnetic α-phase so with its growth the contribution of the magnetic α domains decreases. It has turned out that $\Theta$ can also be changed by irradiation. The effect was reported by Dubiel et al. who studied three Fe-Cr alloys containing 5.8, 10.75 and 15.15 t.% Cr which were irradiated with 25 keV $He^+$ to different doses [61]. The largest effect was discovered for the most Cr-concentrated alloy in which $\Theta$ decreased with the dose exponentially by ~13° in maximum. Dubiel and Żukrowski studied the influence of 250 keV $^4He^+$ ions on a $Fe_{84.85}Cr_{15.15}$ alloy [63]. As illustrated in Fig. 73, the values of $\Theta$ found from the irradiated side of the samples are significantly smaller than those determined from the non-irradiated sides. It is also clear that in the NIR-samples $\Theta$ had different values reflecting thereby a lack of homogeneity in the cold-rolled-induced texture of the studied samples is concerned. Consequently, in order to reveal a genuine effect of the irradiation on $\Theta$, a difference between $\Theta_{IR}$ and $\Theta_{NIR}$, $\Delta\Theta$, was calculated and is plotted in Fig. 74.

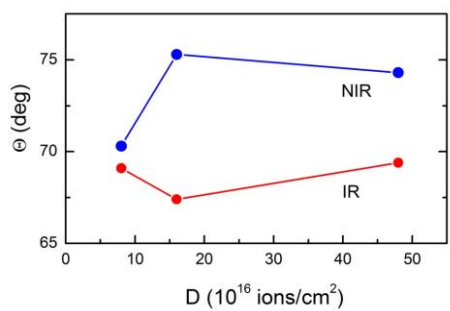

Fig. 73 Dependence of Θ on the dose of irradiation, D, for the non-irradiated (NIR) and irradiated (IR) side of the investigated sample of $Fe_{84.85}Cr_{15.15}$ alloy. The solid lines are a guide to an eye [63].

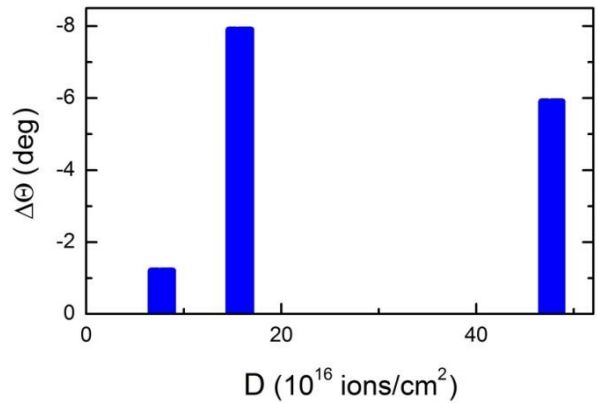

Fig. 74 The difference in Θ, ΔΘ=$Θ_{IR}$-$Θ_{NIR}$, for different doses of the irradiation [63].

## 7.2. Mechanical alloying

Fe-Cr alloys can be also synthesized by so-called mechanical alloying (MA). This technique has been often used to prepare alloys in form of powders, in particular those with nanometer grains. There is a great body of papers relevant to the matter in which the Mössbauer spectroscopy was used as a research tool e.g. [148-162]. MS being sensitive to local arrangements of atoms permits following the alloying process versus milling time. In this way chemical heterogeneities can be detected at various stages of MA. Fe-Cr alloys can be prepared by MA both in form of crystalline disordered bcc solid solutions e. g. [149,150,151] as well as in form of amorphous phase e. g. [152,153].

An example of the spectra for a 30:70 mixture of Fe and Cr powders recorded at RT after different milling time is displayed in Fig. 75. Clearly seen is the disappearance of the magnetic contribution characteristic of the Fe-rich phase in favor of the singlet indicative of the paramagnetic phase containing ≥ 70 at.% Cr [150].

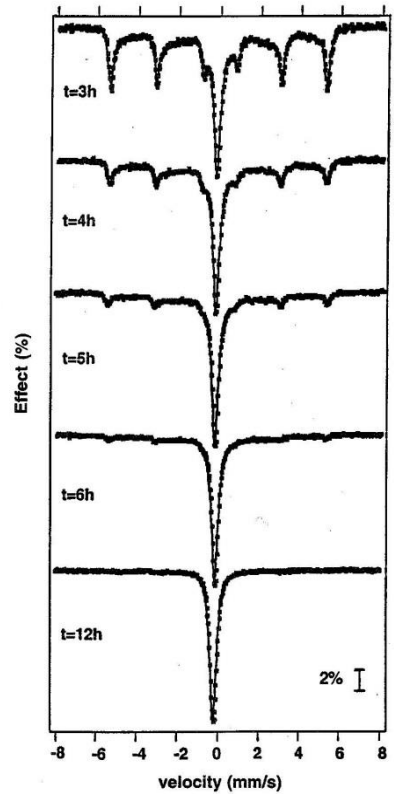

Fig. 75 $^{57}$Fe Mössbauer spectra recorded at RT on powder mixture of $Cr_{70}Fe_{30}$ mechanically milled for different periods shown [150].

Kinetics of the MA was investigated in a wide concentration range [151]. The authors revealed that after 30 h of milling a "stationary mixing state" was achieved i.e. the product was no more mixture of powder but a genuine alloy with average size of grains about 10 nm. Furthermore, based on the distributions of the hyperfine field, the authors concluded that the grains were compositionally homogeneous after 30 h of milling.

Costa et al. investigated a $Fe_{57.8}Cr_{42.2}$ alloy synthesized by MA of the components in Ar atmosphere for 100 h [158]. Based on the analysis of the Mössbauer spectra recorded at 300, 83 and 4.2 K and other applied techniques the authors concluded that the final product was structurally, chemically and magnetically heterogeneous. Two crystalline phases with different compositions viz. (1) $Fe_{55}Cr_{45}$ and (2) $Fe_{86.5}Cr_{13.5}$ and relative abundance of ~50% and ~15%, respectively, and one amorphous phase, but with two different crystallization temperatures, were detected. The existence of the two crystalline phases was interpreted in terms of milling-induced phase decomposition.

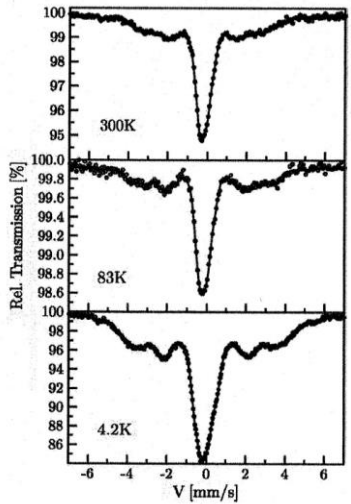

Fig. 76 $^{57}$Fe Mössbauer spectra recorded at different temperatures on the final product of the mechanical synthesis of the Fe$_{57.8}$Cr$_{42.2}$ [158[.

Finidiki et al. investigated the effect of milling time on a Fe$_{60}$Cr$_{40}$ nanocrystalline alloy [155]. They found that the alloy was 100% paramagnetic after 85 h of milling. For longer milling time a de-mixing process was revealed.

Dubiel and co-workers studied by means of several techniques including MS a series of nanocrystalline n-Fe$_{100-x}$Cr$_x$ alloys (0 $\leq$ x $\leq$ 83.8) prepared by mechanical alloying [159]. Mössbauer spectra were recorded in the TRANS mode in the temperature interval of 80-300 K. Examples of the spectra and determined therefrom histograms of the hyperfine field distributions are displayed in Figs. 77. The spectra were analyzed in terms of the hyperfine field distribution (HFD) method.

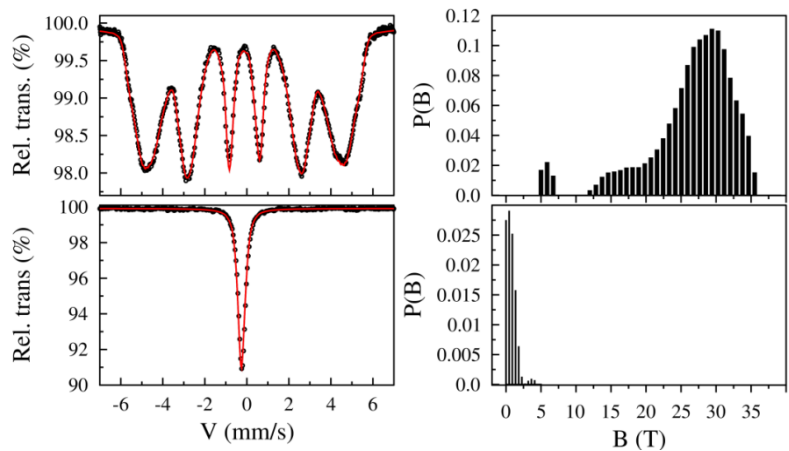

Fig. 77 (Upper panel) $^{57}$Fe Mössbauer spectra recorded on the n-Fe$_{83.2}$Cr$_{16.8}$ sample with the average size of grains <d>=5.1 nm (top), and that on the n-Fe$_{16.2}$Cr$_{83.8}$

sample with <d>=4.6 nm (bottom). The HFD histograms derived from the spectra are illustrated in the right-hand sides in each panel [159].

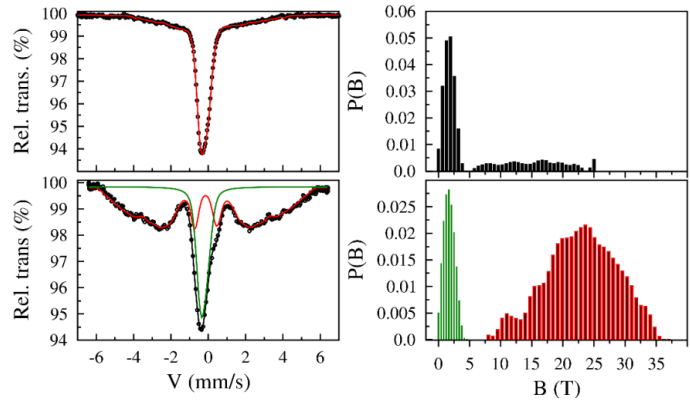

Fig. 77. (Lower panel) $^{57}$Fe Mössbauer spectra recorded on the n-Fe$_{55.5}$Cr$_{44.5}$ sample with the average size of grains <d>=6.5 nm (top) and <d>=10.5 nm (bottom). The corresponding HFD histograms are shown in the right-hand side [159].

The analysis of the spectra also yielded values of the average center shift, <CS>. Its temperature dependence,<CS>(T), was used for determination of the Debye temperature, $\theta_D$. The obtained values of $\theta_D$ were compared with the corresponding values of the Debye temperature found earlier for the bulk Fe-Cr alloys revealing an enhancement relative to bulk $\theta_D$-values in the range of ~40 ≤ x ≤~50 – see Fig. 78.

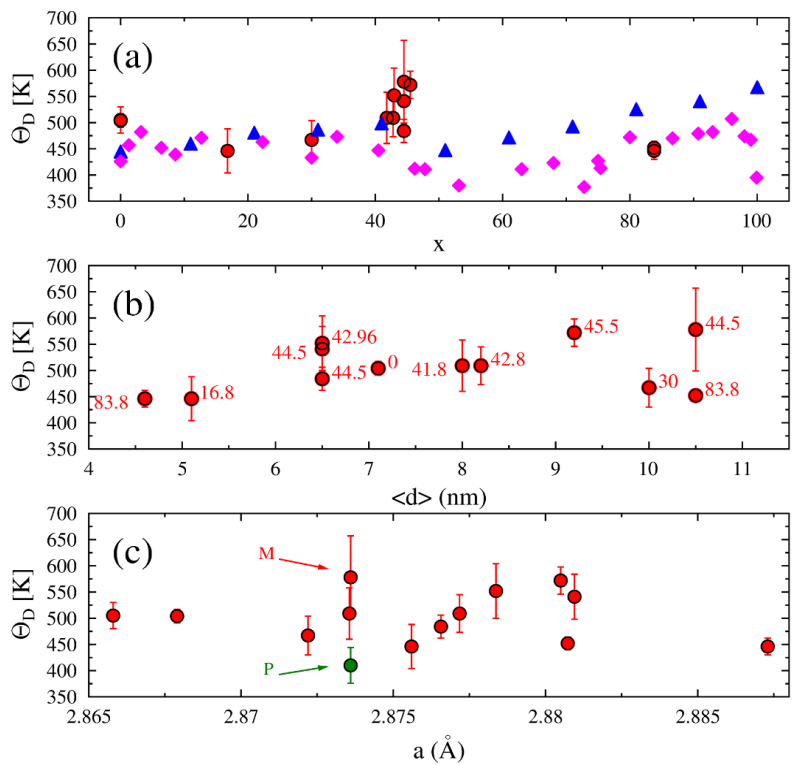

Fig. 78 The Debye temperature, $\Theta_D$, vs: (a) $x$ for the presently studied n-$Fe_{100-x}Cr_x$ samples (circles) and that found previously for the bulk Fe-Cr [154] (diamonds) and MA Fe-Cr alloys [155] (triangles), (b) $<d>$ and (c) $a$ for the n-$Fe_{100-x}Cr_x$ alloys. In the latter case the values of $\Theta_D$ derived from the paramagnetic (P) and the magnetic (M) sub spectra recorded on the n-$Fe_{55.5}Cr_{44.5}$ sample with $<d>$=10.5 nm are correspondingly indicated. Figure reproduced from [159].

Two phases were detected in the sample of n-$Fe_{55.5}Cr_{44.5}$ viz. (1) crystalline and magnetic with $\theta_D$=572(56) K, and (2) amorphous and paramagnetic with $\theta_D$=405(26) K [159].

Baldakhin and Cherdyntsev studied a series of $Fe_{100-x}Cr_x$ (5 ≤ $x$ ≤ 90) prepared by MA [160]. They concluded from the analysis of the Mössbauer spectra that the n-$Fe_{100-x}Cr_x$ alloys with $x$=5 and 10 were chemically and structurally homogenous, the alloy with $x$=20 was chemically heterogeneous i.e. it exhibited some features characteristic of Cr atoms clustering, whereas the alloys containing 35 and 50 at.% Cr underwent the phase separation into Cr-rich and Fe-rich phases. The n-$Fe_{10}Cr_{90}$ alloy was to ~93.5 % paramagnetic and ~6.5 % magnetic.